\documentclass[reqno,12pt]{article}
\usepackage{ae} 
\usepackage[T1]{fontenc}
\usepackage[ansinew]{inputenc}
\usepackage{amsmath}
\usepackage{graphicx}
\usepackage{color}
\definecolor{darkblue}{cmyk}{0.9,0.9,0,0}
\usepackage[colorlinks=true,linkcolor=darkblue,citecolor=darkblue,urlcolor=darkblue]{hyperref}
\usepackage{epsfig}

\newcommand{\beq}{\begin{equation}}
\newcommand{\eeq}{\end{equation}}
\newcommand\beqa{\begin{eqnarray}}
\newcommand\eeqa{\end{eqnarray}}
\newcommand\bea{\begin{array}}
\newcommand\eea{\end{array}}

\def\XXint#1#2#3{{\setbox0=\hbox{$#1{#2#3}{\int}$}
\vcenter{\hbox{$#2#3$}}\kern-.5\wd0}}

\newcommand{\nn}{\nonumber}

\newcommand{\diag}[1]{{\rm diag}(#1)}
\newcommand{\neqa}{\nonumber\end{eqnarray}}
\newcommand{\la}[1]{\label{#1}}

\newcommand{\Tr}{{\rm Tr}}

\renewcommand{\d}{\partial}

\newcommand{\<}{{\langle}}
\renewcommand{\>}{{\rangle}}

\newcommand{\cA}{{\cal A}}

\newcommand{\re}{\relax{\rm I\kern-.18em R}}

\renewcommand{\sp}{p\hspace{-.40em}/}

\def\su2{{SU(2)}}

\def\[{\left[}
\def\]{\right]}

\def\({\left(}
\def\){\right)}
\def\[{\left[}
\def\]{\right]}

\def\<{\langle}
\def\>{\rangle}

\def\i2{\frac{i}{2}}

\def\spi{\relax{\rm \pi\kern-0.5em /}}
\def\sA{\relax{\rm A\kern-0.5em /}}
\def\sp{\relax{\rm p\kern-0.5em /}}
\def\sd{\relax{\rm \d\kern-0.5em /}}
\def\sk{\relax{\rm k\kern-0.5em /}}
\def\sn{\relax{\rm n\kern-0.5em /}}
\def\sl{\relax{\rm l\kern-0.5em /}}
\def\sP{\relax{\rm P\kern-0.7em /}}
\def\sBethe{\relax{\rm \Bethe\kern-0.5em /}}

\def\bk{{\bf k}}
\def\bx{{\bf x}}

\usepackage{varioref}
\usepackage{makeidx}
\makeindex

\usepackage[english]{babel}
\begin{document}


\thispagestyle{empty}

\renewcommand{\thefootnote}{\fnsymbol{footnote}}
\setcounter{footnote}{0}
\setcounter{figure}{0}
\begin{center}
{\Large\textbf{\mathversion{bold} Y-system for Scattering Amplitudes }\par}

\vspace{1.0cm}

\textrm{Luis F. Alday$^a$, Juan Maldacena$^a$, Amit Sever$^b$ and Pedro Vieira$^{b}$}
\\ \vspace{1.2cm}
\footnotesize{

\textit{$^{a}$ School of Natural Sciences,\\Institute for Advanced Study, Princeton, NJ 08540, USA.} \\
\texttt{alday,malda@ias.edu} \\
\vspace{3mm}
\textit{$^b$
Perimeter Institute for Theoretical Physics\\ Waterloo,
Ontario N2J 2W9, Canada} \\
\texttt{amit.sever,pedrogvieira@gmail.com}
\vspace{3mm}
}


\par\vspace{1.5cm}

\textbf{Abstract}\vspace{2mm}
\end{center}

\noindent

We compute ${\cal N} =4$ Super Yang Mills planar amplitudes at
strong coupling by considering minimal surfaces in $AdS_5$ space.
The surfaces end on a null polygonal contour at the boundary of $AdS$.
We show how to compute the area of the surfaces as a function of the
conformal cross ratios characterizing the polygon at the boundary.
We reduce the problem to a simple set of functional equations for the
cross ratios as functions of the spectral parameter. These equations have
the form of Thermodynamic Bethe Ansatz equations. The area is the free
energy of the TBA system. We consider any number of gluons and in any
kinematic configuration.

\vspace*{\fill}

\setcounter{page}{1}
\renewcommand{\thefootnote}{\arabic{footnote}}
\setcounter{footnote}{0}

\newpage





 \def\nref#1{{(\ref{#1})}}


\tableofcontents

\section{Introduction}

In this paper we consider minimal area surfaces in $AdS$ space that end on a   null polygonal contour
at the boundary of $AdS$. Our goal is to compute the area of the surfaces as a function of the
shape of the contour.
Our solution to the problem consists of a system of integral equations of the thermodynamic Bethe ansatz (TBA)
 form \cite{TBApapers}.
The area is given by the TBA free energy of the system.

Our motivation for this investigation is the study of scattering amplitudes in ${\cal N}=4$
super Yang Mills. Planar ${\cal N}=4$ Super Yang Mills is an integrable theory
\cite{integrability}.  This means
that if one finds the appropriate trick, one is going to be able to perform computations
for all values of the 't Hooft coupling $\lambda$ \cite{integrability2,integrability3}.
Finding the appropriate trick is usually tricky.
Via the $AdS/CFT$ correspondence, this problem amounts to solving the quantum sigma model describing
strings in $AdS_5\times S^5$. The classical limit of this theory is simpler to analyze.
This is what we do in this paper. We consider classical solutions for strings moving in $AdS_5$.
In the classical limit we can forget about the worldsheet fermions and the five sphere and
study strings that are in $AdS_5$. We think that the knowledge of these classical solutions will be useful
for solving the full quantum problem. Classical solutions that were useful for the problem of operator dimensions
were considered in \cite{KazakovQF,ArutyunovVX} and several other papers. Here we consider classical solutions relevant
for scattering amplitudes or Wilson loops.

A scattering amplitude at strong coupling corresponds to a surface  that ends on the $AdS$ boundary
on a very peculiar polygonal contour \cite{AMFirst}. When we consider a color ordered amplitude involving $n$ particles
with  null momenta $\bk_1, \cdots, \bk_n$ we get the following contour.
The contour   is specified by its  ordered vertices $\bx_1, \cdots , \bx_n$, with $x^\mu_{i} -x^\mu_{i-1} = k_i^\mu $,
see figure \ref{polygon}.
The problem becomes identical to the problem of computing a Wilson loop with this contour.
In fact, we have a ``dual conformal symmetry'' which acts as the ordinary conformal symmetry on
the positions $\bx_i$ \cite{DrummondConformal}.
 The amplitude has a divergent part and  a finite part. The divergent part has a structure
 that is well understood \cite{BDS}. In addition, a piece of the finite part is also known
 \cite{BDS,DrummondAU}. There is an interesting
  finite piece which has not yet been computed in general. Two loop perturbative computations of this piece include
  \cite{BernAP,DrummondAQ,Anastasiou:2009kna,DelDuca:2009au}, and several subsequent papers.
The interesting part of the amplitude is
a function of conformal cross ratios of the $\bx_i$. If we have $n$ points we have $3(n-5)$ independent
cross ratios.
At strong 't Hooft coupling we can compute this in terms of the area of the minimal surface that ends on the
polygonal contour \cite{AMFirst}.

Our method will use integrability of the sigma model in the following way.
First we define a family of flat connections with a spectral parameter $\theta$.
Sections of this flat connection can be used to define solutions which depend on the
spectral parameter $\theta$. With these, we define a set cross ratios $Y_k(\theta)$.
 We find a functional $Y$ system that constrains the $\theta$ dependence of the
functions $Y_k$.
This system has $3(n-5)$ ``integration constants" which come in when we specify the boundary
conditions for $\theta \to \pm \infty$. We can restate these functional equations in terms of
integral equations, where the $3(n-5)$ parameters appear explicitly.
These integral equations have a TBA form. Schematically they are
\beqa
 \log Y_k(\theta ) =  - m_k \cosh \theta + c_k + K_{k,s} \star \log ( 1 + Y_s )\nn
 \eeqa
 where the $m_k$ and $c_k$ are the $3(n-5)$ parameters we mentioned above and $K_{r,s}$ are some kernels.
 \footnote{  There are $2n-10$ complex "masses" $m_k$ and $n-5$ "chemical potentials" $c_k$ with a precisely reality.}
Moreover, the area has an expression in terms of
the TBA free energy of the system.
\beqa
 {\rm Area}  =  \int {  d \theta \over 2 \pi }  m_k \cosh \theta \log (1 + Y_k(\theta) )\nn
 \eeqa
   Evaluating $Y_k$ at $\theta =0$
  we get the physical values of the cross ratios. However, we can view other values of $\theta$ as a one parameter
  family
  of cross ratios which give the same value for the area. Thus, changing $\theta$  generates
  a symmetry of the problem.

 The case involving a six sided polygon was treated in
 \cite{Alday:2009dv} and the octagon, in a particular kinematic subspace,
 was considered in \cite{Alday:2009ga,Alday:2009yn}.
 Using this method,  the area is computed without
  finding the explicit shape of the minimal surface.

\begin{figure}[t]
\begin{center}
\includegraphics[width=120mm]{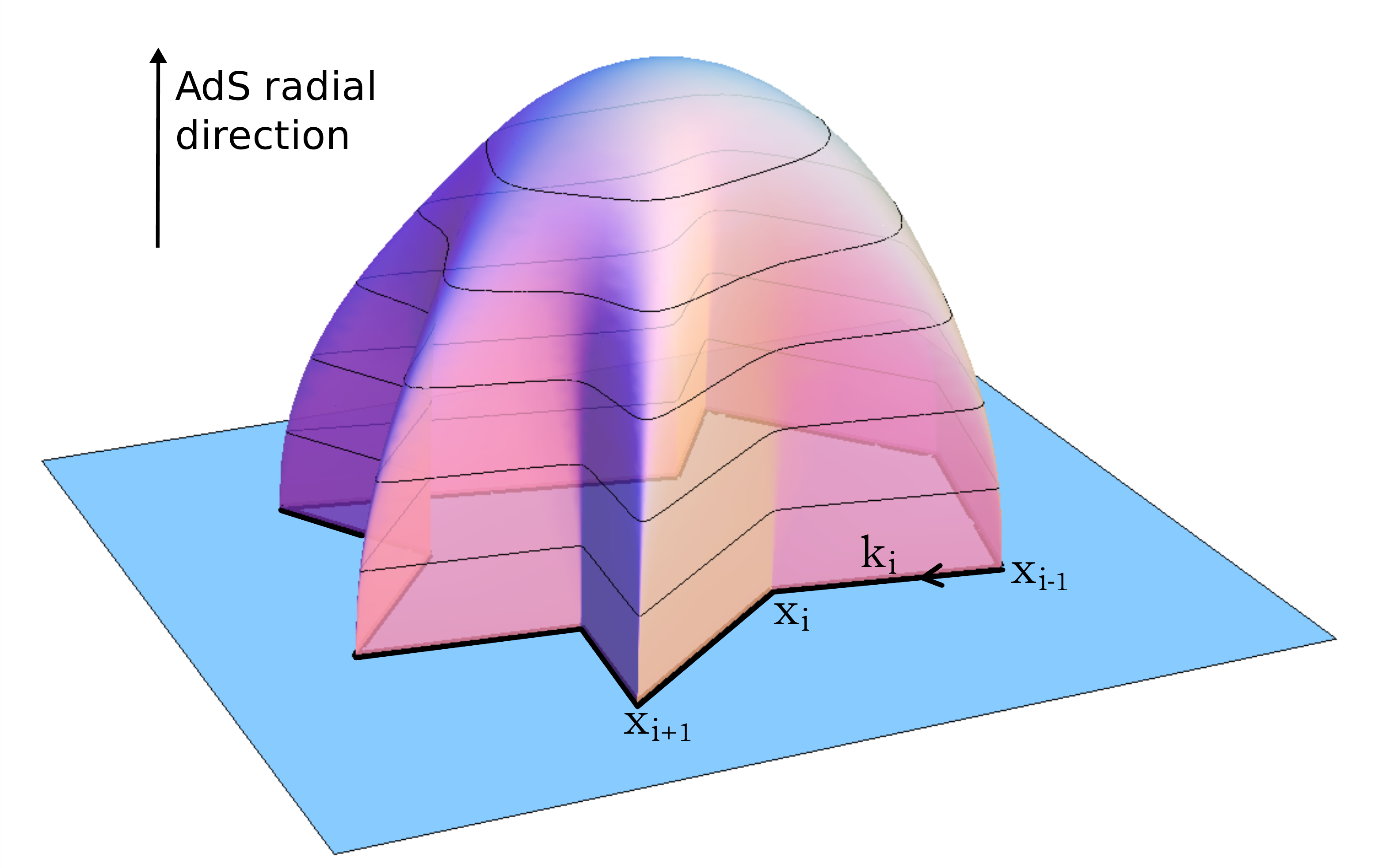}
\end{center}
\caption{The polygon is specified at the $AdS$ boundary by the positions of the cusps $\bx_i$. These positions  are related to an ordered sequence of momenta $\bk_i$ by $\bk_i = \bx_{i} - \bx_{i-1}$. The two dimensional minimal surface streches in the $AdS$ bulk and ends on the polygonal contour at the boundary.
} \la{polygon}
\end{figure}

Our paper is organized as follows.
In section two we recall the connection between sigma models which obey the Virasoro constraints and
Hitchin equations. In section three we discuss the case where the minimal surfaces are embedded in $AdS_3$.
This is a warm up problem, which is simpler than the general problem.
In section four we   solve the full $AdS_5$ problem. We derive the $Y$ system, the integral equations and the area.
We perform some checks. We also compute the exact answer for a one parameter family of regular polygons.
Finally, we present some conclusions.  We also have several appendices with useful details.

\section{The classical sigma model and Hitchin equations }

The classical $AdS_5$ sigma model is integrable.
This can be shown by exhibiting a one parameter family of flat
connections. For our problem, it will be convenient to choose
this one parameter family in a special way which will simplify
its asymptotic behavior on the worldsheet. In fact, to make
this choice we will make use of the Virasoro constraints of the
theory. This has been explained in detail in previous papers
\cite{PohlmeyerNB,DeVegaXC,JevickiAA,GrigorievJQ,MiramontesWT,dornrecent}.
Instead of repeating the whole discussion, we will present a
slightly more abstract and algebraic version here.

\subsection{General integrable theories and Hitchin equations}

Let us assume that we have a   coset space $G/H$. Let us assume
that the Lie algebra $\mathcal{G}$ has a $Z_2$ symmetry that
ensures integrability. In other words, imagine that the Lie
algebra has the decomposition $\mathcal{G} = \mathcal{H} +
\mathcal{K}$ so that   $\mathcal{H}$ is left invariant under
the action of the $Z_2$ generator while elements in
$\mathcal{K}$ are sent to minus themselves. We then write the
$G$ invariant currents $J = g^{-1} d g $. This is a flat
current $dJ + J \wedge J=0$.
 We can decompose $J$ in terms its components along $\mathcal{H}$ and $\mathcal{K}$ as
\beqa \label{flatjdec} J = g^{-1} d g = H + K \eeqa When we
gauge the sigma model we add a gauge field along $\mathcal{H}$,
and we can do local $H$ gauge transformations. The equations of
motion of the system can be written in terms of the $H$-gauge
invariant currents $k = g K g^{-1}$ as $d * k =0$. Notice that
$k$ are the Noether currents of the problem. These equations of
motion together with the flatness condition for $J$ lead to
\beqa \label{hitch}
 && D_z K_{\bar z} =0 = D_{\bar z} K_{z} ~,~~~~~~~~~~~~ [D_z,D_{\bar z} ] + [K_z,K_{\bar z} ] =0
\\
 && {\rm where } ~~~~~D_z X \equiv  \partial_z X+ [ H_z , X ]
 \eeqa
 We can view these as equations for the connection.  Once we solve these,
  we can find  a coset  representative by
 solving the flatness condition
  \beqa \label{flatcondj}
  ( d + J ) g^{-1} = ( d + H + K ) g^{-1} = 0
  \eeqa
  More precisely, we start with a set of independent vector solutions to the equation
   $  ( d + J ) \psi=0$, orthonormalize them, and assemble
  them into $g^{-1}$.  These vector solutions are called flat sections.
  The global $G$-symmetry acts by left multiplication of $g$ and the equations \nref{hitch} are $G$ invariant.
 The  equations \nref{hitch}
 are identical to Hitchin's equations after the identification $\hat \Phi_z = K_z$, $\hat \Phi_{\bar z} =
 K_{\bar z } $, $A= H$ \footnote{Actually, to be a bit more precise, we have a $Z_2$ projection of the
 Hitchin problem based on $G$ by the $Z_2$ symmetry we considered above. Namely, we project on to
 $ \Phi = - s(\Phi)$, and $A = s(A)$ where $s$ is the $Z_2$ transformation that multiplies the elements of ${\cal K}$ by
 minus one. }.
 These equations are equivalent to the flatness of the one parameter family of connections  
 \beqa  \label{flatco}  d + \hat  {\cal A}(\zeta) ~,~~~~~~~~{\rm with} ~~~
 \hat {\cal A}(\zeta)  =
  { K_z dz \over \zeta^2 } + H + { \zeta^2 \, K_{\bar z } d \bar z  }
 \eeqa
 Flat sections of this connection, at $\zeta =1$, give back the group element $g^{-1}$.
 This connection differs from the connection that is often written (e.g. in \cite{KazakovQF})
  by a gauge transformation by the
 group element $g$. Though we will not need it here, let us quote the more usual form of the flat connection
 \beqa
  d + a ~,~~~~~~~{\rm with } ~~~~~~~~a =  g \hat {\cal A}g^{-1} -  d g  g^{-1 }  =
  k_{z}dz ( {  1 \over \zeta^2} -1) + k_{\bar z} d \bar z ( \zeta^2 - 1 )\nn
  \eeqa
  We did not find this form of the flat connection particularly useful for our purposes.

  The equations \nref{hitch} imply that $T(z) =\Tr[K_z^2]$ is holomorphic. This is the usual holomorphicity
  of the stress tensor. For the $SO(n+1)/SO(n)$ type cosets that we are interested in, higher traces
  of $K_z$ vanish, so we do not obtain any other interesting holomorphic quantities. In particular,
  if we are considering a theory obeying the Virasoro constraints, $T=0$, then we do not appear to
  get any interesting holomorphic quantities in this fashion.

  \subsection{Integrable theories with Virasoro constraints and Hitchin's equations}

  As we mentioned above, in the case that $T=0$, we must work a bit harder in order to
  obtain interesting holomorphic quantities. In fact, it is possible to choose a slightly different
  form (or different gauge) for the connection so that we obtain a more interesting Hitchin system.
  This is a small variant of the Pohlmeyer type reduction. For the case with non-zero stress tensor this was
  described in \cite{PohlmeyerNB,GrigorievJQ,MiramontesWT}.
  In the particular case that is going to be of interest to us, which is the
  SO(2,4)/SO(1,5) or $AdS_5$ sigma model this was done in \cite{DeVegaXC,dornrecent,Alday:2009dv}.
  Since we do not want to repeat those derivations here, let us give a more abstract perspective on it.

 We consider cosets of the form  $G/H = SO(n+1)/SO(n)$, or $SO(2,n)/SO(1,n)$.
  Probably, similar considerations are true for other cosets  but we have not checked the details.
 We now have the Virasoro constraints $\Tr[K_z^2]=0$ and $\Tr[K_{\bar z}^2]=0$.
 We assume that $ \Tr[ K_z K_{\bar z}]$ is generically non-zero. This quantity is
  the action density or the area element, so it will be non-zero for our solutions.
   We can then think of $K_z$ and $K_{\bar z}$ as spanning a
  two dimensional subspace of ${\mathcal K}$. We consider a generator $q$ in ${\mathcal H}$ such that $K_{z}$ has charge +1
  and $K_{\bar z}$ has charge -1 under $q$. In other words, we view $K_z$ and $K_{\bar z}$ as two lightcone directions
  in the Lie algebra, and $q$ is the ``boost" generator.
  We can further split the lie algebra ${\mathcal H}$ according to the charges under $q$.
  In our case, we have  $\mathcal{H}^{(0)}$, $\mathcal{H}^{(1)}$ and $\mathcal{H}^{(-1)}$, where the superscript
  indicates the charge under $q$.
  We then take \nref{flatco}  and
  make a global gauge transformation by $\zeta^q$.
  We obtain\footnote{
 In this derivation we have used that $H_{{\bar z}}^{(-1)} =0 = H_{z}^{(1)}$. This follows from
  \nref{hitch} plus the condition
 that $K_z$ is non-vanishing (and the $q$-charges of $H$ and $K$).}
  \beqa \label{flatfin}
 {\cal A}  =  \zeta^q \hat {\cal A} \zeta^{-q} = { 1 \over \zeta} ( K_z + H^{(-1)}_z )dz +
   H_0 + \zeta ( K_{\bar z} + H^{(1)}_{\bar z } )d \bar z   \equiv {  \Phi_z dz \over \zeta } + A + \zeta \,
   \Phi_{\bar z } d \bar z
 \eeqa
 This is the final form of the flat connection that  we will use.
We saw that it is  a simple transformation of the previous one.
Moreover, when $\zeta =1$ the gauge transformation
 is trivial and the flat sections of this connection are still giving us the solution $g^{-1}$, as in \nref{flatcondj}.
  One nice aspect is that now $   P(z) \equiv  { 1 \over 4 } Tr[\Phi_z^4]$ is a non-vanishing holomorphic current.
  One can wonder why we have a spin four holomorphic current.
  In general,  the
  integrable theory has  higher spin conserved currents. These higher spin currents are usually not holomorphic.
   When the stress tensor vanishes, the spin four
  current becomes holomorphic. In terms of embedding coordinates with $X^2 =-1$, we have $P \propto
   \partial^2 X \cdot \partial^2 X $.
  We also have that $Tr[\Phi_z^r] =0$ for $r < 4$.
  Finally, note that if we start from a general $SO(n)$ Hitchin equation, we can specialize into \nref{flatfin} by
  performing a $Z_4$ projection generated by the product of the $Z_2$ transformation we had above times a conjugation
  by $(i)^q$, where $q$ is the $U(1)$ generator we discussed above.
   This combined generator, let us call it $r$,  should then give $r(\Phi_z) = -i \Phi_{z}$,
  $r(\Phi_{\bar z }) = i \Phi_{\bar z} $, $r(A) = A$. In section \ref{ads5pre}
   we give a more explicit form for this generator.

  In our case, we will further use the relation between $SO(2,4)$ and $SU(2,2)$ in order to write an $SU(2,2)$ flat
  connection. If we denote by $\psi$ the flat sections of the $SU(2,2)$ connection, then
  anti-symmetric products of two different sections $\psi$ and $\psi'$ will give a flat section in the vector
  representation of $SO(2,4)$. Schematically $q^A = (\Gamma^{A})^{\alpha \beta} \psi_{[\alpha} \psi'_{\beta]}$, where
  $\psi$ and $\psi'$ are two solutions of the problem in the spinor (or fundamental of $SU(2,2)$) and $q^A$ is a
  solution in the vector representation of $SO(2,4)$.

  Note that the action for the problem, which is equal to the area, is given by
  \beqa
  A &=& \int d^2 z  Tr_{SO(2,4)}[ K_z K_{\bar z}] = 2 \int d^2 z  Tr_{SU(2,2)}[ K_z K_{\bar z}] \nn
  \\  \label{areaform}
  A &=&  \int d^2 z
  Tr[ \Phi_z \Phi_{\bar z}]  + {\rm total~ derivative}
  \eeqa
  The total derivative term is a constant proportional to the degree of the polynomial $P$ (and independent
  of the kinematics). In order to show the last equality in \nref{areaform} we can take the trace of the generator
  $q$ times the second equation in \nref{hitch} and use the Jacobi identity. (Alternatively, one can show it
  via an explicit parameterization as in \cite{Alday:2009dv}.)

  Once we compute this geometric area we can compute the amplitude, or the Wilson loop expectation value as
  \beqa
  {\rm Amplitude } \sim \langle W \rangle \sim e^{ - { R^2 \over 2 \pi \alpha' }A } = e^{ - {
   \sqrt{\lambda} \over 2 \pi } A }\nn
   \eeqa
   Here $A$ is the geometrical area of the surface
    in units where the radius of AdS has been set to one.
   This area is infinite, but it can be regularized in a well understood
   fashion. The central object of this paper is certain
   regularized area, defined by
   \begin{equation}\la{regarea}
   A_{reg}=\int d^2 z \left(
  Tr[ \Phi_z \Phi_{\bar z}]- 4  (P \bar P)^{1/4} \right)
   \end{equation}
namely, we subtract the behavior of $Tr[ \Phi_z \Phi_{\bar z}]$
far away. Since (\ref{regarea}) is invariant under conformal
transformations, it is a function of the cross-ratios. When
using a physical regulator, the area will have additional
terms. These additional terms are well understood and described
in appendix \ref{areacomponents}.

  \subsection{Flat sections, Stokes sectors and cross ratios}

  In this subsection we recall some facts, which were discussed in more detail in \cite{Alday:2009dv}.
  For the amplitude problem the worldsheet is the whole complex plane and $P(z)$ is a polynomial.
  We then study the problem $(d + {\cal A}(\zeta) ) \psi =0$.
  As we go to  large $z$  some flat sections $\psi$ will diverge and some will go to zero.
  The fact that some diverge means that
  the worldsheet goes to the boundary of $AdS$ space.
  For large $z$, the boundary conditions are such that
 we can simultaneously diagonalize $\Phi(z) \sim P(z)^{1/4} {\rm diag}(1,-i,-1,i)$ and
  $\Phi_{\bar z} \sim \bar P^{1/4}(\bar z)
 {\rm diag}(1,i,-1,-i)$. The particular relation between
  eigenvalues is determined by the $Z_4$ symmetry of the problem.
  This determines the
  large $z$ asymptotics  of the four solutions\footnote{$A(z)$ in (\ref{flatfin}) decays as $1/z$ for large $z$ and therefore can be 
  dropped when considering the leading asymptotics that determines the Stokes sectors \cite{Alday:2009dv}. 
  We will have to keep $A$ when we approximate the cross ratios.}
  \beqa
  \psi_a \sim  \exp\left\{ -  i^{-a} { \int P^{1/4}(z) d z \over \zeta}  - i^{a} \zeta \int P^{1/4}(\bar z) d \bar z
  \right\} ~,~~~~~~~a=0,1,2,3\nn
  \eeqa
    The problem displays the Stokes phenomenon at large $z$.
   This means that the previous behavior of solutions is only valid within a given
   Stokes sector. The number of Stokes sectors is determined by the degree of the polynomial. Namely, for large $z$
   we have $\int^z dz' P(z')^{1/4} \sim z^{n/4} + \cdots $, for a polynomial of degree $ n-4$.
    In order to characterize the problem it is convenient to choose the smallest  solution $s_i$ in each of the Stokes
  sectors. This smallest solution is well defined up to an overall rescaling.
   Given any four flat sections, it is possible to construct a gauge invariant inner product  
   $ \langle \psi_1  ,\psi_2, \psi_3, \psi_4 \rangle \equiv \epsilon^{\alpha \beta \gamma \delta }  \psi_{1 \alpha}\psi_{2 \alpha }\psi_{3 \gamma }\psi_{4 \delta } $.
   This inner product is independent of the position where we compute it.
  A full solution of the problem is given by choosing four arbitrary flat sections $\psi_{1}, \cdots \psi_4$ where
  the subindex runs over the four solutions, but each of them is a four component spinor.   We will use greek letters for spinor components and latin letters for labeling different solutions.
  The target space
  conformal group $SU(2,2)$ acts on the latin indices, but not on the greek indices where the flat connection acts. 
   The spacetime embedding coordinates $X^I $ are given in terms of these solutions at $\zeta=1$. More explicitly,  
    $X^I \Gamma^I_{ab} =
   M^{\alpha \beta} \psi_{a\alpha} \psi_{b\beta,}$ where $M$ is a fixed matrix and $a,b$ are spacetime indices.
   As we go to large $z$, some of the solutions diverge. The particular combination of solutions that form the
   two solutions that diverge most rapidly, determines a ray $X^I$. This maps to a point
   on the boundary of $AdS_5$ space. We can find this point in a convenient way by picking the two smallest solutions
   which will be $s_i$ and $s_{i+1}$,  if we are between Stokes sectors $i$ and $i+1$. Then the spacetime direction
   is obtained by taking
   \beqa \label{divX}
    X^i_{ab} \propto \langle \psi_a, \psi_b ,s_i ,s_{i+1} \rangle
    \eeqa
    This determines the direction in which   $X^i_{ab}$ is diverging.
    Recall that we can think of the boundary of $AdS$ as a projective space, given by six coordinates $\hat X^I$, with
    $\hat X^2 =0$ and $\hat X^I \sim \lambda \hat X^I$. Thus the diverging solution determines a point $ \hat X$ in
    projective space, which is the same as saying that it determines a point on the boundary of $AdS$.

    The index $i$ labels the cusp number.
    We can form quantities of the form
    $X^i \cdot X^j \propto \langle   s_i ,s_{i+1},  s_j ,s_{j+1} \rangle $.
    Finally, cross ratios are given by quantities of the form
    \beqa\la{CrossRatios}
     Y_{ijkl}  = { X^i\cdot X^j \, X^k\cdot X^l \over  X^i \cdot X^k \, X^l\cdot X^j } =
     { \langle   s_i, s_{i+1},  s_j ,s_{j+1} \rangle \langle   s_k ,s_{k+1},  s_l, s_{l+1} \rangle
     \over  \langle   s_i, s_{i+1} , s_k ,s_{k+1} \rangle  \langle s_l ,s_{l+1}  ,s_j ,s_{j+1} \rangle }
     \eeqa
   The cross ratios
   do not depend on the normalization of each of the $s_i$.

    These cross ratios are functions of the spectral parameter $\zeta$. In what follows, we will choose
    a convenient basis of cross ratios and study their $\zeta $ dependence. We will write an integral equation
    determining the values of the cross ratios as a function of $\zeta$. Finally we will express the area in terms
    of certain integrals of the cross ratios over $\zeta$.

\section{ Minimal surfaces in $AdS_3$}

 We first consider minimal surfaces that can be embedded in an $AdS_3$ subspace of $AdS_5$. This is a simpler
 problem that illustrates the method that we will use in the $AdS_5$ case. The reader that is only interested in
 the $AdS_5$  case can jump directly to the next section.

\subsection{$AdS_3$ preliminaries}

 When the surface can be embedded in $AdS_3$ the problem simplifies and it   reduces to
  a $Z_2$ projection of an $SU(2)$ Hitchin problem.
  The derivation of this fact is rather similar to what we discussed above and was treated in detail in
  \cite{Alday:2009yn}. We will not repeat the derivation, but we will state the final results.
We have a polynomial $ p = \frac{1}{2}Tr[\tilde \Phi_z^2]$ whose degree determines the
  number of cusps. \footnote{Note that $P  \propto p^2$, where $P= { 1 \over  4 } \Tr(\Phi_z^4)$ in the $AdS_5$ polynomial discussed in the previous section.} We now have $SU(2)$ quantities
  $\tilde \Phi_z , ~\tilde A, \tilde \Phi_{\bar z} $ which are in the adjoint of $SU(2)$ and we have the
  $Z_2$ projection condition
  $ \tilde \Phi_{z } = - U \tilde \Phi_{z } U^{-1}$,  $ \tilde \Phi_{  \bar z} = - U\tilde
  \Phi_{  \bar z } U^{-1}$ and $\tilde A = U \tilde A U^{-1}$ where $U = \sigma_3 $ is the
  usual Pauli matrix.
  This restricts the components of $\tilde A$ and $\tilde \Phi$ that are non-zero.
  General SU(2) Hitchin problems were studied in \cite{GMNtwo} and we are now considering a special
  case of their discussion, though we will rederive some of their formulas in a different way.
  We  study sections of the flat connection which obey
  \beq
   ( d + { \tilde \Phi_z  dz \over \zeta} + \tilde A +  \tilde \Phi_{\bar z } d \bar z \, \zeta ) \psi(\zeta) = 0\nn
   \eeq
  The $Z_2$ symmetry
  relates solutions $\psi(\zeta)$ with different values of the spectral parameter. Namely, if
  $\psi(\zeta)$ is a flat section with spectral parameter $\zeta$, then
   $ \eta(\zeta ) \equiv  U \psi( e^{i \pi } \zeta) $ is a solution of the problem with spectral parameter $  \zeta$.
   We can track how the small solutions change as we change $\zeta$ by looking at them in the large $z$ region.
   In a given Stokes sector  the  small solution contains a factor behaving
    as $e^{- \int^z \sqrt{ p} dz' \over \zeta}
   \sim e^{ - z^{n/4 } \over \zeta} $, where $n$ is determined  by the degree of the polynomial and is
   equal to the number of cusps of the polygon. $n$ is even. There are $n/2$ Stokes sectors and
   thus $n/2$ small solutions
   $s_i$.
    As we change the phase of $\zeta$, the ray in the $z$ plane where this solution is smallest
   rotates accordingly.
      In particular, if we start with the solution $s_i(\zeta)$,  which is the  small in the $i$th Stokes sector,
     we find that
 $s_{i}(e^{ 2 \pi i } \zeta ) \propto  s_{i+2}(\zeta)$ and   $U s_i( e^{i \pi } \zeta ) \propto   s_{i+1}(\zeta)$.
    Note that the solutions do not come back to themselves after a shift by $e^{ 2 \pi i } \zeta$.
   We can choose a solution $s_1$ in the first Stokes sector and
   define all others as $s_{j} = U^{j-1} s_{1}(  e^{ j i \pi } \zeta ) $.
    Then, as we go around we  have that $s_{{n \over 2 } +1} = A(\zeta)  s_{1} $.
 $A(\zeta)$ can be set to one when $n/2$ is odd. When $n/2$ is even it has a simple form that we will discuss
   later.

   The full connection with spectral parameter is an $SL(2)$ connection and thus we can form an $SL(2)$ invariant
   product $\langle \psi \psi' \rangle $ with two solutions.
   Now we have that
   \beqa
    \langle s_i ,s_j \rangle(e^{i \pi } \zeta)   =  \langle s_{i+1} ,s_{j+1} \rangle(  \zeta) \la{ads3shift}
   \eeqa
    We can normalize
   $s_1$ so that $\langle s_1 ,s_2 \rangle =1$. Then \nref{ads3shift}
    also implies that $\langle s_i ,s_{i+1} \rangle =1$.

\begin{figure}[t]
\center\includegraphics[width=70mm]{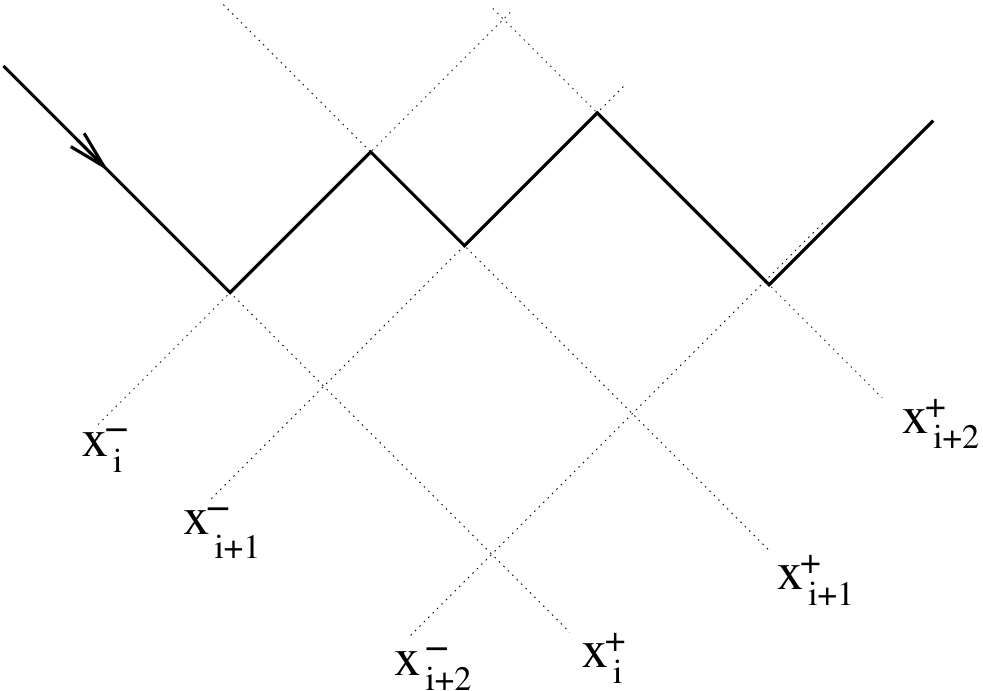}
\caption{ Spacetime positions of the cusps for a polygon that is embedded in $R^{1,1}$, which is the boundary
of $AdS_3$. The positions of the cusps are given by a set of $n/2$ values $x^+_i$ and  a set of $n/2$ values  $x^-_i$.
} \la{AdSthreecusps}
\end{figure}

   We can form  cross ratios by forming quantities like
   \beqa \la{ads3cr}
    \chi_{ijkl}(\zeta)&=& { \langle s_i ,s_j \rangle
   \langle s_k ,s_l \rangle \over \langle s_i ,s_k \rangle
   \langle s_j ,s_l \rangle }
   \eeqa
   These quantities do not depend on the arbitrary normalization of the $s_i$. By construction they
   are also invariant under the conformal symmetries of $AdS_3$.
   They can be related to the conformal invariant cross ratios formed from the positions of the cusps of
   the polygon. Recall that a polygon in $AdS_3$ is given by $n/2$ positions $x_i^+$ and $n/2$ positions $x^-_i$,
   see figure \ref{AdSthreecusps}. We can form spacetime cross ratios from the positions of the points $x_i^\pm$.
   These spacetime cross ratios can be expressed in terms of the  cross ratios
   in \nref{ads3cr} as
   \beqa
    \chi_{ijkl}(\zeta=1) & = &  { x^+_{ij} x^+_{kl} \over x^+_{ik} x^+_{jl} } \la{rightCR}
    \\
   \chi_{ijkl}(\zeta = i ) & = &  { x^-_{ij} x^-_{kl} \over x^-_{ik} x^-_{jl} } \la{leftCR}
   \eeqa
\subsection{The $AdS_3$ functional Y-system} \la{YsystemAdS3}
We will now derive a set of functional equations for the inner products, or Wronskians,
 $\<s_i, s_j\>(\zeta)$ made out of two small solutions of the linear problem.
The starting point is the Schouten identity, $\<s_i ,s_j\> \< s_k, s_l\>+\<s_i ,s_l\> \<s_j ,s_k\>+\<s_i ,s_k\> \<s_l ,s_j\>=0$, applied to a particular choice of small solutions:
\beq \label{shid}
\<s_{k+1}, s_{-k}\> \< s_{k} ,s_{-k-1}\> = \<s_{k+1},s_{-k-1}\> \< s_{k},s_{-k}\> + \<s_{k},s_{k+1}\> \< s_{-k-1}, s_{-k}\> \,.
\eeq
In our normalization the last
two brackets are equal to one. Using (\ref{ads3shift}) we see that this identity becomes the $SU(2)$ Hirota equation
\beqa\la{HirotaLine}
 && T_{s}^+T_{s}^- = T_{s+1}T_{s-1}+1 \,,
\\
 && {\rm where: } ~~~~~~~
T_{2k+1}=\<s_{-k-1}, s_{k+1}\> ~,~~~~~~~~~~T_{2k}=\<s_{-k-1} ,s_{k}\>^+   \label{Tdef}
\eeqa
or more uniformly $T_s = \langle s_0 ,s_{s+1} \rangle( e^{ - i (s+1) \pi/2 } \zeta)$.
The superscripts $\pm$ indicate a shift in spectral parameter,
 $f^{\pm} \equiv f(e^{\pm i{\pi\over2}}\zeta)$. Actually, from \nref{shid} we get \nref{HirotaLine} for $s=2k$. For $s$
 odd we need to start from a slightly different choice of indices in \nref{shid}.
  $T_s$ is non-zero for $s=0,\dots,n/2-2$.
Finally, we introduce the $Y$-functions $Y_{s} \equiv T_{s-1}T_{s+1}$.
 Being a product of two next-to-nearest-neighbor $T$-functions,
 the $Y$-functions are non-zero in a slightly smaller lattice parametrized by $s=1,\dots,n/2-3$.
 The number of $Y$-functions coincides with the number of independent cross ratios.

The Hirota equation \nref{HirotaLine}  implies the  Y-system for these new quantities:
\beq
\la{Yads3}
Y_{s}^+ Y_s^-=(1+Y_{s+1})(1+Y_{s-1}) \,.
\eeq
These equations are of course not enough to fix the Y-functions.
 After all they came from a trivial determinant identity without
 any information on the dynamics! To render them
 more restrictive we need to supplement them with the analytic properties of the
  Y-functions. This will then pick the appropriate solutions to these equations.
  Furthermore, to make  these solutions useful   we must relate them
  to the actual expression for the area. Before considering these points
  let us comment on some general properties of Hirota equations and their corresponding Y-systems.

\subsection{Hirota equation, gauge invariance and normalization of small solutions} \la{GenHirota}

The general form of the Hirota system of equations --
which generalizes the $SU(2)$ case derived above --
 is a set of functional equations for functions
 $T_{a,s}(\zeta)$.\footnote{Typically these relations arise in the
 study of quantum integrable models and describe the fusion relations
  for the eigenvalues $T_{a,s}$ of  transfer matrices in rectangular representations
   parametrized by Young tableaus with $a$ rows and $s$ columns   \cite{Kuniba}.}
The  indices $a,s$ take integers values and can be thought of as parametrizing
a two dimensional lattice. At each point of this lattice we have a function $T_{a,s}(\zeta)$ of the spectral parameter $\zeta$. Then, for each site $o=(a,s)$ we have an Hirota equation
\beq
T_o^+ T_o^-=T_\leftarrow T_\rightarrow + T_\uparrow T_\downarrow \la{Hirota}
\eeq
involving the function at that site and the four $T$-functions at the four nearest-neighbor sites,
 $T_{\rightarrow}\equiv T_{a,s+1}$, $T_{\uparrow}=T_{a+1,s}$, etc. Recall that $T_o^\pm = T_o(e^{\pm i \pi/2} \zeta)$.
This equation has a huge gauge redundancy
\beq
T_{a,s} \to \prod_{\alpha,\beta=\pm } g_{\alpha \beta}\(e^{\frac{i\pi}{2} (\alpha a +\beta s) }\zeta\) T_{a,s}(\zeta)\nn
\eeq
where $g_{\alpha\beta}(\zeta)$ are four arbitrary functions. It is therefore instructive to construct a set of gauge invariant quantities
\beq
Y_o=\frac{ T_{\leftarrow} T_{\rightarrow} }{   T_\uparrow T_\downarrow  }  ~~~~~~~~~~~~{\rm or}
 ~~~~~~~Y_{a,s} = { T_{a,s-1} T_{a,s+1}
\over T_{a+1,s} T_{a-1,s} }
 \la{Ydefinition}
\eeq

It is  instructive to think of the gauge invariant quantity $Y_o$  as a field strength made of the gauge dependent gauge field $T_o$.
Suppose the $T$-functions are non-zero in some rectangular domain in the $(a,s)$ lattice.\footnote{Strictly speaking the $T$-functions can not be non-zero \textit{only}
inside a finite rectangle: by analyzing Hirota at the upper right corner $(a^*,s^*)$ of the rectangle we would conclude that $T_{a^*,s^*}=0$. This would then imply that the neighbors of this point,  $T_{a^*-1,s^*}=T_{a^*,s^*-1}=0$ which will then imply that $T_{a^*-2,s^*}=T_{a^*,s^*-2}=0$ etc; at the end we would be left with $T_{a,s}=0$ everywhere. What we \textit{can} have, for example, is $T_{a,s} \neq 0$ in a rectangle  and on the two infinite lines containing the upper and lower edges of the
rectangle, see figure \ref{TandYfunctions}.  At these lines $T_{a,s}$ are trivial (pure gauge) but they are non-zero. This is what we mean in the text.} At the edges of the rectangle
 either the first or the second term in the right hand side of  (\ref{Hirota}) is zero.
 We are left with a discrete Laplace equation for (the logarithm of) the $T$-functions and therefore  they become pure gauge.
At these boundary points the $Y$-functions are trivial (either zero or infinity) as expected from the analogy. The $Y$-functions are non-trivial in a smaller rectangle obtained from by removing the first and last columns and rows of the original domain.

The Hirota equation \ref{Hirota} then translates into the $Y$-system
\beq
\frac{Y_o^+ Y_o^-}{Y_{\uparrow} Y_{\downarrow}}=\frac{(1+Y_{\leftarrow})(1+Y_{\rightarrow})}{(1+Y_{\uparrow})(1+Y_{\downarrow})} \la{Ysystem}
\eeq
for these gauge invariant quantities. Different domains in $a,s$ where $Y$'s are nontrivial together with different boundary condition and analytic properties describe different integrable models.

In the treatment of the previous section we considered the case where the $T$-functions
live in a finite strip with three rows and $n/2-1$ columns, where $n$ is the number of gluons, see figure
\ref{TandYfunctions}.
The functions denoted by $T_{s}$ in that section are the T-functions in the non-trivial middle row,
$T_s =T_{1,s}$. Similarly $Y_s = Y_{1,s}$. The $T$-functions introduced in that section are inner products
of small solutions and are therefore sensitive to their normalization. This arbitrariness is a manifestation
of the gauge freedom in Hirota equation. The normalization $\< s_i,s_{i+1}\>=1$ corresponds to the gauge choice where
$$
\begin{array}{l}T_{0,2k}\equiv \<s_{-k-1},s_{-k} \> \,, \\T_{2,2k}\equiv \<s_{k},s_{k+1} \>\,,
\end{array}~\qquad\begin{array}{l}T_{0,2k+1}\equiv \<s_{-k-2},s_{-k-1} \>^+ \,, \\T_{2,2k+1}\equiv
\<s_{k},s_{k+1} \>^+\,, \end{array}
$$
are gauge fixed to one.
 We could of course opt not to fix a normalization for the T-functions but then we should
 use the gauge invariant combination (\ref{Ydefinition}) when defining the Y-functions:
\beqa\la{Ydef2}
Y_{2k}&=&\frac{T_{1,2k-1}T_{1,2k+1}}{T_{0,2k}T_{2,2k}}=
{\<s_{-k},s_{k}\>\<s_{-k-1},s_{k+1}\>\over\<s_{-k-1},s_{-k}\>\<s_{k},s_{k+1}\>}\\
Y_{2k+1}&=& \frac{T_{1,2k}T_{1,2k+2}}{T_{0,2k+1}T_{2,2k+1}}=
\[{\<s_{-k-1},s_k\>\<s_{-k-2},s_{k+1}\>\over\<s_{-k-2},s_{-k-1}\>\<s_{k},s_{k+1}\>}\]^+ ~.\nn
\eeqa
We see that they are now manifestly independent of the choice of the normalization of the small solutions.
At spectral parameter $\zeta=1$, or $\zeta = e^{\pm i \pi/2}$
 they yield physical space-time cross-ratios as in \nref{rightCR} \nref{leftCR}.

 A particularly interesting quantity is
 \beq\la{TheB}
 B(\zeta)=\frac{Y_{n/2-3}}{Y_{n/2-5}}\ \frac{Y_{n/2-7}}{Y_{n/2-9}}\ \frac{Y_{n/2-11}}{Y_{n/2-13}}\ \dots~
\eeq
It follows from the $Y$-system equations (\ref{Yads3}) that $B^+B^-=1$.  $B(\zeta)$ is constructed from the $Y$-functions and is therefore gauge invariant. Using the definition of the $Y$-functions we see that $B(\zeta)$ is given by
a bunch of boundary  $T$-functions. In our normalization all these functions except
for the rightmost one are gauged to one.  We find therefore
 $B=T_{1,n/2-2}=\<s_1,s_{n/2}\>\( e^{ - i \pi (n/2 +1)/2} \zeta\)$. This means that $B(\zeta)$ is the function that governs
the monodormy $s_{n/2} = - B(\zeta e^{ i \pi (n/2 +1)/2} ) s_0 $ of the small solutions.

\begin{figure}[t]
\center\includegraphics[width=100mm]{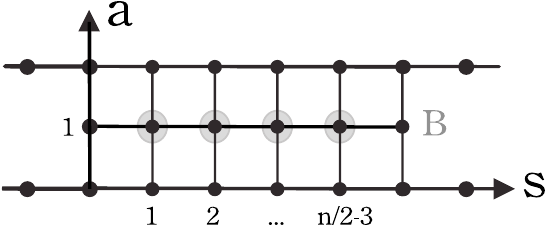}
\caption{ In this figure we have summarized the structure of the $T$'s and the $Y$'s in a gauge where
 we simplified the $T$'s that can be simplified. The small solid black dots represent non-zero $T$-functions. They are equal to one unless $a=1$ and $s=1,\dots,n/2-2$. At the rightmost point in this line we have $T_{1,n/2-2}=B$ where $B$ is a   function which cannot be set to one only in the case that $n/2$ is even. In fact, in our case it is
  $B =-   e^{ m/\zeta + { \bar m }\zeta}$. This is the function that governs
the monodormy $s_{n/2} = - B(\zeta e^{ i \pi (n/2 +1)/2} ) s_0 $. The $Y$-functions are finite in the points indicated by  fat gray shaded balls. At all other points they are either zero or infinity. }
 \la{TandYfunctions}
\end{figure}

\subsection{Analytic properties of the $Y$-functions}

For finite values of $\zeta$ it is clear from \nref{Tdef} that
the $T_{s}$ are analytic functions of $\zeta$, for $\zeta \not
= 0, \infty$. Generically, they will not be periodic under
$\zeta \to e^{ 2 \pi i } \zeta$. In general, the $Y$'s will be
meromorphic functions. However, in our case, since we can
choose to set the denominators to one, we see that the $Y$s
have no poles and are thus analytic away from $\zeta =0,
\infty$. For $\zeta \to 0$ and $\zeta \to \infty$ they will
have essential singularities. In this section we analyze the
behavior in these two regions.

When $\zeta \to 0$ we can solve the equations for the flat
sections by making a WKB approximation, where $\zeta$ plays the
role of $\hbar $. This is explained in great detail in
\cite{GMNtwo}, here we will summarize that discussion and apply
it to our case.
 The final result is that, for an appropriate choice of the polynomial $ p$,
we have the standard boundary conditions in TBA equations. We
will later discuss what happens for more general polynomials.

We are considering the equation \beqa
 \( d + { \Phi_z dz \over \zeta } + A + \zeta\,{ \Phi_{\bar z } d\bar z  }  \) s =0\nn
 \eeqa
 When $\zeta \to 0$, it is convenient to make a similarity transformation that diagonalizes $\Phi_z \to \sqrt{ p}\,
 {\rm diag}(1,-1) $. The solutions in this approximation go like
 $\exp \( \pm \frac{1}{\zeta} \int \sqrt{ p} dz \) $ times constant vectors.
 The WKB is  a good approximation  if  we are following the solution along a line of steepest descent.
 This is a line where the variation of the exponent, is real,   ${\rm Im} ( \sqrt{ p(z)} \dot z/\zeta ) =0$.
 This condition is an equation which determines the WKB lines.
 Through each point in
 the $z$ plane we have one such line going through. At the single zeros of $ p$ we have three lines coming in.
 The WKB approximation fails at the zeros of $p$ (which are the turning points).
 From each Stokes sector we have WKB lines that emanate from it. These lines can end in other Stokes sectors or, for
 very special lines, on the zeros of $ p$.
  If a line connects two Stokes sectors, say $i$ and $j$,
  then we can use it to approximate reliably the inner product $\langle s_i , s_j \rangle$.
  This estimate is  good in a sector of width $\pi$ in the phase of $\zeta$, centered
 on the value of $\zeta$ where the line exists.
 As we change the phase of $\zeta$ the pattern of flow lines changes. It also changes when we change the polynomial
 $ p$.
  We first select a polynomial with all zeros along the real axis and such that $ p(z) > 0$ for large enough values
 of $z$ along the real line.

\begin{figure}[t]
\center\includegraphics[width=70mm]{WKBlineszetaone.pdf}
\center\includegraphics[width=70mm]{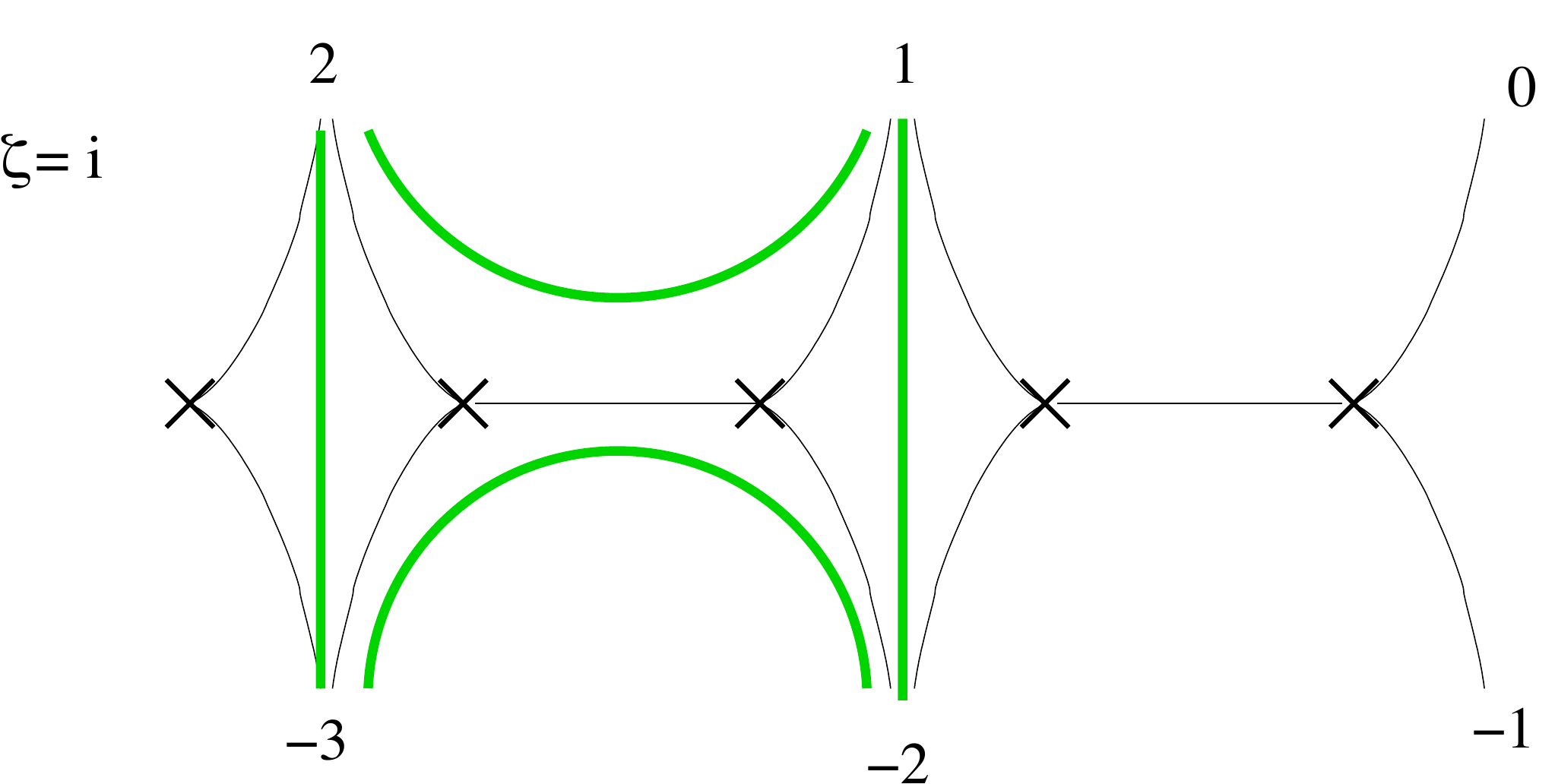}
\center\includegraphics[width=70mm]{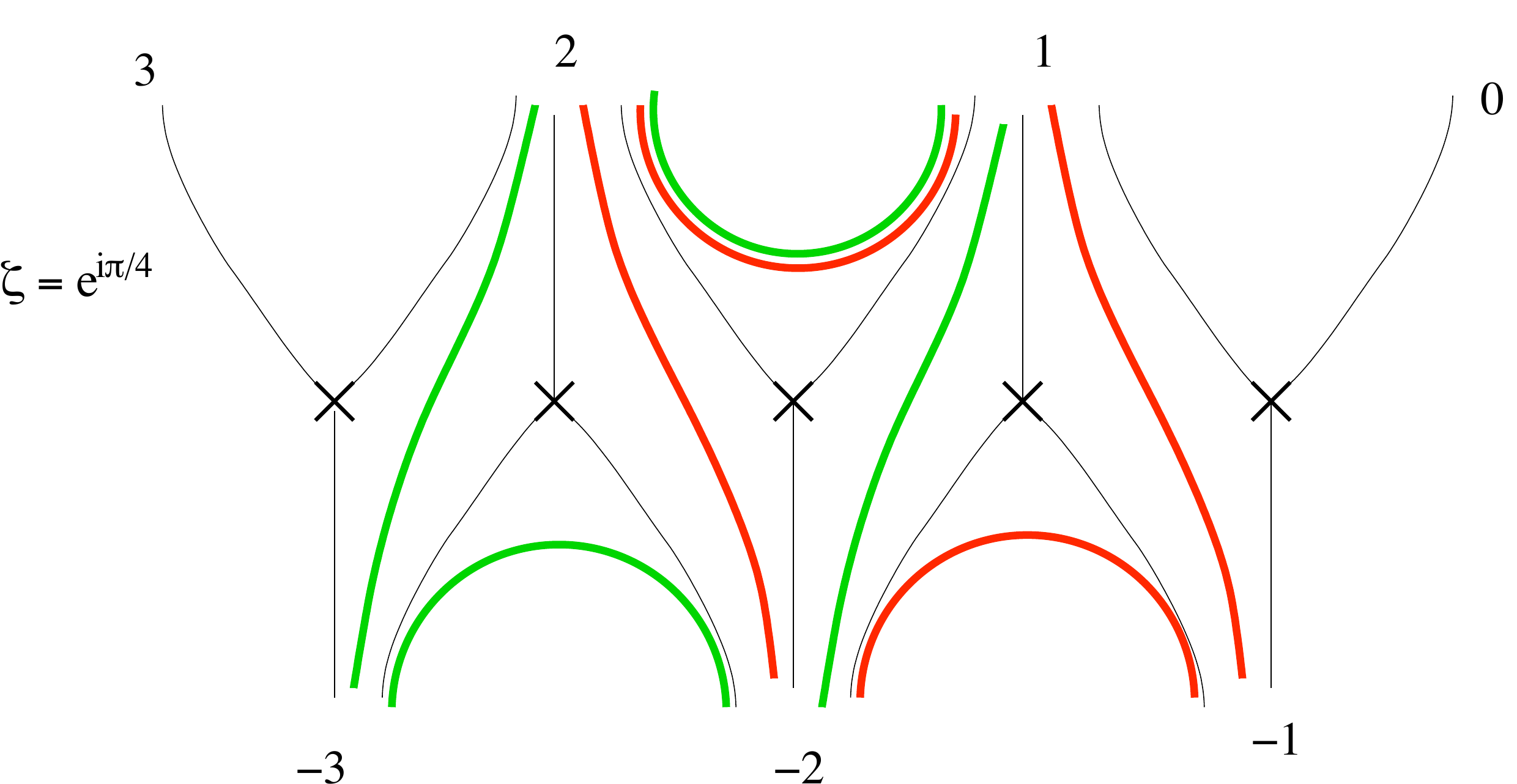}
\caption{ Sketch of the pattern of WKB lines for various phases of $\zeta$.
 The crosses denote the various zeros of $ p(z)$. The numbers indicate the various Stokes sectors.
  The black thin lines end on the zeros and
separate different classes of WKB lines.
The thick colored lines are the WKB lines that
 we use to evaluate cross ratios. Here we have
indicated only the ones used to evaluate $Y_{2}$ and $Y_3$. Finally, note that by setting the phase of
$\zeta $ to $e^{ i \pi/4}$ we have WKB lines that enable us to evaluate all the $Y_s$.
} \la{WKBlines}
\end{figure}

 With this choice the pattern flows for WKB lines is shown in figure \ref{WKBlines}, for some values of the phase of
 $\zeta$.
 The WKB lines ending on zeros  separate  regions where lines flow between different Stokes sectors.
 In our problem we have some inner products evaluated at $\zeta$ and some and $i \zeta$.
 The flows for $\zeta  = 1$ and $\zeta = i  $
  are displayed on the top of figure
  \ref{WKBlines}, and they can be used
 to evaluate the various inner products. Alternatively, we can set $\zeta = e^{ i \pi/4}  $,  evaluate them
 all and then continue them from this region.  The resulting
 flow pattern is sketched on the bottom of figure
 \ref{WKBlines}.

 Using those flow patterns it is a simple matter to evaluate various inner products.
   It turns out that the inner
 products in the definitions of the Y-functions (\ref{Ydef2})
 combine to give a contour integral around a certain cycles.
  See figure \ref{Ycycles}.
 Thus, each $Y_s$, is estimated by the integral of $ \sqrt{p}$ along a cycle $\gamma_s$.
 We can call
 \beqa
 Z_s = - \oint_{\gamma^s}  \sqrt{ p} dz\nn
 \eeqa
  and the corresponding $Y$ functions have the small
 $\zeta $ behavior
 \beqa
 \log Y_{2k} \sim { Z_{2k} \over \zeta } +\cdots ~,~~~~~~~~~~~\log Y_{2k+1} \sim { Z_{2k+1} \over i \zeta} + \cdots\nn
 \eeqa
In figure \ref{Ycycles} we display the cycles corresponding to
each of the $Y_s$.
 \begin{figure}[t]
\center\includegraphics[width=70mm]{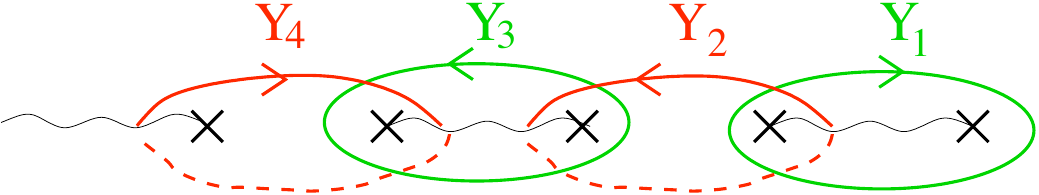}
\caption{  Cycles along which we need to integrate $\sqrt{p} dz$ in order to determine
the asymptotic form of $Y_s$. By being careful about the sheet selected by the various small solutions
one can determine the cycle orientations shown here. }
 \la{Ycycles}
\end{figure}
 It is convenient to define the parameters $m_s$ via
 \beqa \label{massesz}
 m_{2k} =- 2Z_{2 k} ~~~~~~~~~~~~ m_{2k-1} = - 2Z_{2k+1}/i
 \eeqa
For our choice of polynomial the $m_{s}$ are all real and
positive. In order to check the positivity of the $m_s$ we need
to be careful with the choice of branch when we evaluate the
cross ratios.
 The
two branches correspond to the two eigenvalues of $\Phi_z$ and
differ by an overall sign. Taking the same cycles but on
different branches is equivalent to changing the sign of $Z$.
The correct branch is determined by the behavior of the various
small solutions, each of which goes like $e^{ \pm \int \sqrt{
p} dz }$. After taking this into account we can check that the
$m_s$ are indeed positive for the polynomial we chose.

 A similar computation at large $\zeta$ gives a similar result, with $\log Y_{2k} \sim \zeta \bar Z_{2k}$.
 Thus, we have shown that all the $Y$ functions have the asymptotic behavior
 \beqa
 \log Y_s = - m_s \cosh \theta + \cdots\nn
 \eeqa
 for large $\theta$, $\zeta = e^\theta$.

 Furthermore, this behavior is good over a range of $(- \pi ,\pi)$ in the imaginary part of $\theta$\footnote{Recall that
 the functions are not periodic in ${\rm Im} (\theta)$.}.
 The reason is the following. For each $Y_s$, the region of $Im(\theta)=0$ corresponds to the center of
 the region where the WKB line exists. Furthermore, the corresponding WKB lines exist for a sector of
 angular size $\pi$ around this line. In addition, we have mentioned that the WKB approximation continues
 to be good for a further sector of $\pi/2$ on each side.
  In fact, this is  more than enough for deriving the integral equations.

\subsection{Integral form of the equations}
 \la{AdS3TBA}

The analytic properties derived above together with the functional
equations (\ref{Yads3}) uniquely determine the $Y$-functions.
 However, for practical purposes -- specially for numerics --
 it is useful to have an equivalent formulation of these $Y$-system
 equations in terms of TBA-like integral equations.

To derive them we follow the usual procedure which we
 briefly review for completeness. We note that
  $l_s\equiv \log (Y_s/e^{-m_s \cosh \theta})$ is analytic  in
  the strip $|{\rm Im} (\theta)|\le  \pi/2$, vanishes as $\theta$ approaches
   infinity in this strip, and obeys
\beq
l_s^+ + l_s^- = \log(1+Y_{s+1})(1+Y_{s-1})\nn
\eeq
which is nothing but the logarithm of the Y-system equations. Now we convolute this equation with the kernel
\beq
K(\theta)=\frac{1}{2\pi \cosh \theta}\nn
\eeq
to get
\beq
K \star (l_s^+ + l_s^-) \equiv
\int\limits_{-\infty}^{+\infty} \frac{dy}{2\pi} \, \frac{l_s(y+i \pi/2)+l_s(y-i \pi/2)}{\cosh(x-y)}=
 \oint\limits_{\gamma} \frac{dy}{2\pi i} \, \frac{l_s(y)}{\sinh(x-y)}=l_s(x)\nn
\eeq
where $\gamma$ is the rectangle made out of the boundaries of the physical
 strip together with two vertical segments at ${\rm Re}(\theta) \to \pm \infty$.
 In order to be able to add these extra segments to the integral thus making it a contour integral, it was important to use the $l_s$ instead of $\log Y_s$. This is why this quantity was introduced in the first place. Furthermore, in the last step we used the fact that $l_s$ has no singularities inside the physical strip, this is an important input on the analytic properties of the $Y$-functions.
 Rewriting $l_s$ in terms of $Y_s$ leads therefore to the desired form of the integral equations:
\beq
\log Y_s = -m_s \cosh\theta + K\star \log(1+Y_{s+1})(1+Y_{s-1})   \,. \la{YAdS3}
\eeq
For a given choice of masses $m_s$, the solution to these integral equations is
 unique and a basis of cross ratios can be read from evaluating the $Y_s(\theta)$'s
  at $\theta=0$.  These equations are of the form of those appearing in the study of the
   TBA \cite{TBApapers} for quantum integrable models in finite volume. Furthermore, as explained in the
    next section, the (regularized) area of the minimal surfaces turns out to be given
    in terms of the $Y$-functions as the free energy of the corresponding integrable model.

Up to now we have discussed the case where the zeros of the
polynomial are along the real axis. Let us briefly discuss what
happens as we start moving the zeros of the polynomial away from the
real axis. Notice that the functional $Y$ system equations
\nref{Yads3} do not depend on the polynomial. Thus, these equations
continue to be true, regardless of its form. What changes are the
asymptotic boundary conditions.

Let us consider first the case were we start from the above polynomial and we move the zeros around a little bit.
 Then the above derivation of the asymptotic form of the $Y$ functions goes through with only one change.
 Namely the quantities $Z_{s}$ and $m_s$ are now more general complex numbers and the asymptotic behavior
 is
 \beqa
 \log Y_s & \sim & - { m_s \over 2 \zeta} ~,~~~~~~~~{\rm for}~~~~~~ \zeta \to 0\nn
 \\   \log Y_s  & \sim &  - {\bar m_s  \over 2 } \zeta ~,~~~~~~~~{\rm for}~~~~~~ \zeta \to \infty\nn
 \eeqa
  In this case, when we derive the integral equation, it is convenient to shift the line where the $Y$ functions
  are integrated to be along the direction where $ m_s/\zeta\equiv |m_s|e^{i\varphi_s}/\zeta$ is real and positive, which also makes
  $\bar m_s \zeta$ real and positive. Then, defining $\tilde  Y_s(\theta) = Y_s(\theta + i \varphi_s)$
  we find that the integral equations have the form
  \beqa\la{tildeYAdS3}
  \log \widetilde Y_s = - |m_s|\cosh \theta + K_{s,s+1} \star \log (1 + \widetilde Y_{s+1}) + K_{s,s-1}
  \star \log(1 + \widetilde Y_{s-1} )
  \eeqa
  where now $K_{s,s'} =1/\cosh( \theta - \theta' + i \varphi_s - i \varphi_{s'} ) $.

As long as $| \varphi_s - \varphi_{s+1}| < \pi/2$, the integral equations conserve the form that we have derived. If we deform the phases beyond that regime we will have to change the form of the integral equations by picking the appropriate pole contributions from the kernels (which become singular for $| \varphi_s - \varphi_{s+1}| = \pi/2, 3\pi/2, \dots$).
Of course, the integral equation changes but the $Y$'s and therefore the area are continuous.
 The pattern under which the  integral equation changes is explained in appendix
 \ref{WallCrossing}.\footnote{A very similar kind of manipulations is outlined in section
 \ref{CRsec} when we explain in detail how to compute the $Y$-functions in the $AdS_5$ case
 for large values of the imaginary part of $\theta$.}  This is the wall crossing phenomenon d
 iscussed in \cite{GMN,GMNtwo}. 

These integral equations are a special case of the general case discussed in \cite{GMN}. In fact, the equations in \cite{GMN} are true for an arbitrary ${\cal N}=2$ theory, and a Hitchin problem is just a special case.
 Due to the $Z_2$ projection we have that the $\mathcal{X}_{\gamma}$ quantities in \cite{GMN} obey the additional property $\mathcal{X}_{-\gamma}(\zeta) = \mathcal{X}_{\gamma} (-\zeta)$. Using this, we can easily map the kernel in \cite{GMN} to the ${ 1 \over \cosh } $ found here.

\subsection{Area and free energy}
\la{Areatofreeenergy}

 As we mentioned above, the interesting part of the area is given by
 the integral
 \beqa \label{areastar}
 A = 2 \int d^2 z Tr[\tilde\Phi_z \tilde\Phi_{\bar z} ]
 \eeqa
 By definition, the area is independent of $\zeta$.

  It is convenient to think again about the small $\zeta $ regime and the WKB approximation that
  we did for small $\zeta$.
 In fact, we can improve on the WKB approximation and find the next couple of terms by systematically expanding
 the expressions for the inner products. We take complex masses but with small enough phases so that the WKB approximations that
we did before continue to be valid, with the same cycles.\footnote{In general,  the cross ratios that have a simple WKB approximation
will change as we change the phase of the masses beyond a certain point, see \cite{GMNtwo}. For us, it is enough to do the  derivation  for some range of masses. Then we can analytically continue
the final formula, as explained in appendix \ref{WallCrossing}.}. The final
 At this point it is convenient to use slightly different functions
 defined by
\beqa \la{Yhat}
\widehat Y_{2k}(\zeta) = Y_{2k}(\zeta)~,~~~~~~~~ \widehat Y_{2k+1} = Y_{2k+1}^- = Y_{2k+1}(\zeta e^{- i \pi/2 } )
\eeqa
 in order to undo
 the shifts in \nref{Ydef2}.
 With these definitions we find that
 \beqa \label{chiexpan}
 \log \widehat Y_k  \ &\sim&  - \left[ { \oint_{\gamma_k} \lambda  \over \zeta } +  \oint_{\gamma_k} \alpha
 + \zeta \, \oint_{\gamma_k}  u   + \cdots \right]
  \\  && \lambda = x d z ~,~~~~~~~~~ x^2 = p(z) \nn
  \eeqa
  where $u$ is an exact one form. Here $\alpha$ given by the   diagonal components of the connection $A$.
   In our
  case $\alpha =0$ due to the $Z_2$ projection. (But even if $\alpha$ were non-zero,
   it would not affect what we say below.)
   We know that $u$  is exact because we can deform the
  contour and $\log Y_s$ should not change. It has a $u_z$ and a $u_{\bar z}$ component.
  For our purposes, it will only be important to compute the $u_{\bar z}$ component which is
  \beqa \label{phizcom}
   u^i_{\bar z} = \tilde \Phi^{ii}_{\bar z}
   \eeqa
   where the index $i$ is not summed over.
   In other words, we get the diagonal components of $\tilde\Phi_{\bar z}$ and we
   get the first or second diagonal component depending on whether we are on the first or second sheet of the Riemann
   surface. That is, $u$ is a one form on the Riemann surface, not on the $z$ plane.
  In the basis where $\tilde\Phi_z$ is diagonal, we can thus rewrite \nref{areastar}, using \nref{phizcom},
  as  
  \beqa
  A = { i  } \int   \lambda \wedge u  =  - i \sum_{r,s} w_{rs}
  \oint_{\gamma^r} \lambda \oint_{\gamma^s} u\nn
  \eeqa
  where ${\gamma^r}$ are a basis of cycles\footnote{This formula looks suspicious because the left hand side is infinite
  while the right hand side is finite!. Here we have implicitly used a regularization which puts a cutoff in the
  $w$ plane for large values of $|w|$, where $dw = \sqrt{p} dz$. We have then subtracted the same integral but
  with a polynomial whose zeros are all at the origin. This procedure works well when $n/2$ is odd.}.
    We will take this basis to be the basis of cycles that gives
  the WKB approximation to the $Y_s$, see figure \ref{Ycycles},
   and $w_{rs}$ is the inverse of the intersection form of the cycles. The matrix of cycle intersections
   can be read off from figure \ref{Ycycles} and it is summarized in
   figure \ref{PBAdS3}.

   \begin{figure}[t]
\center\includegraphics[width=160mm]{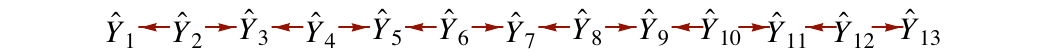}
\caption{Intersection form $\theta^{r,s}$ for all the cycles associated to
the $Y$ functions. This is computed from the cycles in figure \ref{Ycycles}.
If an arrow points from $Y_s$ to $Y_r$, then we have $\langle \gamma_{s},
\gamma_{r} \rangle =1$,
otherwise the intersection vanishes.
} \la{PBAdS3}
\end{figure}

   We can also compute the small $\zeta$ behavior of $\widehat Y_k$ by
   expanding the integral equations  \nref{YAdS3}.
   We get
   \beqa \label{integreq}
   \log \widehat Y_r  = Z^r/\zeta  + \zeta \left[ \bar Z^r + \sum_s \theta^{rs} { 1 \over  \pi i }
   \int { d \zeta' \over \zeta'} { 1 \over \zeta' }
   \log (1 +\widehat Y_s) \right]
   \eeqa
   It turns out that $\theta^{rs}$ is given by the intersection form of the cycles involved.
    This follows from the general theory
   in \cite{GMN}, but it can be easily checked in
   this case by examining the integral equations  \nref{YAdS3}  and remembering that the $\widehat Y_s$ differ
   by simple shifts in the argument from the $Y_s$.

   Thus, we obtain that $A = A_{periods} + A_{free}$ with
   \beqa
   A_{periods} &=&- i w_{rs} Z^r \bar Z^s \nn
  \\
   A_{free} &=&  -  { 1 \over  \pi}  \sum_s \int {  d \zeta \over \zeta' }
    {Z_s \over \zeta'}  \log(1 + \widehat Y_s) \label{areafreefirst}
\\
 A_{free} &=&  \sum_s \int {  d \theta \over 2 \pi }  |m_s| \cosh \theta
  \log(1 +   \widetilde Y_s) \label{areafree}
  \\ &&
  {\rm with}~~~~~~~~\widetilde Y_s(\theta) = Y_s(\theta + i \varphi_s) ~,~~~~~~
  m_s = |m_s| e^{ i \varphi_s} \nn
 \eeqa
In order to obtain \nref{areafree}
 we have averaged the result from  \nref{areafreefirst} with the result we obtain from the
large $\zeta$ expansion.
 The fact that large $\zeta$ and small $\zeta$ should give the same answer translates
into the statement that the total momentum of the TBA system
should be zero.  The explicit form of $A_{periods}$ in terms of the masses is
given in \nref{Aperiods3}.

This derivation has assumed that $n/2$ is odd, because we said
that $\theta^{rs}$ was invertible. If $n/2$ is even, then we start from $n/2 +1$ and we take away
one zero of the polynomial. Then the result contains two
pieces, one piece has the form of $A_{free}$ discussed above
and the other contains an extra term that was discussed in
detail in \cite{Alday:2009yn}.

Also, in this derivation, we have assumed that the intersection
form of the cycles associated to the $Y_{\gamma}$ that appear
in the integral equation is invertible. While this is true in
our case, it would cease to be true once we cross walls and we
get extra cycles \cite{GMN}.
 One can slightly modify the above derivation
and the final answer continues to be \nref{areafree}, see appendix \ref{WallCrossing}.

Note that, in the end, we do not need to know the
polynomial, or the Riemann surface, or the cycles. That is only
needed for the derivation. Ultimately,
 everything is expressed in terms of
the (complex) masses $m_s$ appearing in the integral equations, see
\nref{areafree}, \nref{Aperiods3}.

\subsection{The octagon, or $n=8$}

Here we rederive some of the results in \cite{Alday:2009yn}
from this point of view. In this case there is only one $Y$
function and the functional equation is $Y^+Y^-=1$, whose
solution is just $Y = e^{ Z/\zeta + \bar Z \zeta }$. The free
energy is \beqa A_{free} = \frac{1}{2\pi}\int d\theta\, 2|Z|
\cosh \theta \log (1 + e^{ -2 |Z| \cosh \theta } )\nn \eeqa which
agrees with what was called $A_{sinh}$ in \cite{Alday:2009yn}.

The full result in \cite{Alday:2009yn} contains an extra piece
$A_{extra}$ which is related to an extra complication that
arises in the case that $n/2$ is even. In this case we will
also need the Hirota variable $T_{1}$.

\section{Minimal surfaces in $AdS_5$}
\subsection{$AdS_5$ preliminaries} \la{ads5pre}

As we mentioned in section two, the worldsheet theory describing strings in $AdS_5$ can
be reduced to a $Z_4$ projection of an $SU(4)$, or $SU(2,2)$, Hitchin system.

After some gauge choices we can represent the action of the $Z_4 $ in the following way
\beqa
 r( X ) = - C X^t C^{-1}  ~,~~~~~~~~~~~~~ C^{-1}  =
 \(\begin{array}{cccc}\  0 & 1 & 0& 0 \\  0 & 0 & 1& 0 \\ 0 & 0 & 0& 1 \\  - 1 & 0& 0 & 0   \end{array}\) \la{projection}
 \eeqa
 where $X$ is $\Phi$ or $A$.  Recall that we will be imposing the projection conditions
 $r(\Phi_z) = - i \Phi_z $, $r(A) =   A$ and $ r(\Phi_{\bar z} ) =  i \Phi_{\bar z }$.

 This $Z_4$ symmetry relates solutions to the problem at $i \zeta$ with solutions to a related
 problem at $ \zeta$. More precisely, they relate solutions to the ``inverse'' problem at $\zeta$.
 More explicitly, we have
 \beqa
  \bar \psi (   \zeta ) = C^{-1}   \psi (i \zeta ) \nn
 \eeqa
 where $\bar \psi(\zeta)$ is a flat section of $ ( d - {\cal A}^t ) \bar \psi =0$. Note that the bar { \it does not }
 denote complex conjugation. Given a solution $\psi(\zeta)$ of the straight problem and a solution
 $\bar \psi'(\zeta)$ of the inverse problem, one can form an inner product of the form
 $\langle \bar \psi'(\zeta) \psi(\zeta) \rangle $. This inner product can be computed at any point on the worldsheet,
 and it will be independent of the point where it is computed.

 Another property that we will use is the following. Imagine we start with three different solutions of the
 linear problem $\psi_1, ~\psi_2, ~\psi_3$. Then the following combination is a solution of the inverse problem
 \beqa
 \bar \psi_{123}^\alpha = \epsilon^{\alpha \beta_1 \beta_2 \beta_3 } \psi_{1 \beta_1} \psi_{2 \beta_2} \psi_{3 \beta_3 }
 \equiv \psi_1 \wedge \psi_2 \wedge \psi_3 \nn
 \eeqa
 where the last equality is simply the definition of the wedge product.
 This gives something interesting when it is applied to small solutions $s_i$ for three consecutive Stokes sectors
 \beqa
   s_{i-1} \wedge s_i \wedge s_{i+1} \propto \bar s_i \nn
 \eeqa
 In other words, this product of three solutions gives us a solution to the inverse problem that is small in
 Stokes sector $i$.
  This follows from the asymptotic behavior of the solutions in Stokes sector $i$.

  We show in appendix \ref{MoreDetails}
   that we can choose normalizations for all solutions, $s_i$, $\bar s_i$,
    so that the following equalities are true  
   \beqa
    1 &=& \langle s_i  ,s_{i+1} ,s_{i+2} ,s_{i+3} \rangle \nn \\
     \bar s_i &=& s_{i-1} \wedge s_i \wedge s_{i+1}  \la{norsol} \\
     \bar s_{i+1}(\zeta) &=& C^{-1} s_{i}(e^{i\pi/2} \zeta ) \nn \\
       s_{i+1}(\zeta) &=& C^{T} \bar s_{i}(e^{i\pi/2} \zeta ) \nn
   \eeqa
Using these formulas, plus identities involving $\epsilon$ symbols, it is possible to show that
\beqa
  \langle s_{k},  s_{k+1} , s_{j} , s_{j+1}\rangle(\zeta) &=& \langle
s_{k-1},  s_{k} , s_{j-1}  ,s_{j}\rangle \(e^{i\pi/2}\zeta\) \nn\\
 \langle s_{j}  ,s_{k}  ,s_{k+1}  ,s_{k+2}\rangle(\zeta)  & = &
  \langle s_{j}  ,s_{j-1}  ,s_{j-2}, s_{k}\rangle (e^{i\pi/2}\zeta) \la{useful2} \\
\langle s_j ,\bar s_{k+1} \rangle (\zeta) &=&
\langle \bar s_{j-1} ,s_{k} \rangle \(e^{i\pi/2}\zeta\) \nn
\eeqa
Where the last two lines are stating the same result in two alternative notations.

Finally, if the problem  involves $n$ Stokes sectors, we expect that $s_{i+n} \propto s_i$,
where $s_{i+n}$ has been obtained by going around the $z$ plane and normalizing the solutions via
\nref{norsol}. We will not need the proportionality constant to derive the $Y$ system. In fact, one
can calculate it from the $Y$ system.
However, it is possible to compute it just from the behavior of the solutions at infinity,
see appendix \ref{MoreDetails}.
The result is that for $n$ odd we can normalize the solutions so that $s_{n} = s_0$.
When $n$ is even this is not possible.
When $n= 4 k $ one has
\beqa
s_{n} = \mu \,  e^{ { w_0 \over \zeta } + {\bar w}_0 \zeta } \,  s_0 \,\, , \qquad \,\,\,\,  s_{n-1}  =  { 1 \over  \mu }  \,   e^{ { i  w_0 \over \zeta } -i  {\bar w}_0 \zeta } \, s_{-1} \la{asympt} \,\, , \qquad \,\,\,\,  s_{n-2}  =  \mu \,  e^{ -{ w_0 \over \zeta } - {\bar w}_0 \zeta } \, s_{-2}
\eeqa
where $w_0$ is a constant.
For $n = 4 k +2 $,  $w_0 =0$ and  only $\mu$   is allowed.

\subsection{The $AdS_5$ Y-system.}

The basic identity which we will use to derive the $Y$ and
$T$-system functional
 equations for the spectral parameter dependent products, or Wronskians,
  is a particular case of Plucker relations: Let $(x_1,\dots, x_N)$ be the determinant of the $N\times N$ matrix whose columns are $N$-vectors $x_i$. Then, for $N+2$ generic vectors $x_{1},\dots,x_N,y_1,y_N$ we have the following identity
\beqa \(x_1, \,\,\star\,\, ,x_N\)\(y_1, \,\,\star\,\, ,y_N\)=
\(y_1, \,\,\star\,\, ,x_N\)\(x_1, \,\,\star\,\, ,y_N\)- \(y_N,
\,\,\star\,\, ,x_N\)\(x_1, \,\,\star\,\, ,y_1\) \,,
\la{plucker} \eeqa
where $\star= x_2,\dots, x_{N-1}$. 

For us, $N=4$ and $(x_1,\star,x_4)=\<s_i,s_j,s_k,s_l\>$ are
determinants of $4\times4$ matrices composed from four sections
of the linear problem. Using the small solutions $\{s_{m},
s_{m+1},s_{m+2}, s_{m+3},s_{-1},s_{0}\}$,
$\{s_{-2},s_{m+1},s_{-1},s_m,s_{0},s_{m+2} \}$ and $\{
s_{-2},s_{-1},s_0,s_1,s_{m+1} ,s_{m+2} \}$ for
$\{x_1,x_2,x_3,x_4,y_1,y_4\}$ we obtain three Hirota relations
\beqa
T_{1,m}^+T_{3,m}^-&=&T_{1,m-1}T_{3,m+1}+ T_{0,m} T_{2,m}\,,\nn \\
T_{2,m}^+T_{2,m}^-&=&T_{2,m+1}T_{2,m-1}+T_{1,m}T_{3,m} \,,\nn\\
T_{3,m}^+T_{1,m}^-&=&T_{1,m+1}T_{3,m-1}+T_{4,m} T_{2,m} \,.\nn
\eeqa In a more compact notation these read
\beq\la{Hirotatwist} T_{a,m}^+\ T_{4-a,m}^-\ =\ T_{4-a,m+1}\
T_{a,m-1}\ +\ T_{a+1,m}\ T_{a-1,m}~, \eeq where $a=1,2,3$
correspondingly. The $T$-functions are given by
$$T_{0,m}(\zeta)=\<s_m,s_{m+1},s_{m+2},s_{m+3}\>^{[-m-1]}=1\, , \qquad T_{4,m}(\zeta)=\<s_{-2},s_{-1},s_0,s_1\>^{[-m-1]}=1\,,$$
 and most importantly
\beqa
T_{1,m}(\zeta)&=&\<s_{-2},s_{-1},s_0,s_{m+1}\>^{[-m]} \,, \nn\\
T_{2,m}(\zeta)&=&\<s_{-1},s_0,s_{m+1},s_{m+2}\>^{[-m-1]} \,, \la{naive}\\
T_{3,m}(\zeta)&=&\<s_{-1},s_{m},s_{m+1},s_{m+2}\>^{[-m]}\nn  \,.
\eeqa Here the superscripts $\pm$ and $[m]$ indicate shifts in
the spectral parameter, $f^{\pm} \equiv f(e^{\pm
i{\pi\over4}}\zeta)$ and $f^{[m]} \equiv
f(e^{i{\pi\over4}m}\zeta)$.  Note that these shifts are half
the ones appearing in the $AdS_3$ case.\footnote{Whether we are
discussing the $AdS_3$ or the $AdS_5$ case will be clear from
the text. We hope that the difference between the shifts in the
two cases will not cause a confusion.} When deriving the Hirota
equations from the Plucker identity we use the relations
(\ref{useful2}).

Equation (\ref{Hirotatwist}) is an exotic form of the Hirota
equation (\ref{Hirota}). It is exotic
because of the appearance of $T_{4-a,s}$ instead of the usual $T_{a,s}$.

For $s=-1$ and $s=n-3$ we see that $T_{a,s}=0$. This follows
simply from the defintion of the $T$-functions together with
$s_{k+n}\propto s_k$. Thus, the functions $T_{a,s}$ live in a
strip of five rows ($a=0,1,2,3,4$) and $n-3$ columns
($s=0,1,\dots,n-4$), see figure \ref{stripads5}. As explained
in section \ref{GenHirota}, the Hirota equation has a huge
gauge symmetry. The same is also true for the more exotic form
considered here.\footnote{For example, if $g(a,s,\theta)$ is a
gauge symmetry transformation of Hirota (\ref{Hirota}), then
$g(a,s,\theta)g(4-a,s,\theta)$ is a gauge symmetry of the more
exotic form (\ref{Hirotatwist}).} Our choice of normalization
$\<s_is_{i+1}s_{i+2}s_{i+3}\>=1$, corresponds to the gauge
choice where the $T$-functions at the left, top and bottom
bondaries of the strip are set to one, \beq\la{boundary}
T_{a,0}=T_{0,s}=T_{4,s}=1~. \eeq At the right boundary of the
strip, in our normalization (\ref{norsol}), the $T$-functions
are related to the formal monodromy which arise when comparing
$s_i$ with $s_{i+n}$ (\ref{asympt}), \beq \label{Tform}
T_{1,n-4}=\mu^{-1} \(e^{ -{ w_0 \over \zeta } - {\bar w}_0
\zeta}\)^{[-n + 2]}, T_{2,n-4}=\(e^{ - \sqrt{2}({ w_0 \over
\zeta } + {\bar w}_0 \zeta)}\)^{[-n + 2]},T_{3,n-4}=\mu\(e^{ -{
w_0 \over \zeta } - {\bar w}_0 \zeta}\)^{[-n + 2]} \eeq

 \begin{figure}[t]
\center\includegraphics[width=100mm]{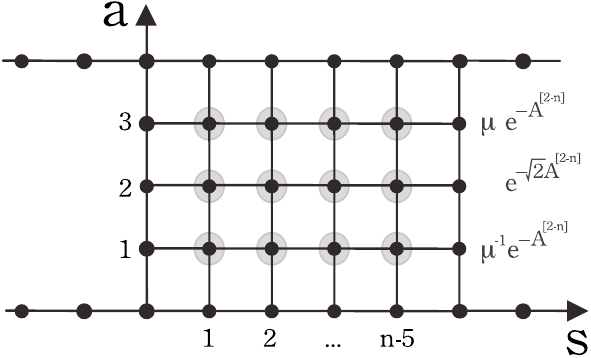}
\caption{ Strip where the $T$ and $Y$-functions live in the $AdS_5$ case. Small solid black dots represent $T$-functions. At the boundary the $T$-functions are equal to one except at the three nodes in the right boundary; there they take the values indicated in the figure. In all the points of the boundary the $Y$-functions are either zero or infinity. They are non-trivial in the smaller domain indicated by the  fat shaded gray circles. }
 \la{stripads5}
\end{figure}

Next, we introduce the $Y$-functions \beqa \la{Ydef}
Y_{a,m}\equiv{T_{a,m+1}T_{4-a,m-1}\over T_{a+1,m}T_{a-1,m}}\,.
\eeqa
Being composed from next-to-nearest-neighbor $T$-functions, the
$Y$-functions $Y_{a,s}$ are finite in a slightly smaller
lattice parametrized by $a=1,2,3$ and $s=1,\dots,n-5$. The
number of $Y$-functions coincides with the number of
independent cross ratios. The Hirota equation then implies the
following Y-system for these quantities:\footnote{Similar
exotic $Y$-systems recently appeared in a very different
context \cite{cp3}}
\beqa {Y_{2,m}^-Y_{2,m}^+\over
Y_{1,m}Y_{3,m}}&=&{(1+Y_{2,m+1})(1+Y_{2,m-1})\over(1+Y_{1,m})(1+Y_{3,m})}
\nn
\\
 {Y_{3,m}^-Y_{1,m}^+ \over
Y_{2,m}}&=&{(1+Y_{3,m+1})(1+Y_{1,m-1})\over
1+Y_{2,m}}\la{Yone}\\ {Y_{1,m}^-Y_{3,m}^+ \over
Y_{2,m}}&=&{(1+Y_{1,m+1})(1+Y_{3,m-1})\over 1+Y_{2,m}} \nn
\eeqa or in a more compact notation
\beqa
{Y_{a,m}^-Y_{4-a,m}^+\over
Y_{a+1,m}Y_{a-1,m}}&=&{(1+Y_{a,m+1})(1+Y_{4-a,m-1})\over(1+Y_{a+1,m})(1+Y_{a-1,m})}
\,\, \,\,\, ,\,\, \,\,\,
\begin{array}{l}
a=1,2,3 \,, \\
s=1,\dots,n-5
\end{array}
\label{Ytwo} \eeqa
To recover \nref{Yone} we need to use that $Y_{0,s} = Y_{4,s} =
\infty$.

\subsection{Analytic properties of the $Y$-functions}

To derive the integral form of the Y-system equations it is
important to identify the large $\theta $ asymptotics.
 They are fixed by a WKB analysis. The method is very similar to the one used for the $AdS_3$ case,
  but a bit more involved.
 We will leave the details for appendix \ref{AdSfiveAsymptotics} and state here the final results.

 We choose the polynomial $P$ to be such that all zeros are on the real axis and $P(z) >0$ for sufficiently large $z$. Then
the large $\theta $ behavior of the $Y$-functions is \beqa
&&\log Y_{1,s} \to -m_s \cosh\theta - C_s \pm D_s  \,\,\, , \,\,\, \theta \to \pm \infty  \nn  \\
&&\log Y_{3,s} \to -m_s \cosh\theta + C_s \mp D_s  \,\,\, , \,\,\, \theta \to \pm \infty \la{asymconone}\\
&&\log Y_{2,s} \to -\sqrt{2} \,m_s \cosh\theta  ~,~~~~~~
\,\,\,  \,\,\, \theta \to \pm \infty \nn \eeqa where
$\theta=\log(\zeta)$. The constants $C_s$ and $D_s$ arise from
the components of the connection $A$ that survive the $Z_4$
projection. For loops in signature $(1,3)$ or $(3,1)$ the
$D_s$'s are real
 while the $C_s$'s are purely imaginary, see appendix \ref{realco}.
In fact, we have the general reality condition\footnote{In
(2,2) signature the reality condition  is $( Y_{a,s}(\zeta) )^*
= Y_{a,s}(1/\zeta^* ) $. } \beqa
\(Y_{a,s}(\zeta)\)^*=Y_{4-a,s}(1/\zeta^*) \nn
\eeqa In particular, for large $\theta=\log \zeta$, we see that
\beq
\[\log\(Y_{1,s}/Y_{3,s}\)(+\infty)\]^*=-\log\(Y_{1,s}/Y_{3,s}\)(-\infty) \nn
\eeq which indeed implies what we said above regarding $D_s$
and $C_s$. It turns out that the $m_s$ can be promoted to
complex constants by changing the position of the zeros of the
polynomial\footnote{In this case the conditions
\nref{asymconone} get modified to $ \log Y_{1,s} \sim - { m_s
\over 2 } e^{-\theta } - C_s - D_s $ for $\theta \to -\infty$
and $ \log Y_{1,s} \sim - { \bar m_s \over 2 } e^{\theta } -
C_s + D_s $ for $\theta \to +\infty$. And similarly for the
other $Y$ functions.}. These $n-3$ complex constants,  together
with the purely imaginary
 $C_s$,  constitute
the $3(n-5)$ parameters of the problem.

\subsection{TBA equations in $AdS_5$}

To derive a set of integral equations from the functional
Y-system we follow again the same route as in the $AdS_3$ case.
As explained in the $AdS_3$ case, a big advantage of the
integral form of the Y-system equations is the straightforward
numerical implementation of these equations. We first consider
the case where all masses are real and positive.
 We then introduce a set of functions
$l_{a,s}=\log(Y_{a,s})+m_{a,s} \cosh \theta$ which are
meromorphic in the strip $-\pi/4< Im(\theta) < \pi/4$ and
bounded as we approach infinity inside this strip.
Let us first assume that these functions are actually holomorphic in the strip and then
we will mention what happens when there are poles.
 Equally
important they obey \beq l_{a,s}^+ + l_{4-a,s}^-  - l_{a+1,s} -
l_{a-1,s} = \log Y_{a,s}^+ + \log Y_{4-a,s}^-  - \log Y_{a+1,s}
- \log Y_{a-1,s}  \,.\nn \eeq The right hand side of this equality
is the logarithm of the left hand side of the Y-system
functional equations (\ref{Yone}) derived above. Now we go to
Fourier space where we have
 \beq {\cal F}(  l_{a,s}^\pm )(\omega)= e^{\mp \frac{\pi
\omega}{4}} {\cal F}( l_{a,s})(\omega)\nn \eeq where ${\cal F}$
denotes the Fourier transform. When writing this relation we
are making use of the analytic properties of $l_{a,s}$
mentioned above. The $Y$-system equations can then be cast as
\beq A_{aa'}(\omega)  {\cal F } (l_{a',s})(\omega) = {\cal
F}\left(
\log\frac{(1+Y_{a,s+1})(1+Y_{4-a,s-1})}{(1+Y_{a-1,s})(1+Y_{a+1,s})}
\right)(\omega)\nn \eeq For $w\ne0$, the $3\times 3$ matrix
$A_{aa'}(\omega)$ is invertible and we can multiply this
relation by $A^{-1}(w)$ to extract ${\cal F}(l_{a,s})$. For
$w=0$ the matrix is not invertible and therefore when doing
this operation we should allow for the constant zero modes in
the final result. Finally, we rewrite the corresponding
expression for ${\cal F}(l_{a,s})$ in position space. We obtain
in this way the final set of integral equations
\beqa \log Y_{2,s}&=&-m_{s} \sqrt{2}  \cosh\theta
-K_2\star
\alpha_s-K_1 \star \beta_s \nn \\
\log Y_{1,s}&=&-m_{s}  \cosh\theta -C_s-\frac{1}{2}
K_2\star \beta_s-K_1 \star \alpha_s - \frac{1}{2}K_3 \star
\gamma_s
\label{Integraleq}\\
\log Y_{3,s}&=&-m_{s}  \cosh\theta+C_s-\frac{1}{2}
K_2\star \beta_s-K_1 \star \alpha_s + \frac{1}{2}K_3 \star
\gamma_s \nn \eeqa where $\alpha,\beta,\gamma$ are short-hand
for \beqa &&\alpha_{s} \equiv \log \frac{\(1+  Y_{1,s}\)  \(1+
Y_{3,s}\) }{\(1+Y_{2,s-1}\)\(1+Y_{2,s+1}\)} \,\, , \,\,
\gamma_{s} \equiv \log \frac{ \(1+ Y_{1,s-1}\)\(1+
Y_{3,s+1}\)}{ \(1+  Y_{1,s+1}\)\(1+ Y_{3,s-1}\) } \,,  \nn \\
&&\beta_{s} \equiv \log \frac{\(1+Y_{2,s}\)^2 }{\(1+
Y_{1,s-1}\)\(1+ Y_{1,s+1}\)\(1+ Y_{3,s-1}\)\(1+ Y_{3,s+1}\)}
\la{integrands} \,, \eeqa and the kernels read \beq K_1\equiv
\frac{1}{2\pi \cosh\theta} \,\,  , \qquad K_2=\frac{\sqrt{2}\cosh
\theta}{\pi \cosh 2\theta} \,\, ,\qquad K_3=\frac{i}{\pi} \tanh
2\theta \,.\nn \eeq The unusual appearance of a kernel which does
not decay at infinity ($K_3$) is a direct consequence of the
singular behavior of $A(w)$ at $w=0$.

Comparing the large $\theta$ asymptotics following from these
equations with those predicted from the WKB analysis we see
that the zero modes $C_s$ correspond precisely to the constants
$C_s$ in (\ref{asymconone}) while the $D_s$ in the WKB
asymptotics are given by $D_s = \frac{i}{\pi} \int d\theta \,
\gamma_s(\theta)$.

A more straightforward exercise, compared with deriving the
integral equations, is to check that they indeed yield the
functional relations. To do so we simply compute the left hand
side of the functional equations using the integral equations.
When doing this we should use \beq f^{\pm}=f(\theta \pm i \pi
/4 \mp i0) \nn \eeq in order not to touch the lines where the
kernels $K_2$ and $K_3$ become singular. Then, simple
identities such as  $K_1^++K_1^-- K_2=0$ and
$K_2^++K_2^--2K_1=\delta(\theta)$ eliminate all the kernels in
the right hand side of the integral equations and the
functional equations are indeed reproduced.

Up to now we have discussed the case where all masses
are real and positive. To consider the case of complex masses
$m_s=|m_s| e^{i\varphi_s}$, we proceed in exactly
 the same way as described in section \ref{AdS3TBA} for the $AdS_3$ TBA.
 That is, for small phases $\varphi_s$ the integral equations take the same form as in (\ref{Integraleq}) with
$$
m_s\to|m_s|~,\quad Y_{a,s}(\theta)\to Y_{a,s}(\theta+i\varphi_s)~,\quad K_{s,s'}^{a,a'}(\theta-\theta')\to K_{s,s'}^{a,a'}(\theta-\theta'+i\varphi_s-i\varphi_{s'})~,
$$
where $K$ stands for the three different kernels. At
$|\varphi_s-\varphi_{s+1}|=\pi/4,\pi/2,3\pi/4,\dots$
 we pick the poles from the appropriate kernels
 (see section \ref{CRsec} and appendix \ref{WallCrossing} for illustration).
  All in all, the Y's and therefore the area are continuous
  whereas the apparent jumps in the integral equations are
   just an issue of the choice of contour.

\subsection{Simple combinations of $Y$ functions and $s_{n+1} \to s_1$ monodromies}
\la{monodromies}

When we normalize the solutions as in \nref{norsol} it can happen that $s_{n+1}$ is not equal to
$s_1$. Of course, they have to be proportional to each other. The proportionality constant is
called a ``formal monodrom''. For $n$ odd,  this constant can be removed, by rescaling the solutions
appropriately. For $n$ even, there is some non-trivial gauge invariant information in this constant.
In fact, this non-trivial information is a particular combination of $Y$ functions which is particularly
simple.


This is  most interesting for $n=4k$ so let us consider that case first.
We introduce three combinations of $Y$ functions
\beqa
\nn B_1&=&\(Y_{1,n-5} Y_{2,n-6} Y_{1,n-7}\) \(Y_{3,n-9} Y_{2,n-10} Y_{3,n-11}\)^{-1} \(Y_{1,n-13} Y_{2,n-14} Y_{1,n-15}\) \dots \\
B_2 &=& ( Y_{2,n-5} Y_{1,n-6} Y_{3,n-6} Y_{2,n-7} )( Y_{2,n-9} Y_{1,n-10} Y_{3,n-10} Y_{2,n-11} )^{-1} \dots
\nn \\
B_3 &=&\(Y_{3,n-5} Y_{2,n-6} Y_{3,n-7}\) \(Y_{1,n-9} Y_{2,n-10} Y_{1,n-11}\)^{-1} \(Y_{3,n-13} Y_{2,n-14} Y_{3,n-15}\) \dots  \nn
\eeqa
They are the analogue of $B(\zeta)$, equation (\ref{TheB}), in the $AdS_3$ case.
The $Y$-system alone, without recurring to the definition of the $Y$-functions, implies the following equations for $B_a$:
\beqa
\frac{B_2^+ B_2^-}{B_1 B_3}=\frac{B_1^- B_3^+}{B_2}=\frac{B^-_3 B^+_1}{B_2}=1\,.\nn
\eeqa
They have the form of a discrete Laplace equation for $\log(B_a)$ with a non-diagonal metric. Given the expected analytic properties of this functions the solution to these equations is
\beq \nn
B_1=\mu^{-1} \(e^{ -{ w_0 \over \zeta } - {\bar w}_0 \zeta}\)^{[-n + 2]},\quad B_2=\(e^{ - \sqrt{2}({ w_0 \over \zeta } + {\bar w}_0 \zeta)}\)^{[-n + 2]},\quad B_3=\mu\(e^{ -{ w_0 \over \zeta } - {\bar w}_0 \zeta}\)^{[-n + 2]}
\eeq
The value of $w_0$ can be found by computing the large $\theta$ asymptotics in the definiton of the $B_a$'s. We find that $$
w_0=(m_{n-5}+\sqrt 2 m_{n-6}+m_{n-7})-(m_{n-9}+\sqrt 2m_{n-10}+m_{n-11})+\dots
$$
is the cycle around infinity and $\bar w_0$ its complex conjugate. In  our normalization these quantities turn out to be equal to $B_a = T_{a,n-4}$ hence we have not only derived the form (\ref{Tform}) but we also computed $w_0$ from the $Y$-system equations.

For $n = 4k + 2 $ we only have one simple combination which is
\beq\la{muRelation}
\prod\limits_{k=1}^{\frac{n-4}{2}}\frac{ Y_{3,2k-1}}{Y_{1,2k-1}} =\mu^2 \,.
\eeq
This relation should be thought of as a gauge invariant \textit{definition} of $\mu$.
The fact that $\mu$ is constant also follows directly from the integral equations.

These relations imply some identities on the constants appearing in
the WKB asymptotics. By considering the small and large $\zeta$
limit  of \nref{muRelation} we find the expression for $\mu$ in
terms of the $C_s$   \beqa \log \mu^2 = \sum_{k=1}^{\frac{n-4}{2}}
C_{2k-1}\nn \eeqa In addition, we find a relation on the $D_{s}$ of the
form \beq \la{constD} 0= \sum_{k=1}^{\frac{n-4}{2}} D_{2k-1} \eeq
Note that we are viewing the $C_s$ as arbitrary constants, while the
$D_s$ are determined
by solving the integral equation.
The results in this subsection can also be obtained by a direct
analysis of the solutions at infinity, as is explained in appendix
\ref{MoreDetails}.

\subsection{Area and free energy}

In this subsection we show that  the area is related to the
free energy of the TBA system. The final result is
\nref{finalarea}. The reader that is not interested in this
derivation can jump to the final result.

 Note that the area is independent of $\zeta$. However, our objective is to relate the
cross ratios to the area. For this purpose, it is convenient to
consider the small
 $\zeta$ expansion of the $Y$  functions. We have already encountered the
 small $\zeta $ expansion when we looked at the asymptotic conditions.
 This expansion can be viewed as a WKB expansion where $\zeta$ is playing the role of $\hbar$.
We take complex masses but with small enough phases so that the WKB approximations that
we did before continue to be valid, with the same choice of cyl
For our purpose, we need  to expand these functions to first
order in  $\zeta$. For small $\zeta$ we diagonalize $\Phi_z =
{\rm diag}(x_i )$, where $x_i$ are roots of
 $det(\Phi_z -x) =0$. In our case, we have $x^4 - P(z)=0$.
 The $x_i$ are different sheets of a Riemann surface over the $z$ plane.
 Hitchin's equations imply that $A_{\bar z }$ is also diagonal in the same basis. By a diagonal gauge
 transformation we can set $A_{\bar z }\to 0$.
 In general, $A_{ z}$ is not diagonal in this basis.
 To order $\zeta^0$, only the diagonal components are relevant.
 Let us call these diagonal components by $\alpha^i_z$. Again, we can think of $\alpha$
as a one form on the Riemann surface. This is a closed form $d
\alpha =0$ due to Hitchin's equations.

 When we evaluate the cross ratios we will be evaluating integrals of the form
 \beqa \la{innertoeval}
 \langle i | U |i \rangle = \langle i | P e^{ - \int { x_i dz \over \zeta } + A_z dz +
 \zeta \Phi_{\bar z} d \bar z } |i\rangle
 \eeqa
 Here $U$ is the ``evolution operator'' taking us between two points in the $z$ plane.
 The states $|i\rangle$ indicate that we follow a given branch, since changing branches is suppressed by
 an exponential amount. In other words,
 in the WKB approximation, we are following the ``ground state" and the
 excited states are integrated out and  change the evolution  (at order $\zeta$ and higher) of  the ground
 state.  Expanding to order $\zeta$ we find that  \nref{innertoeval} has the
 form
   \beqa \la{innertoevalexp}
\log \langle i | U | i \rangle  =   - \left[ \int { x_i dz
\over \zeta } + \alpha^i_z dz +
 \zeta  ( u^i_{z} d z + u^i_{\bar z } d \bar z   )  + o (\zeta^2 )   \right]
 \eeqa
where $u^i$ is a certain one form. This one form is closed
because the connection is flat. The $u_{\bar z}$ components of
this one form are simple, they only come from the diagonal
components of $\Phi_{\bar z }$,
 $ u^i_{\bar z} = \Phi_{\bar z}^{ii} $.
 The $u_z$ component comes from a term that involves the off diagonal components of $A_z$, but we will not need
the explicit expression here. \footnote{ If you are curious,
the expression is
 $u^i_z = \sum_{k \not = i } { A_{z}^{ik} A_{z}^{ki}
 \over (x_i - x_k) } $. This is a familiar formula from second order perturbation theory (the $x_i$ are the
 energy levels). From Hitchin's equations one can directly check that
 $ d u =0$. }

 Once we form cross ratios, we find   integrals over closed cycles on the Riemman surface of certain closed one form
 forms. Thus we obtain
 \beqa \la{expansionx}
  \log \widehat Y_{\gamma } = - \left[ { \oint_\gamma x dz \over \zeta }  + \oint_\gamma \alpha + \zeta  \oint_\gamma u  +
   o (\zeta^2) \right]
   \eeqa
   where $\alpha$ and $u$ are viewed now as forms on the Riemann surface (in the $i^{\rm th}$ sheet they are $\alpha^i$ or
   $u^i$ respectively).
   Since they were constructed from a flat connection, we see that the quantities in \nref{expansionx} do not
  depend on the precise choice of contour within its cohomology class. Thus, we can view them as conserved quantities
  of the Hitchin equation. In fact, we could expand to higher powers in $\zeta$ and find extra conserved quantities,
   though \nref{expansionx} is  enough for our purposes.

 In the gauge where  we diagonalize $\Phi_z$,  we can  rewrite the expression for the area as  
  \beqa
   A &=&
   \int d^2z Tr[ \Phi_z \Phi_{\bar z } ] =    \int d^2z \sum_i x_i \Phi_{\bar z}^{ii} =  { i \over 2 } \int
    \sum_i x^i dz \wedge u^i_{\bar z} d \bar z
   \nn \\ \la{areainter}
    A &=&  { i \over 2 }  \int \,   x dz \wedge u  = - { i \over 2} w_{\gamma , \gamma'} \oint_{\gamma} x d z \oint_{\gamma' }  u
    \eeqa
  In the second line we expressed the integral as an integral over the Riemann surface. We then used a complete
   basis of cycles indexed by $\gamma$, and we used  a formula that is valid
   when we integrate products of closed forms. Here $w_{\gamma, \gamma'}$ is the inverse
  of the intersection form for the cycles $ \langle \gamma , \gamma' \rangle $.
  This step is valid when the number
   of cycles is even (odd number of gluons), otherwise we will not find an inverse.
    In our
   particular problem, we will choose the basis of cycles to consist of all the WKB cycles associated
   to the $Y_{a,s}$ functions.
   It is convenient to undo the shifts that
  define the $Y$ functions and define $\widehat Y_{a,s}$ functions which consist of various cross ratios evaluated
  at the same value of $\zeta$, see  \nref{explicitY} in appendix \ref{explicit}.
  We then do the WKB expansion for a $\zeta$ with
 phase $e^{i \pi /8}$, so that all the $\widehat Y_{a,s}$ are associated to a corresponding cycle.

 It is useful to compute the small $\zeta$ expansion of the $\widehat Y_{a,s}$ functions from the
  integral equations. We obtain
    \beqa \la{Yexpint}
 \log \widehat Y_{a,s}  =   { Z_{a,s} \over \zeta} + a_{a,s} + \zeta \left[ \bar Z_{a,s}  +
  M^{a,s ; a',s' }  \int
     { d \theta' \over 2 \pi }  e^{ -\theta'}
  \log (1 +  \widehat Y_{a,s}(\theta')  ) \right]
  \eeqa
  where  $M^{a,s ; a',s'}$, or $M^{\gamma , \gamma'}$,
  is  a certain matrix that results from the small $\zeta$
  expansion of the integral equation for $\widehat Y_{a,s}$.
    By comparing \nref{Yexpint} with \nref{expansionx} we can read off the values of $ \oint_{\gamma_{a,s}} u$.
     We obtain
     \beqa
     - \oint_{\gamma_{a,s}} x d z   & = & Z_{a,s}\nn
     \\
     -  \oint_{\gamma_{a,s} } u  &=& \bar Z_{a,s} + M^{a,s; a',s'}  \int
     { d \theta' \over 2 \pi }  e^{ -\theta'}
   \log (1 +  \widehat Y_{a,s}(\theta')  )\nn
   \eeqa
   When we insert these expressions in
     \nref{areainter} we get two terms $A = A_{periods} + A_{free} $ which are equal to
     \beqa
     A_{periods} &=& - { i \over  2 } w_{\gamma , \gamma'} Z_\gamma \bar Z_{\gamma' } \nn
     \\
     A_{free} &=& - { i \over  2} Z_{\gamma} w_{\gamma , \gamma' } M^{\gamma' , \gamma''} \int
     { d \theta' \over 2 \pi }  e^{ -\theta'}
   \log (1 +  \widehat Y_{a,s}(\theta')  ) \nn
   \\
   A_{free} &=&  - 2
        \sum_{a,s}  \int { d \theta \over 2 \pi } Z_{a,s} e^{-\theta} \log [
        1 + \widehat Y_{a,s}(\theta ) ] \la{freealmost}
        \eeqa      Here we have used that $Z_\gamma w_{\gamma , \gamma' } M^{\gamma' , \gamma''} =
         - 4 i Z_{\gamma''} $
     for
        our case,  due to the relation between the $Z_\gamma$ (or $m_{a,s}$) with various values of $a$
        implicit in \nref{asymconone}.
    The matrix of cycle intersections is given in appendix \ref{AdSfiveAsymptotics}, figure \ref{PB}.
       We can take the average of \nref{freealmost} with a similar expression which we
        obtain if we did the large $\zeta$ expansion to obtain
               \beqa
        A_{free} &=&  - \sum_{a,s}  \int { d \theta \over 2 \pi } [Z_{a,s}
        e^{-\theta} + \bar Z_{a,s} e^\theta ]   \log [
        1 + \widehat Y_{a,s}(\theta ) ] \la{alfinalarea}
        \\
        A_{free} &=&  \sum_{s}  \int { d \theta \over 2 \pi } | m_s|\cosh \theta
        \log \left[(1 + Y_{1,s})(1 + Y_{3,s}) (1 + Y_{2,s})^{\sqrt{2} } \right](\theta + i \alpha_s)
       \la{finalarea}
       \eeqa
       where $e^{i \alpha_s}$ is the phase of $m_s$. In the last equation we have written the final answer in
       terms of the $Y$ functions (as opposed to the $\widehat Y$ functions). We have also used the relation between
        $m_s$ and $Z_{a,s}$, implicit in \nref{asymconone} and explicit in
         appendix \ref{AperiodsSec}, eqn. \nref{massZAdS5}. The explicit value of $A_{periods}$ for
         in terms of the masses $m_s$ is given in appendix \ref{AperiodsSec}.

One can give an alternative derivation of this formula in the
spirit of the one given in \cite{Alday:2009dv} for the $n=6$
case. This alternative derivation starts with the observation
that the area $A$ can be viewed as the generating function of
transformations that change $\zeta$. This is the generating
function when we define a Poisson bracket for the Hitchin
system that makes $\Phi_z$ and $\Phi_{\bar z}$ conjugate
variables and similarly for $A_z$ and $  A_{\bar z} $. Then one
uses that the $\widehat Y $ are variables whose Poisson brackets
are computable and related to the cycle intersections. Finally
one uses the integral equations as above to expand the $\widehat Y$
functions for small $\zeta$, to find the Poisson brackets of
the quantities involved in this expansion. In this way one can
check that the final expression for the area does indeed
generate the desired transformation.

 Note that if we view the Hitchin system as arising from an ${\cal N}=2$ supersymmetric theory,
 then  the $Z_4$ projection that we had would break
 ${\cal N}=2$ to ${\cal N}=1$ supersymmetry. This ${\cal N}=1$ theory
has a global symmetry for rotations of $\Phi_z$ and $\Phi_{\bar
z}$ in opposite directions.
 The area is then  the $D$ term potential (or momentum map) for this symmetry.
 This connection between a four dimensional supersymmetric theory and two dimensional quantum integrable
models is in the spirit of \cite{Nekrasov}. It would be
interesting to find the precise relation.  Note however, that
in our $AdS_5$ problem we do not have supersymmetry, we have
{\it integrability}.

\subsection{The geometrical meaning of the Y-functions}
\la{CRsec}

In the previous sections we saw how to determine the area of the
minimal surfaces for a given choice of masses and chemical
potentials in the $Y$-system equations. To identify which polygons
correspond to a given choice of masses and chemical potentials we
must compute the space-time cross ratios (\ref{CrossRatios}) for
these solutions.
 This can be done by evaluating the $Y$-functions at special values of the spectral parameter
 as we explain in this section. In fact, the $Y$ functions themselves are cross ratios, but written in
 terms of somewhat exotic variables,  introduced in \cite{Hodges},
  as we explain in the next subsection, \ref{TwistorCR}. In subsection \ref{normalCR} we explain
 how to obtain the more conventional cross ratios defined in terms of distances between cusps.

\subsubsection{Relation between the $Y$ functions and twistor cross ratios}
\la{TwistorCR}

 As we explained in the introduction, we can recover
the form of the coset representative, $g^{-1}$ by picking a set of
independent solutions of the linear problem, see \ref{flatcondj}. In
our case, this is a set of four solutions $\psi_a$ of the linear
problem. Each of these solutions is a spinor and $a=1,2,3,4$ labels
the solution number. We can orthonormalize them so that $g^{-1}$ is
a proper group element. These solutions then determine the shape of
the string worldsheet in spacetime. Rotations of the $a$ indices by
a group element $g_0$ corresponds to the $SO(2,4)$ $AdS$ isometries, $g \to g_0 g $.
A given solution, say $\psi_a$, can be expanded in terms of four other solutions
 as\footnote{It turns out that $\tilde \eta^i = - \lambda^{i+1} $. This follows by continuing this formula
to the $i+1$ Stokes sector and reexpressing $ s_{i-1}$ in terms of the $s_j$ appearing in \nref{psiexp} for that Stokes sector. }
 \beqa \la{psiexp}
\psi_a
= \lambda^i_a s_{i+2} + \eta^i_a s_{i+1} + \tilde \eta^i_a s_{i-1} +
\hat \eta^i_a s_i
 \eeqa
where $s_{i+2}$ is the big solution in the
Stokes sector $i$, and $s_{i\pm1 }$ are the intermediate solutions
in Stokes sector $i$. As we a go to infinity within this Stokes sector $s_{i+2}
\to \infty$. Thus, this is the dominant solution and it determines the
behavior of $g^{-1} $ and eventually, also for the spacetime solution. We
can calculate $\lambda^i_a$ as
 \beqa \la{expr}
 \lambda^i_a = \langle   s_{i-1}  , s_i ,  s_{i+1} ,\psi_a \rangle = \langle \bar s_i ,\psi_a \rangle
 \eeqa
Here $\lambda^i$ is a spinor of the spacetime conformal group, which acts on the index $a$.

Now, imagine that we want to evaluate inner products of the $\lambda^i$. In these inner products
we contract the $a$ indices of the four spinors with an $\epsilon$ symbol. Notice that these are
indices transforming under global conformal transformations.
We obtain
\beqa
\langle \lambda^i ,\lambda^j, \lambda^k, \lambda^l \rangle =
\epsilon^{a,b,c,d} \lambda^i_a \lambda^j_b \lambda^k_c \lambda^l_c  = \langle \bar s_i ,\bar s_j,
\bar s_k ,\bar s_l \rangle\nn
\eeqa
Here we have used that, by performing a local gauge transformation and a global transformation,
 we
can set the $\psi_{\alpha , a} = \delta_{\alpha , a}$ at some point on the worldsheet.

Using this formula plus the identities \nref{re1}, \nref{re2} we can show that
\beqa
 \langle s_i , s_{i+1} ,  s_j ,  s_{j+1} \rangle &=& \langle \lambda^i , \lambda^{i+1},
 \lambda^{j} , \lambda^{j+1} \rangle\nn
\\
\langle \bar s_i , s_j \rangle =
 \langle s_{i-1}, s_i , s_{i+1} ,  s_j   \rangle &=& \langle \lambda^i , \lambda^{j-1},
 \lambda^{j} , \lambda^{j+1} \rangle  \equiv \langle \lambda^i , \bar \lambda^j
  \rangle\nn
 \eeqa
 where ${\bar \lambda}^i = \lambda^{i-1} \wedge \lambda^i \wedge \lambda^{i+1}$.
 Using these relations we can express the $Y$ functions in terms of the $\lambda^i$, which are
 spacetime quantities. These ``momentum twistors"   can be introduced  just from the
 knowledge of the position of the cusps \cite{Hodges}.
 In order to introduce them, we only need to know that
 we have a null sided polygon. When we introduce the $\lambda^i$ from this latter point of view, the
 $\lambda^i$ are defined up to an overall rescaling. In \nref{expr} we have picked a particular normalization.
 However, in the final expressions for the $Y$ functions in terms of $\lambda$, the overall normalization of
 each $\lambda^i$ drops out, for the same reason that the overall normalization of the $s_i$ drops out.
 Each $\lambda^i$ is associated to a null side of the polygon, and a pair of consecutive $\lambda$s determine
 the position of the cusp $X^i_{ab} = \lambda^i_{[a} \lambda^{i+1}_{b]}$, where $X^i_{ab}$ are six coordinates
 defined up to a rescaling obeying $X^2 =0$. Thus, they define a point on the boundary of $AdS$ space.

Note that the $Y_{1,s}$ and $Y_{3,s}$ only differ by $s
\leftrightarrow \bar s$, or $\lambda \leftrightarrow \bar \lambda $,
see appendix \ref{explicit}. This operation is simply target space
parity.

\subsubsection{ Traditional cross ratios from the $Y$ functions }
\label{normalCR}

In this subsection, we explain how to obtain traditional cross ratios from the $Y$ function.
By a ``traditional" cross ratio we mean one constructed from physical distances as in (\ref{CrossRatios}).
These can be introduced via
\beqa
U_{s}^{[r]} \equiv \left.1+ { 1\over   Y_{2,s} } \right|_{ \theta = i\pi r /4 }   =
\left.{T_{2,s}^+ T_{2,s}^- \over T_{2,s+1} T_{2,s-1} } \right|_{ \theta=i\pi r/4 }\,.\nn
\eeqa
where we combined   the definition of $Y$-function in terms of the $T$-functions with Hirota equation
\nref{Hirotatwist}.
This ratio has been constructed so that it   only involves the functions $T_{2,s}$.
These are determinants of four small solutions of the form $\<s_{i}s_{i+1}s_{j}s_{j+1}\>$.
 We recall that the physical cross ratios are ratios of four such quantities (\ref{CrossRatios}).
  For example,
\beq\la{crosseven}
U_{2k-2}^{[0]} ={\<s_{-k},s_{-k+1},s_{k},s_{k+1}\>\<s_{-k-1},s_{-k},s_{k-1},s_k\>\over \<s_{-k-1},s_{-k},s_k,s_{k+1}\>\<s_{-k},s_{-k+1},s_{k-1},s_k\>}={\bx_{-k,k}^2\bx_{-k-1,k-1}^2\over \bx_{-k-1,k}^2\bx_{-k,k-1}^2}~,
\eeq
\begin{figure}[t]
\center\includegraphics[width=100mm]{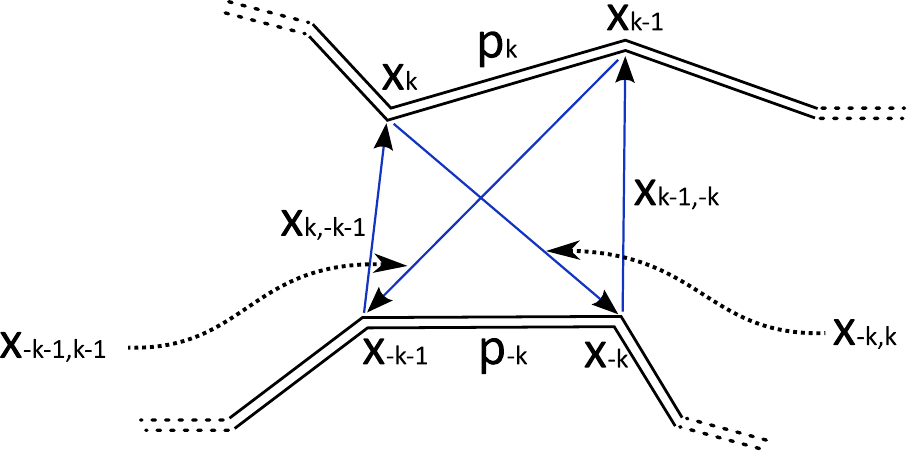}
\caption{
The cross ratio $U_{2k-2}^{[0]}={\bx_{-k,k}^2\bx_{-k-1,k-1}^2\over \bx_{-k-1,k}^2\bx_{-k,k-1}^2}$.
} \la{CRfig}
\end{figure}
where $\bx_{i,j}\equiv \bx_i-\bx_j$, see figure \ref{CRfig}.
 If we consider $U_{2k-2}^{[2p]}$ then we will shift the position of the small solutions by $p$ sectors, i.e. we will get a cross ratio involving the cusps $\{\bx_{-k-1+p},\bx_{-k+p}\}$ and $\{\bx_{k-1+p},\bx_{k+p}\}$ of the polygon.
We see that the index $2k-2$ is
the number of cusps between $\bx_{-k+p}$ and  $\bx_{k+p-1}$ (counted from the side of the cusp $\bx_0$). Similarly, the cross ratio involving sides $\{\bx_{-k-1+p},\bx_{-k+p}\}$ and $\{\bx_{k+p},\bx_{k+1+p}\}$, which are separated by an odd number of cusps, are given by $U_{2k-1}^{[2p+1]}$.

A generic cross ratio involving four non-consecutive cusps of the polygon can easily be constructed by multiplying appropriate factors of $U_{s}^{[m]}$, e.g.
\beq
{\bx_{-2,1}^2  \bx_{-1,q+1}^2 \over  \bx_{-2,q+1}^2 \bx_{-1,1}^2}= \prod_{l=1}^{q} U_{l}^{[l]} \,.\nn
\eeq
We see that once we solved the integral equations it suffices to evaluate the $Y$-functions $Y_{2,s}$ at some specific values of $\theta$ to read off the corresponding cross ratios.

A subtlety of pragmatic nature is the following. When solving the integral equation we usually do it numerically by iterating the integral equations. At the end of the iteration cycle we are left
with the functions $Y_{a,s}(\theta)$ in the real axis which is precisely what we need to compute the
free energy (\ref{finalarea}).  On the other hand, to get the physical cross ratios we will generically need to evaluate $Y_{2,s}$ at some imaginary values. Using the integral equations, these cross ratios can be written in terms of (integrals of) the $Y$-functions evaluated in the real axis. When continuing the integral equations (\ref{Integraleq}) out of the physical strip we will have to pick some extra pole contributions from the several kernels as we cross the lines ${\rm Im}(\theta)= \pm i \pi/4, \pm i\pi/2$ etc.

To make this procedure clear let us illustrate how to compute $Y_{2,s}(i \pi/2)$. First we notice that $Y_{a,s}\(i\pi/4\)$ can still be expressed in terms of integrals of the $Y$-functions over the real axis using the
original undeformed equations (\ref{Integraleq}).
The only thing we must be careful about is that the kernels $K_2$ and $K_3$ have a pole singularity for $\theta=i\pi/4$ so that we should either interpret the integration contour to go slightly above the real axis or equivalently we should plug $\theta=i\pi/4-i0$ in the right hand side of these equations. To reach values of $\theta$ with an even larger imaginary part (such at $\theta=i\pi/2$) we can simply pick the pole singularities of these to kernels. For example
\beqa
&\log\(Y_{2,s}\)(\theta)=-m_{s} \sqrt{2}  \cosh(\theta) -K_2\star \alpha_s-K_1 \star \beta_s+\alpha_s(\theta-i\pi/4)\, \,\, ,\,\,\, \pi/4 < {\rm Im}(\theta) < \pi/2 \nn
\eeqa
We can now evaluate the right hand side of this equation at $i\pi/2-i0$ to compute $Y_{2,s}(i\pi/2)$. The last term becomes $\alpha_s(i\pi/4)$ and contains a bunch of $Y$-functions evaluated at $\theta=i\pi/4$. We already explained how to get those in terms of integrals of the $Y$-functions in the real axis using the original equations. Hence we are done.

There   is an even more efficient way of computing these cross ratios which goes as follows. Suppose we need some Y-function $Y_{a,s}(\theta)$ where $\theta$ is outside the physical strip, i.e. $| {\rm Im} (\theta) |>\pi/4$. For concreteness let us suppose $ {\rm Im} (\theta) < - \pi/4$. Then, by repeatedly using the functional equations (\ref{Ytwo}) as
\beq
Y_{a,m}={(1+Y_{a,m+1}^{[1]})(1+Y_{4-a,m-1}^{[1]})\over Y_{4-a,m}^{[2]}(1+1/Y_{a+1,m}^{[1]})(1+1/Y_{a-1,m}^{[1]})} \,, \nn
\eeq
we can express $Y_{a,s}(\theta)$ purely in terms of $Y$-functions inside the physical strip. For those we can use the original integral equations (\ref{Integraleq}) as explained above.

\subsubsection{  $\zeta$ symmetry}
We explained how to get the physical cross ratios by evaluating the $Y$-functions at $\theta=0,\pm i \pi /4,\pm i \pi /2$ etc.
Actually we can construct a family of polygons parametrized by a
 complex number $\theta_0$ which will all have the same area!
  These polygons are obtained by evaluating the $Y$-functions 
  at $\theta=\theta_0,\theta_0\pm i \pi/4 ,\theta_0 \pm i \pi/2$ etc.
   The reason why we can evaluate the $Y$-functions around any $\theta_0$
    is that flat sections of the linear problem can be assembled into a physical 
    solution for general $\zeta=e^{\theta_0}$ at not only 
    for $\zeta=1$.\footnote{Because 
    $\zeta\neq 1$ can be absorbed into a redefinition of 
    $P(z)\to \zeta^4 P(z)$ together with a $\zeta $ dependent global gauge transformation.} 
     For purely imaginary $\theta_0$ the new polygons are real, 
     for generic $\theta_0$ they are complexified. 
     Thus we can view changes of $\zeta$ as a symmetry of
     the problem. Namely, we have a one parameter family of polygons labeled by $\zeta$ which have the 
     same area. 
     It would be very interesting to understand this symmetry 
     in greater detail and specially to see if/how it manifests itself at the quantum level.
     
    Note that this is intimately related to the fact  that changes in $\zeta$ are 
     generated by the area when one defines a certain Poisson bracket that is natural in the theory 
     of Hitchin systems \cite{Alday:2009dv}. 
     This Poisson bracket is  different from  the one one gets from the 
     sigma model \cite{Mikhailov:2006uc}.

\subsection{High temperature limit}

In this section we will focus on a particular kinematical regime. From the Y-system point of
 view we want to consider the limit when the $Y$-functions are approximately constant
 (in some large region of $\theta$).
To find the constant values of the $Y$-functions
 we solve the $Y$-system equations (\ref{Ytwo}) and
  (\ref{muRelation}) dropping the $\zeta$ dependence, $Y_{a,s}^{\pm}\to Y_{a,s}$,
\beqa\la{HighTY}
{Y_{a,m} Y_{4-a,m} \over
Y_{a+1,m}Y_{a-1,m}}&=&{(1+Y_{a,m+1})(1+Y_{4-a,m-1})\over(1+Y_{a+1,m})(1+Y_{a-1,m})}
\la{highTY}
\eeqa
For this approximation to be valid, the sources in the integral
 equations must become independent of the spectral parameter which means $m_s\to 0$.
 In the TBA context this arises in the high temperature limit, so
 will adopt that terminology here. Of course, in the $AdS$ problem there is no temperature since we are just
 solving classical equations.
  The $m_s\to0$ condition is however not sufficient. Namely we do \textit{not} have the freedom to chose the values of the constants $C_s$ if we want the solution to the Y-system to be given by constant $Y$-functions. Instead these constants are \textit{found} from the $Y$-system.\footnote{In this sense the nomenclature \textit{high temperature} is slightly abusive. The genuine high temperature limit would correspond to $m_s \to 0$, with the chemical potentials $C_s$ arbitrary.} To derive this,
note that the ratio of the equations (\ref{highTY}) for $a=1$ and $a=3$ implies that $\gamma_s=0$
where $\gamma_s$ is defined in (\ref{integrands}). Thus, from equations  (\ref{Integraleq}), we see that
\beq
e^{2C_s} =  { Y_{3,s} \over Y_{1,s} } \,.\nn
\eeq
We will first consider the richer case of an even number of
gluons. The $n$ odd case is discussed afterwards.  The equations
\nref{highTY} admit a one parameter family of solutions.
We can parameterize this family in terms of
\beqa
\prod\limits_{k=1}^{\frac{n-4}{2}}\frac{
Y_{1,2k-1}}{Y_{3,2k-1}} = \frac{1}{\mu^2} \,.
\la{muHighT}
\eeqa
For a fixed value of $\mu$,    (\ref{HighTY}) admits
a discreet set of solutions. At $\mu=1$ there is a unique solution such
that $Y_{a,s}>0$ for any $a=1,2,3$ and $s=1,\dots,n-5$. We will
consider a solution valid for arbitrary $\mu$ and continuously
connected to that unique solution.
 This one parameter family describes a family of regular polygons with a $Z_n$ symmetry.
 This is shown in more detail  in appendix \ref{RegularPolygons}.

The polygon can be described in terms of the $n$  twistors
\beq \bar \lambda_k=\{  \(1+ a_k \) (1+ b_k )e^{\frac{3 i \pi
k}{n}}, (1+a_k)e^{ \frac{ i \pi k}{n}}, (1+b_k ) e^{-\frac{ i
\pi k}{n}},  e^{-\frac{3 i \pi k}{n}}\} \la{twistor} \eeq where
$\mu=e^{i \phi}$ and
$$a_k=(-1)^k  \tan  \frac{2\pi}{n} \tan \frac{\phi}{n}\,\,\, , \,\,\,
\qquad b_k=(-1)^k  \tan \frac{\pi}{n}  \tan  \frac{\phi}{n} \,.$$
Then the constant solution to the $Y$-system is simply obtained
by plugging the twistor $\bar \lambda_k$ in place of $s_k$ in the
relations (\ref{YAppendix}) in  appendix \ref{explicit}.

When $\phi=0$ we have $a_k=b_k=0$ and the solution simplifies
to \beq\la{Ysmuone}
Y_{a,s}+1=\frac{\sin\(\frac{\pi}{n}\(4-a+s\)
\)\sin\(\frac{\pi}{n}\(a+s\) \)}{\sin\(\frac{\pi}{n}\(4-a\) \)
\sin\(a\frac{\pi}{n}\)} \,\, . \eeq Actually, in this case we
have $Y_{a,s}=Y_{4-a,s}$ and hence the Y-system reduces to the
standard $Y$-system and the solution (\ref{Ysmuone}) can be
found in the literature \cite{kirillov}. In fact, this is a regular polygon
that can be embedded in $AdS_4$.  Another limit where we
find significant simplification is the limit where $\phi \to
\pi (n-4)/2$. In this limit we find $Y_{1,2k+1}=Y_{3,2k+1}=-1$,
$Y_{1,2k}=Y_{3,2k}=0$, $Y_{2,2k+1}\to \infty$ and
\beq
Y_{2,2k} \to \sin\(\pi\frac{k+2}{\hat
n}\)\sin\(\pi\frac{k}{\hat n}\)/\sin^2\(\frac{\pi}{\hat n}\)
\,\,\, , \,\,\, ~~~~  \hat n \equiv { n \over 2} \,.  \la{YAdS5AdS3}\eeq
This corresponds to a regular polygon that can be embedded in $AdS_3$.
In fact,  the curious pattern
of zeros and infinities for the $Y$-functions is a generic feature of the $AdS_3$ limit.
In the next section \ref{reductionsec}, this is
precisely how the $Y$-system in the $AdS_5$ strip reduces to
the $Y$-system in the $AdS_3$ line.
As we move $\phi $ between $\phi =0$ and $\phi = { n-4 \over 2 } \pi$ we
interpolate between these two cases.

It turns out that the free energy can be computed exactly in
this limit, as shown for instance in \cite{Fendley:1991xn}, and
reads
\beq\la{HighTFEe} F_n=-\frac{1}{2\pi}\sum_{a,s}\[
\log(e^{(2-a)C_s}Y_{a,s})\log(1+Y_{a,s})+2
Li_2(-Y_{a,s})\]~. \eeq
Plugging the analytic expressions above into
this expression we find the remarkably simple result
\beq\la{HighTFE} F_n =\frac{\pi}{2n} \( (n-4)(n-5) - \frac{4
\phi^2}{\pi^2}\) \,.
\eeq
As already mentioned, for $\phi=0$ we recover the regular
polygons of second class that can be embedded into $AdS_4$. One
can actually see that the free energy for $\phi=0$ exactly
reproduces the numerical prediction (\ref{Afreetwo}).
Similarly, for $\phi=\frac{n-4}{2}\pi$ we can see that
(\ref{HighTFE}) exactly reproduces the $AdS_3$ result (\ref{threefromfive}).

In addition to the solutions described above, there can be
discrete families of solutions. For instance, let us focus in
the $n$-odd case. In appendix \ref{RegularPolygons} we have
constructed a set of solutions parametrized by the number of
sides $n$ and an extra integer $r$, with $r=2,...,(n-1)/2$. The
spinors characterizing such solutions are
\begin{equation}
\lambda^{k}=(e^{i \pi(r+1)\frac{k}{n}},e^{i \pi(r-1)\frac{k}{n}},e^{-i \pi(r-1)\frac{k}{n}},e^{-i \pi(r+1)\frac{k}{n}}) \nn
\end{equation}
where we have reorganized the expression appearing in the
appendix. From these spinors, one can easily compute the
cross-ratios $Y_{a,s}$. As these polygons can be embedded into
$AdS_4$, we expect $\mu^2=1$ and indeed we find
$\frac{Y_{1,s}}{Y_{3,s}}=1$. In addition, one can explicitly
check that this cross-ratios give a solution of the $Y-$system
equations. This constitutes a non trivial check of the
$Y-$system for the case in which $n$ is odd.

 Let us comment on an interesting feature of \nref{HighTFE}. Note
that $\lambda$ in \nref{twistor} is a periodic function of $\phi$ with period $ n \pi$.
This means that the cross-ratios describing the polygon are periodic functions of $\phi$.
On the other hand we see that \nref{HighTFE} is not periodic! We thus have a family of solution all ending on the same polygon. If the right prescription is to sum over all these surfaces, then the full result will have no monodromy. In that sum, one solution will dominate while the others are non-perturbative corrections (in $1/\sqrt\lambda$). On the other hand, in terms of amplitudes we expect non-trivial analytic continuation properties. For example, it is well known that amplitudes have interesting monodromies as we analytically continue the external parameters. This is an essential tool in weak coupling computations, see \cite{smatrix,Dixon:1996wi} for example. If the right prescription is then \textit{not} to sum over all these surfaces, then it would be interesting to study this particular kinematic configuration and understand in a better way the physical meaning of the monodromy in \nref{HighTFE}.

Such non-trivial monodromies for the free energy are a common occurrence
in TBA systems. After an analytic continuation in the parameters we do not end up with the ground
state, but we end up in an excited state \cite{polesTBA}. In fact, the high temperature TBA typically
corresponds to a CFT. Then the chemical potential simply translates into a winding condition for
the scalar that bosonizes the corresponding $U(1)$ current and thus giving the $\phi^2$ contribution
to the energy,  see e.g. \cite{Fendley:1997ys,Fendley:1993jh}.
This, in particular, shows that in our problem some excited states will typically appear as we
analytically continue the parameter \cite{Fendley:1997ys}. These are related to poles in the physical region and need to be taken into account according to a well understood procedure \cite{polesTBA}. Namely, we need to
deform the contours and pick up the corresponding poles, etc.

\subsubsection{High temperature limit of $AdS_3$ $Y-$system
} \la{highT3sec}

We can consider the high temperature limit of the TBA equations  corresponding to scattering amplitudes on $AdS_3$ for $n=2\hat n$ gluons. In this case the $Y$-system equations (\ref{Yads3}) becomes simply $Y_s^2=(1+Y_{s+1})(1+Y_{s-1})$ and the constant solution reads
\begin{equation}
Y_s= \sin\left(\frac{\pi(s+2)}{\hat{n}} \right)\sin \left(\frac{\pi s }{\hat{n}} \right)/\sin^2\left(\frac{\pi}{\hat{n}}\right) \nn
\end{equation}
These are precisely the $Y$-functions found in the previous section in equation (\ref{YAdS5AdS3}). This is not surprising since we had already anticipated that for $\phi \to (n-4)\pi/2$ the polygons described by the previous solution become $AdS_3$ solutions.   The $AdS_3$ free energy can be then evaluated, exactly as in (\ref{HighTFEe}), and we obtain
\beq\la{HighTFEadsthree} F_n = -\frac{1}{2\pi}\sum_{s}\[
\log(Y_{s})\log(1+Y_{s})+2 Li_2(-Y_{s})\] = \pi
\frac{(n-6)(n-4)}{12n} \eeq
where we have reinstated $n=2\hat n$.
 This result is however \textit{not} the same as (\ref{HighTFE})
 for $\phi=(n-4)\pi/2$ even though we are describing the same solution.
  This is actually not a contradiction: as explained in appendix \ref{direct}
  the regularization of the area in the high temperature limit amounts to
  subtracting the area of $n-4$ regular pentagons in the $AdS_5$ case and
  the area of $n/2-2$ regular hexagons in the $AdS_3$ case. Taking this into account,
   the difference between two free energies is precisely as expected!

\subsection{$AdS_4$ and $AdS_3$ reductions}
\la{reductionsec}

Minimal surfaces that can be embedded in an $AdS_4$ or $AdS_3$
subspaces of $AdS_5$ are more restricted and as a result, the
problem simplifies. The reduction of the $AdS_5$ flat
connection was done in \cite{Alday:2009dv}. In this section, we
will consider the implications of this simplification to the
Y-system.

The worldsheet theory describing strings moving in an $AdS_4$
subspace is obtained from the parent $AdS_5$ by an additional
projection. This projection relates $\cA$ to $\cA^t$ via a
gauge transformation $F$. That gauge transformation therefore
relates solutions to the problem with solutions to the inverse
problem. More precisely we have  $\bar s_i(\zeta)=Fs_i(\zeta)$,
where in our normalization $\det(F)=1$. As a result,
$Y_{1,s}=Y_{3,s}$ (and hence $\mu^2=1$). The $Y$-system
equations can then be written in the form \beq
{Y_{2,m}^-Y_{2,m}^+\over
Y_{1,m}^2}={(1+Y_{2,m+1})(1+Y_{2,m-1})\over(1+ Y_{1,m})^2} \,\,
, \qquad {Y_{1,m}^-Y_{1,m}^+ \over Y_{2,m}}={(1+
Y_{1,m+1})(1+ Y_{1,m-1})\over 1+Y_{2,m}} \,.\nn \eeq

A solution in $AdS_4$ that can be embedded in $AdS_3$ must have
even number of gluons. The linear problem splits into two
decoupled problem denoted by \textit{left} and \textit{right}
problems in \cite{Alday:2009yn}. In an appropriate gauge we can
write \beq s_{2k}=\(\begin{array}{c}
s_{R,k} \\
0\\
\end{array}\) \,, \qquad s_{2k+1}=\(\begin{array}{c}
 0\\
s_{L,k+1} \\
\end{array}\)\nn
\eeq where $s_L$ and $s_R$ are the small solution of the left
and right $AdS_3$ problems respectively. Because of this
factorization the $T$-functions can be dramatically simplified. We choose a normalization where
the $AdS_3$ solutions obey $\langle s_{L,i} , s_{L,i+1} \rangle =1 $, and the same for the right.\footnote{
Note that with this choice, the $AdS_5$ solutions obey $\langle s_i , s_{i+1} ,s_{i+2}, s_{i+3} \rangle = -1$.}
The left and the right problems are related by a rotation in
the spectral parameter
$\<s_{R,a},s_{R,b}\>=\<s_{L,a},s_{L,b}\>^{[2]}$. We can use this
relation to translate all inner products into inner products of
the left problem. We also recall the definition,
$T_{k}\equiv\<s_{L,0},s_{L,k+1}\>^{[-2k-2]}$, of the Hirota functions
for the $AdS_3$ problem in   section
\ref{YsystemAdS3}. We then find that the the Hirota variables $T_{a,s}$ of the
$AdS_5$ problem have the form
\beqa
 T_{1,2k+1}(\zeta) &=&   0~,~~~~~~~~~~~~~ T_{0,s }  = T_{4,s} =-1 ~, \nn
 \\
 T_{1,2k}(\zeta) & =&  - \<s_{L,0},s_{L,k+1}\>^{[-2k]} = - T_k^{[2]} \,, \nn \\
T_{2,2k}(\zeta)  & = & - \<s_{R,0} , s_{R,k+1}\>^{[-2k-1]}
\<s_{L,0},s_{L,k+1}\>^{[-2k-1]}  = - T_k^{[3]} T_k^{[1]} \,\, , \nn \\
T_{2,2k+1}(\zeta)&  =&
\<s_{R,0},s_{R,k+1}\>^{[-2k-2]}\<s_{L,0},s_{L, k+2}\>^{[-2k-2]} = T_k^{[2]} T_{k+1}^{[2]}
\,, \nn \eeqa and of course $T_{3,s}(\zeta)=T_{1,s}(\zeta)$.

 \begin{figure}[t]
\center\includegraphics[width=110mm]{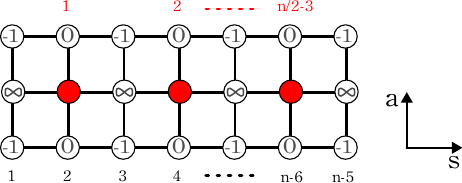}
\caption{
In the $AdS_3$ reduction most of the $3n-15$ functions  $Y_{a,s}$ become trivial (either zero, one or infinity as indicated in the corresponding node in the figure). The $n/2-3$ nontrivial functions $Y_{2,2k}\equiv Y_k$ are indicated by red circles. The $AdS_5$ $Y$-system in the original strip (with shifts of $i\pi/4$)   imply the $Y$-system equation for the line of red nodes $Y_k$ (with spectral parameter shifts of $i\pi/2$). }
 \la{reduction}
\end{figure}
The $AdS_5$
Hirota equations \nref{Hirotatwist} becomes identically satisfied in all nodes
except $s=2k$ and $a=2$. For these, it
becomes\footnote{We actually obtain this expression at
$e^{i\pi/2}\zeta$.} \beqa\la{Hirotareduction}
\( T_{k}  \)^2 \(  T_{k}^{[2]}T_{k}^{[-2]}-T_{k-1} T_{k+1} -1\)
=0~,\nn \eeqa  Inside the parentheses we recognize Hirota
equation (\ref{HirotaLine}) in $AdS_3$. Recall that in the
$AdS_3$ treatment supercripts denoted shifts in the spectral
parameter which were twice as large compared to the ones we are
using now, e.g. \beq \(f^{[2]}\)_{here}=\(f^+\)_{AdS_3
\,\,section} \,. \nn \eeq The $3n-15$ nodes in the  $AdS_5$
strip reduces to the $n/2-3$ nodes in the $AdS_3$ line in a
very funny way which we represent in figure \ref{reduction}.
Basically the only non-trivial functions become $Y_{2,2k}^{[-2]} =
T_{k+1} T_{k-1} \equiv Y_k$ and these obey the $AdS_3$ Y-system
equation (\ref{Yads3}).

\section{Conclusions}

In this paper we have given a way to compute the area of minimal surfaces that end
on a null polygon at the boundary of $AdS_5$. The method uses the integrability\footnote{ 
 The Yangian charges - responsible for integrability - are also visible at weak coupling \cite{Yangian}.
  Still, to this moment it is not clear how exploit integrability efficiently in that regime. 
  At strong coupling the connection between conformal and dual conformal symmetry and integrability was worked out in \cite{Yangian2}.} of the classical equations in an essential fashion.

We have used the map between the integrable sigma model and a Hitchin system.
The Hitchin system is an $SU(4)$, or $SU(2,2)$, or $SL(4)$ system depending on the signature.
More precisely, it is a certain $Z_4$ projection of this system.
Alternatively, we can simply say that we have chosen a specific form for the one
parameter family of flat connections. This family is parametrized by a spectral parameter
$\zeta = e^\theta$.

For this problem the worldsheet is the complex plane and we have an irregular singularity at
infinity. This means that there are Stokes sectors as we approach infinity.
Each of these Stokes sectors has an associated small solution $s_i$ and a large solution.
The large solution determines a four-spinor $\lambda_i$, which specifies the direction in which
the large solution is pointing. These spinors are associated to the sides
 of the polygon and are the same
as the momentum twistors introduced by Hodges \cite{Hodges}. Alternatively, we can say that consecutive
Stokes sectors determine a cusp or a vertex  of the polygon. Using this cusp positions, or the
momentum twistors $\lambda$, we can construct cross ratios. We can introduce a family of cross
ratios depending on the spectral parameter $\zeta$.
If $\zeta=1$   we recover the physical cross ratios of the original problem.

A particular set of cross ratios, denoted by the $Y$ functions $Y_{a,s}$, obeys a
set of functional equations, or $Y$ system equations \nref{Ytwo}.
The number of $Y$ functions is $3 (n-5)$, which is the same as the number of independent
cross ratios of the problem.
 These functional equations,
together with some input regarding their asymptotic properties, imply a set of integral
equations for the $Y_{a,s}$ functions \nref{Integraleq}.
These integral equations involve $3 (n-5)$ real parameters.
These parameters were not present in the $Y$ system function equations but they appeared in the
specification of the boundary conditions for the $Y$ functions. The solution to these equations
relates these parameters to the physical cross ratios. Roughly speaking these parameters are related
to the values of $Y$ at $\theta = \pm \infty$, while the physical cross ratios are $Y(\theta =1)$.
These functional equations have the form of Thermodynamic Bethe Ansatz equations for a certain
quantum system which is not associated in any obvious way to our initial problem, which was a
classical problem. Similar relations were observed in \cite{PDorey}. Moreover, the area is given by the
free energy of the TBA system.
In practice, the TBA equations can be solved numerically.
As in \cite{Alday:2009dv}, we have been able to to solve the
equations in a specific ``high temperature'' regime.
 This gives a one parameter family of solutions (for $n$ even).
  These describe a family of regular
 polygonal contours at the $AdS$ boundary. This is a one dimensional
 line in the $3(n-5)$ dimensional space
 of cross ratios.
 The answer for the area is surprisingly simple, it is just $A \propto  \phi^2$, where $\phi$ parametrizes the
family and it appears as  a
 chemical potential of the TBA system. The cross ratios
 are periodic functions of $\phi$. This simple form is expected from the TBA perspective, since it is
 describing the UV limit of the 2d quantum integrable theory which is a CFT, and $\phi$ appears as
 a chemical potential \cite{Fendley:1993jh}.
 One interesting aspect of this family of solutions is that, since we know it analytically, we can
 analytically continue in the space of parameters and find interesting monodromies.
 Namely, the cross ratios come back to themselves while the area does not.  It would be interesting
 to study them further. One thing we can say, is that this shows that we will get TBA equations for
 excited states as we do analytic continuations. By now, this is a familiar phenomenon \cite{polesTBA}.
 Thus the full problem involves not only the TBA ground state but also some excited states.

 There are several   interesting problems for  the future.
 One of them is to take the large $n$ limit in order to obtain arbitrary (spacelike or timelike) contours.
 This would effectively solve the strong coupling form of the loop equations \cite{Progress}.

 Of course, the most interesting open problem is the extension of this to the full quantum theory.
 This will probably require, as a first step, the knowledge of the classical solutions
 for the full $AdS_5 \times S^5$ sigma model.

 We have emphasized that we get the physical values of
 the cross ratios by evaluating the $Y$ functions at $\zeta =1$. However, we also get equally nice,
 but different, physical values by taking $\zeta = e^{i \varphi}$. By varying $\varphi$ we move in the
 space of cross ratios. All these values of the cross ratios have the same area!. Thus, changing
 $\zeta$ corresponds to a symmetry of the problem. Namely, by changing $\zeta$ we change the
 cross ratios in a way that does not change the area. Other values of $\zeta$, with $|\zeta | \not =1 $
 correspond to generically complex values of the cross ratios and represents an analytic continuation
 of the problem, an analytic continuation that keeps the area fixed. Recall also that, with a
 certain definition of Poisson brackets, the area is precisely the generating
  function for  this symmetry \cite{Alday:2009dv}.

 One curious observation is the following. We have the formula for the amplitude as:
 Amplitude $ = e^{ - { \sqrt{\lambda} \over 2 \pi } ( {\rm Area } ) } $. Since the area is the free energy, this
 formula looks like we are computing the partition function of the  system on a torus, where one
 of the sides has length proportional to $\sqrt{\lambda}$. For large $\sqrt{\lambda}$ only the
 ground state contributes, which is what we computed. The overall sign is not quite right for this
 interpretation. It is nevertheless suggestive.

\section*{Acknowledgments}

We thank N. Arkani-Hamed, D. Gaiotto, N. Gromov and G. Moore for useful discussions. The
work of L.F.A. and J.M. was supported in part by the U.S.
Department of Energy grant $\#$DE-FG02-90ER40542. The research of A.S. and P.V. has been supported in part by the Province of Ontario through ERA grant ER 06-02-293. Research at the Perimeter Institute is supported in part by the Government of Canada through NSERC and by the Province of Ontario
through MRI. A.S. and P.V. thanks the Institute for Advanced Studies for warm hospitality.

\appendix

\section{Numerics} \la{numerics}
\begin{figure}[t]
\includegraphics[width=80mm]{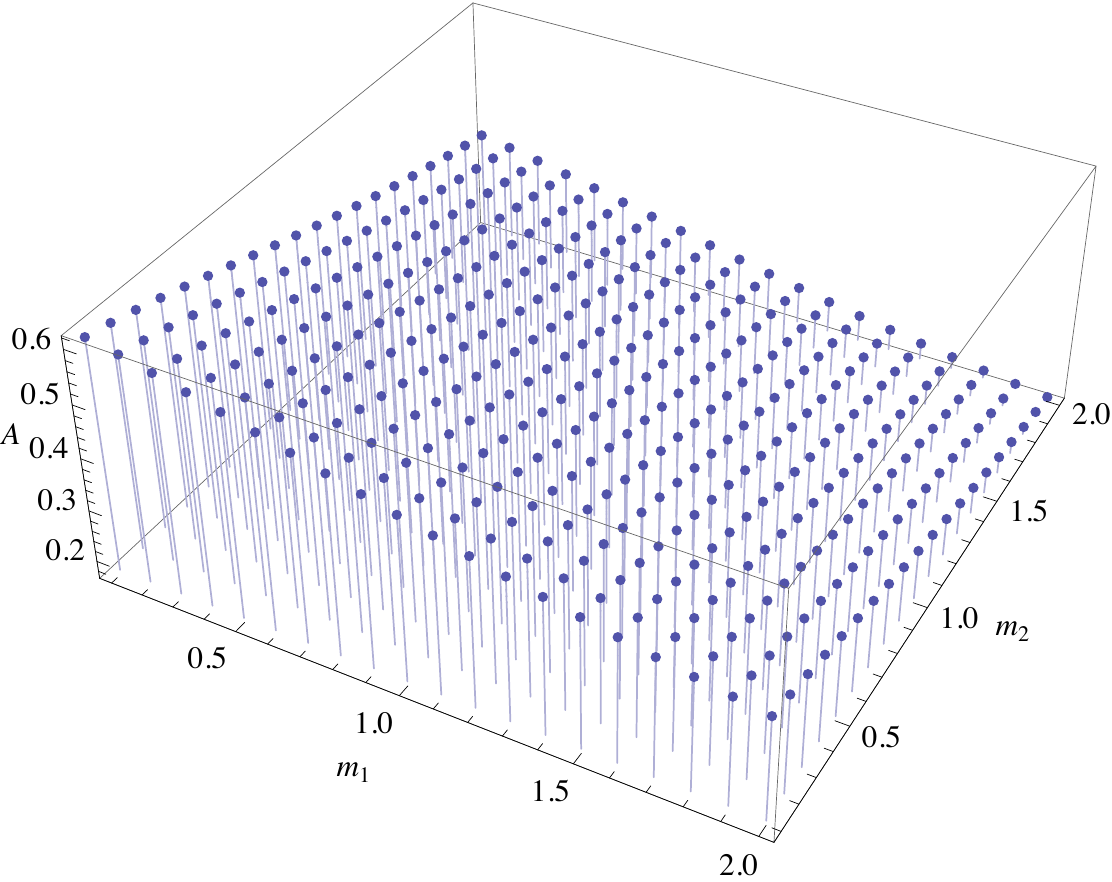}
\includegraphics[width=80mm]{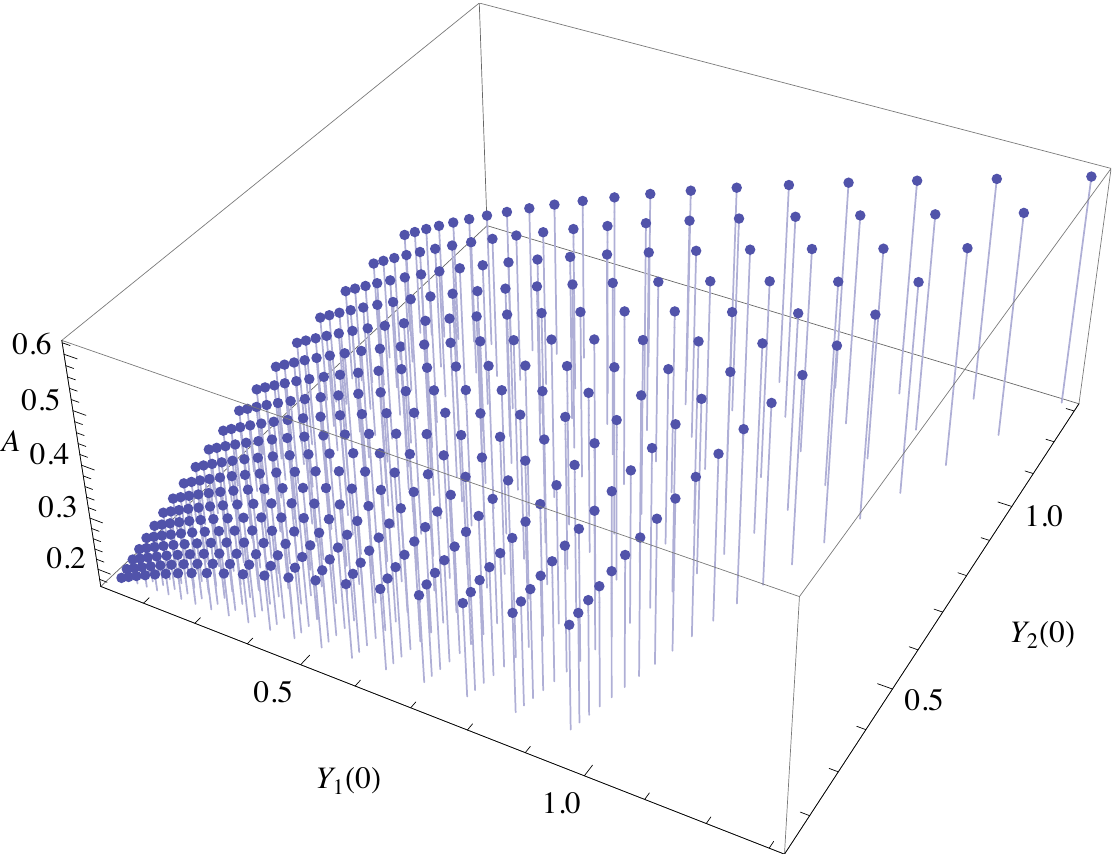}
\caption{Left: Regularized area as a function of two real masses
$m_1$ and $m_2$ for the decagon with $n=10$. When the masses are
very large the $Y$-functions are exponentially suppressed and the
free energy vanishes. When they are very small the free energy tends
to the analytic prediction $F=\pi/5 \simeq 0.63$ from the high
temperature limit.  Right: For those values of $m_1$ and $m_2$ the
cross ratios $Y_1(0)$ and $Y_2(0)$ cover a diamond shaped region.
The regularized area is plotted as function of these cross ratios. }
\la{decagonnumerics}
\end{figure}
In this section we explain how to implement equations (\ref{YAdS3})
numerically in \verb"Mathematica" in a very simple way (the code is
quite similar to the one used in \cite{GKV}). The algorithm is
trivial, we simply iterate the integral equation plugging the
$Y$-functions at iteration $k-1$ in the right hand side of
(\ref{YAdS3}) and reading from the left hand side their values at
the $k$-th iteration.
We start by defining the kernel appearing in the integral equations, \\ \\
\verb"K[x_]=1/(2 Pi Cosh[x]);" \\ \\
and specify how many gluons we want to consider. We do that by
introducing a list of masses which appear in the integral equations.
For example, we set the following
numerical values of the masses \\ \\
\verb"m={1.,2.};M=Length[m];" \\ \\
would correspond to $M=2$ nodes, i.e. to a polygon with $2M+6=10$ sides.  We also introduce a cut off for the several integrals and set the number of iterations,\footnote{This cut-off is chosen so that for the smallest mass we have $e^{-m \cosh ({\rm Cut})}=10^{-8}$. We could do something fancier (and more efficient of course) and introduce a cut-off which would be different for different integrals.} \\ \\
\verb"cut = ArcCosh[8Log[10]/Min[m]]; ni = 8; " \\ \\
At each iteration step we compute the new values of the
$Y$-functions at a discrete set of points and construct the new
$Y$-function as the function which interpolates through these
points. For that we need an interpolating function and a function to
preform the several integrals
\footnote{The last command  in the integration function ensures that we are really integrating something which is numerical once the integration variable $y$ takes some random numerical value. We chose $0.1$ but anything would work.}, \\ \\
\verb"F[S_] := FunctionInterpolation[S,{x,-cut,cut}, InterpolationPoints->100]"\\
\verb"int[S_] := NIntegrate[S,{y,-cut,cut}]/;NumberQ[S/.y->.1]"\\ \\
This is all the machinery we need. The iterations are simply implemented by  \\ \\
\verb"Y[a_,k_] := (0#&)/;a==0||a==M+1"\\
\verb"Y[i_,1] := Exp[-m[[i]]Cosh[#]]&"\\
\verb"Y[i_,k_] := Y[i,k] ="\\
\verb"F[Exp[-m[[i]]Cosh[x]+int[K[x-y]Log[(1+Y[i-1,k-1][y])(1+Y[i+1,k-1][y])]]]]" \\ \\
In the first line we set $Y_0=Y_{M+1}=0$, in the second line we set
the seed values for the iteration procedure to be simply the
asymptotic expressions for the $Y$-functions and in the last command
we compute the $Y$-functions at the iteration step $k$ using the
right hand side of the integral equations with the $Y$-functions at
step $k-1$.
 We can now compute the free energy from (\ref{areafree}) and see how it converges to a particular value as we iterate: \\ \\
\verb"energy=Table[int[Sum[Cosh[y]m[[i]]Log[1+Y[i,s][y]]/Pi/2,{i,M}]],{s,ni}]"  \\
\verb"ListPlot[energy]" \\ \\
For the masses chosen above we see that the free energy converges to $0.27$. Finally, if we want to read the cross-ratios of the polygon whose area we have just computed, we simply need to evaluate the $Y$-functions at $\theta=0$,  \\ \\
\verb"Table[Y[i,ni][0],{i,M}]" \\ \\
To compute the the free energy again for a different choice of
masses we should first run \verb"Clear[Y]".

The results of the numerics are shown for several values of $m_1$
and $m_2$ in figure \ref{decagonnumerics}. Here we have considered
real values of $m_1$, $m_2$.  One can similarly do computations for complex
values of $m_1, ~m_2$, using the kernels after \nref{tildeYAdS3}, and the discussion in
appendix \ref{WallCrossing}, if necessary.

\section{Wall crossing and TBA}
\la{WallCrossing}

In this appendix we describe the pattern under which the $AdS_3$
integral equation (\ref{tildeYAdS3}) changes as the phases of the
masses $\varphi_s=\arg(m_s)$ are deformed beyond the region where
$|\varphi_s-\varphi_{s+1}|<\pi/2$.
 The $AdS_5$ integral equations (\ref{Integraleq}) follows an analogous pattern.
  In the context of the works \cite{GMN,GMNtwo} this corresponds to the wall crossing phenomenon.

As explained in section \ref{AdS3TBA}, as long as
$|\varphi_s-\varphi_{s+1}|<\pi/2$, the integral equation reads
\beq\la{integralsmallphase} \log \widetilde Y_s(\theta) = -
|m_s|\cosh \theta + K_{s,s+1} \star \log (1 + \widetilde Y_{s+1}) +
K_{s,s-1}\star \log(1 + \widetilde Y_{s-1} ) \eeq where $\widetilde
Y_s(\theta)=Y_s(\theta+i\varphi_s)$ and
$$
K_{s,s'} \star \log (1 + \widetilde Y_{s'}) =\int{\log(1+\widetilde
Y_{s'}(\theta'))\over2\pi\cosh(\theta-\theta'+i\varphi_{s,s'})}d\theta'
$$
and $\varphi_{s,s\pm 1}\equiv\varphi_s-\varphi_{s\pm 1}$. Let us
consider the simplest possible situation where $\varphi_{1,2}$
crosses the value $+\pi /2$ (and hence $\varphi_{2,1}$ crosses
$-\pi/2$) while all other $|\varphi_{s,s+1}|<\pi/2$. Any other
possibility can be treated similarly.

When  $\varphi_{1,2}$ crosses the value $+\pi /2$ we pick the pole
contribution from the kernel and the TBA equation for $s=1$ becomes
\beq\la{integraligphase1} \log \widetilde Y_1(\theta) = - |m_1|\cosh
\theta + K_{1,2} \star \log (1 + \widetilde Y_{2}) +\log (1 +
\widetilde Y'_{2}(\theta))~, \eeq where
$$
\widetilde Y'_{2}(\theta)=\widetilde Y_{2}(\theta+i\varphi_{1,2}-i\pi/2)
$$
is a ``new" Y-function which arose in this process. Such
$Y$-functions arise in pairs. Indeed, at the same time
$\varphi_{2,1}$ crosses the the point $-\pi/2$, so that the equation
for $\widetilde Y_{2}$ becomes \beq\la{integraligphase2} \log
\widetilde Y_{2}(\theta) = - |m_{2}|\cosh \theta +\sum_{s=\{1,3\}}
K_{2,s} \star \log (1 + \widetilde Y_{s}) +\log (1 + \widetilde
Y'_1(\theta))~, \eeq where
$$
\widetilde Y'_1(\theta)=\widetilde
Y_1(\theta+i\varphi_{2,1}+i\pi/2)~.
$$
These equations are correct for  $\pi/2 <\varphi_{1,2}<\pi$.  We evaluate
(\ref{integraligphase1}) and (\ref{integraligphase2}) at $\theta\mp
i(\varphi_{1,2}-\pi/2)\pm i0$ to get
\beqa
\log \widetilde Y'_1(\theta) &=& - |m_1|
\cosh(\theta- i\varphi_{1,2}+i\pi/2) +
\widetilde K_{1,2} \star \log (1 + \widetilde Y_{2}) +\log (1 + \widetilde Y_{2}(\theta))\la{primeeq} \\
\log \widetilde Y'_2(\theta) &=& - |m_2|\cosh(\theta+
i\varphi_{1,2}-i\pi/2) + \sum_{s=\{1,3\}}\widetilde K_{2,s} \star
\log (1 + \widetilde Y_{s})  +\log (1 + \widetilde
Y_{1}(\theta))~,\nn \eeqa where $\widetilde K$'s are the kernels
shifted accordingly.\footnote{ We can absorb the
$\log(1+\widetilde Y_s)$'s in the right hand side of these equations
by flipping the sign of $i0$ in $\widetilde K_{1,2}$ and $\widetilde
K_{2,1}$. This $i 0$ prescription is necessary to define these shifted kernels precisely.
Otherwise there is singularity along the integration contour. }

These equations, together with (\ref{integraligphase1}),
(\ref{integraligphase2}) and (\ref{integralsmallphase}) for $s>2$
constitute a closed set of $\hat n-1$ equations for the functions
$\widetilde Y_1,\widetilde Y_2,\dots,\widetilde Y_{\hat n-3},
\widetilde Y'_1,\widetilde Y'_{2}$.
 If more phases become large we get in a
similar way more extra functions $\widetilde Y'_s$.

As usual in integrable models, nothing singular happens when the
phases cross the values of $\pm \pi/2$ and new singularities hit the
integration contour.
 After all the mechanism of picking the poles is precisely designed to keep everything
 smooth!. In particular there is no reason to change the expression of the area
  (\ref{areafree}) which therefore  keeps its form. Notice in particular that the tilded $Y$-functions are
   designed to be always exponentially suppressed as $\theta \to \pm \infty$.

Wall crossing in the $AdS_5$ system can be treated exactly in the
same way.

\subsection{Relation to Wall Crossing in \cite{GMN} }

In this same circumstance,  the authors of  \cite{GMN}  obtained  only one extra function rather
than two functions as we obtained above. This single extra function is associated to
a new cycle on the Riemann surface (in the WKB approximation) which corresponds\footnote{Here we will use
$1+2$ as an index, hopefully, this will not cause confusion.} to
$ \gamma_{1+2} = \gamma_1 + \gamma_2$. Thus, $ Z_{1+2} = Z_1 + Z_2$, or $  m_{1+2} = m_1 +  i m_2 =
|m_{1+2}| e^{i  \varphi_{1+2} } $, where
the $m_i$ are the complex $m$'s. In the language of \cite{GMN} a new hypermultiplet has appeared.

In fact, we can define a new set of functions
\beqa
&& \widetilde Y^n_1(\theta)  = {  \widetilde Y_1 \over 1 + \widetilde Y_2' } ~,~~~~~~~\widetilde Y^n_2 = {
\widetilde Y_2
\over 1 +  \widetilde Y'_1 }
\cr
&& \widetilde Y^n_{1+2} (\theta) = {  \widetilde Y_1( \theta + i   \varphi_{1+2} - i \varphi_1)
\widetilde Y_2 ( \theta +
i   \varphi_{1+2} - i \varphi_2  - i \pi/2 ) \over 1 +   \widetilde Y_1( \theta + i   \varphi_{1+2} - i \varphi_1) +
 \widetilde Y_2 ( \theta +
i   \varphi_{1+2} - i \varphi_2 - i \pi/2 ) } \la{yonetwon}
\eeqa
The first two variables are obviously designed to absorb the $\log (1 + \widetilde Y')$ terms in the
right hand side of \nref{integraligphase1} \nref{integraligphase2}.
These relations look simpler when expressed in terms of the $Y$ functions (without the tildes)\footnote{
Of course, they are even simpler in terms of the $\hat Y_s$ functions defined in \nref{Yhat}, since the shifts disappear
 \cite{GMN}.}
\beqa
   Y^n_1   = {    Y_1 \over 1 +   Y_2^- } ~,~~~~~~~  Y^n_2 = {
  Y_2
\over 1 +    Y^+_1 }
   ~,~~~~~~~~  Y^n_{1+2}   = {  Y_1
  Y_2^-   \over 1 +    Y_1  +
  Y_2^-   }\nn
\eeqa
where the $\pm$ index is the usual shift by $ \pm i \pi/2$.
 The function $Y^n_{1+2}$ was defined
so that
\beqa \label{1plusydef}
( 1 +  Y_1) &= &( 1 +  Y_1^n ) ( 1 +
  Y_{1+2}^n  )
\\ \nn
( 1 +  Y_2) &= &( 1 +   Y_2^n )
( 1 +  Y_{1+2}^{n + }   )
\eeqa
which transforms  \nref{integraligphase1} \nref{integraligphase2} into equations that look
more like the equations in \cite{GMN}
\beqa
&& \nn  \log \widetilde Y^n_1  = - |m_1|\cosh
\theta + K_{1,2} \star \log (1 + \widetilde Y^n_{2})  + K^+_{1,1+2} \star \log (1 +\widetilde Y^n_{1+2} )
\\
&& \nn \log \widetilde Y^n_2  = - |m_1|\cosh
\theta + \sum_{s=1,3} K_{2,s} \star \log (1 + \widetilde Y^n_{s})  +
K_{2,1+2} \star \log (1 +\widetilde Y^n_{1+2} )
~,
\\ && \nn
\log \widetilde Y_3 = - |m_3| \cosh \theta  +   \sum_{s=2,4} K_{3,s} \star \log (1 + \widetilde Y^n_{s}) +
 K^+_{3,1+2} \star \log (1 +\widetilde Y^n_{1+2} )
\eeqa
where $Y_{s}^n = Y_s$ for $s>2$. The rest of the equations remains the same.
Finally,  the new equation for $\widetilde Y_{1+2}^n$
 follows from considering  the sum of \nref{integraligphase1} and \nref{integraligphase2} evaluated at the appropriate
 values, which are the ones in the numerator of \nref{yonetwon}. One of the kernels gives a delta function
  and produces a factor of $\log (1 + \widetilde Y_{1+2}^n)$. This
 combines with other terms to give
\beqa
\la{integraligphase12w}
\log \widetilde Y^n_{1+2}(\theta)&=& - |m_1 +i m_2|\cosh
\theta +  K^-_{1+2,1} \star \log (1 + \widetilde Y^n_{1})   +
K_{1+2,2} \star \log (1 + \widetilde Y^n_{2})\nn \\
&&+ K^-_{1+2,3} \star \log (1 + \widetilde Y_{3})
\eeqa

In the previous subsection, the equation for the free energy did not change, it was still given by
$\widetilde Y_1$ and $\widetilde Y_2$, with no appearance of two extra functions
$\widetilde Y_1'$, $\widetilde Y_2'$.
Here, however, due to \nref{1plusydef}, we have a change in the expression of the free energy to
\beqa
 F \to F^n = \int{ d \theta \over 2 \pi } \cosh \theta
 \left[ |m_1| \log(1 + \widetilde Y_{1}^n ) +|m_2| \log(1 + \widetilde Y_{2}^n ) +|m_1 +i  m_2|
 \log(1 + \widetilde Y_{1+2}^n )  \right] \nn
 \eeqa
plus the usual terms for $s> 2$.
From our point of view, it is not clear whether this way of writing the integral equation has any
advantage with respect to the one we wrote above.

\section{Asymptotic behavior of the solutions at large $z$}
\la{MoreDetails}

In this appendix we prove some formulas that are necessary to derive
the $AdS_5$ $Y$-system equations.

We will start by defining the small solutions $s_i$ and $\bar s_i$
in a certain normalization which simplifies some of the formulas. Of
course, the final $Y$ system involves cross ratios and is
independent of such normalization details. Up to a normalization,
$s_i$ is a solution of $( d + {\cal A}(\zeta) ) s =0$ that is small
in Stokes sector $i$. Similarly, $\bar s_i$ is a solution of $( d -
{\cal A}^t(\zeta) ) \bar s =0$ which is small in Stokes sector $i$.
Our first goal is to set a normalization of all these solutions so
that \nref{norsol} are valid.

It is convenient to introduce the $w$ ``plane'' via $dw = P^{1/4}
dz$. For large $z$, or large $w$ we cover the $w$ plane $n/4$ times
as we go around the $z$ plane, since for large $z$ we have $w
\propto z^{n/4}$, where $n$ is the number of gluons.

Our boundary conditions for the connection are such that, for large
$w$ the connection is \beqa \la{BCcon} d + {\cal A}(\zeta) \sim  d
+  {\rm diag} (1,-i,-1,i) { dw \over \zeta } + {\rm diag}
(1,i,-1,-i)\zeta  d\bar w \eeqa We can now choose a basis of
approximate solutions of the form \beqa
 \psi_1 &=& e^{- ( w/\zeta + \bar w \zeta ) } (1,0,0,0)^t \nn
\\
\psi_2 & =& e^{ - ( -i w/\zeta + i \bar w \zeta ) } (0,1,0,0)^t \nn
\\
 \psi_3 &=& e^{- ( - w/\zeta - \bar w \zeta ) } (0,0,1,0)^t  \la{approxbasis}
\\
\psi_4 & =& e^{ - ( i w/\zeta - i \bar w \zeta )  } (0,0,0,1)^t \nn
\eeqa

For zero phase of $\zeta$,  the above solutions are the small
solutions in consecutive sectors starting from the one centered on
the positive real axis, followed by the one centered on the positive
imaginary axis, and so on.

We normalize the small solutions $s_i$ by saying that    $s_i =
\psi_a $ with $a = i \, {\rm mod}(4)$ in the corresponding Stokes
sector.\footnote{ In principle, we could analyze $s_i$ in the Stokes
sector $j$, but then $s_i$ will be, in general, an arbitrary linear
combination of \nref{approxbasis}.} Similarly, we choose a basis for
the bar solutions \beqa \bar \psi_1 &=& e^{ ( w/\zeta + \bar w \zeta
) } (1,0,0,0)^t\nn
\\
\bar \psi_2 & =& e^{  ( -i w/\zeta + i \bar w \zeta ) } (0,1,0,0)^t\nn
\\
 \bar \psi_3 &=& e^{ ( - w/\zeta - \bar w \zeta ) } (0,0,1,0)^t\nn
\\
\bar \psi_4 & =& e^{  ( i w/\zeta - i \bar w \zeta )  } (0,0,0,1)^t\nn
\eeqa The normalization of $\bar s_i$ is fixed by setting
$$ \bar s_i = \bar \psi_{a+2} ~,~~~~ i = a +2 \, \,  {\rm mod}(4) $$
in the corresponding Stokes sector.

We can now check that
 \beqa
&& \langle s_{i},   s_{i+1} ,  s_{i+2} , s_{i+3} \rangle  =
(-1)^{i+1} \nn
\eeqa
by   evaluating this expression at infinity in sectors $i+1$ or $i+2$ where the asymptotics (\ref{approxbasis}) of all these small solutions are still reliable.
Similarly,  
\beqa
 && \la{norapp}
 \bar s_i = (-1)^{i+1}  s_{i-1} \wedge s_i \wedge s_{i+1}
 \\
&& s_{k+1}  (\zeta) =   (\hat C)^{T}   \bar s_{k} ( i \zeta) \nn \\
 && \bar s_{k+1}  (\zeta) =   (\hat C)^{-1}   s_{k} ( i \zeta)  ~;~~~~~~~~~~~~~~~~
 \hat C^{-1} =
 \(\begin{array}{cccc}\  0 & 1 & 0& 0 \\  0 & 0 & 1& 0 \\ 0 & 0 & 0& 1 \\  1 & 0& 0 & 0   \end{array}\) \nn
 \eeqa
 This is not exactly as in the main text, so we perform a further redefinition
 of the small solutions
\beq\la{shift} s_{4k+a}\to(-1)^ks_{4k+a}~,\qquad\bar
s_{4k+a+2}\to(-1)^k\bar s_{4k+a+2}~,\qquad a=1,2,3,4~, \eeq This
transforms the relations \nref{norapp} into the ones in the main
text \nref{norsol}. It also transforms the matrix $\hat C$ into the
matrix $C$ in the main text, see \nref{projection}, with ${\rm det } \, C=1$.

Using epsilon symbol identities one can check (easily with
\verb"Mathematica") that
 \beqa
&&\langle \bar s_{k},  \bar s_{k+1},  \bar s_{j}, \bar s_{j+1}
\rangle =  \langle
s_{k}, s_{k+1}, s_{j}, s_{j+1} \rangle \la{re1}\\
&&\langle \bar s_{k} ,\bar s_{k+1}, \bar s_{k+2}, \bar s_{m} \rangle
= \langle s_{k+1}, s_{m-1}, s_{m}, s_{m+1} \rangle \la{re2} \eeqa
This identities, together with \nref{norsol}, imply that \beqa
\langle s_{k}, s_{k+1},s_{j}, s_{j+1}\rangle(\zeta)&=&
\langle  s_{k-1}, s_{k},s_{j-1},s_{j} \rangle \(e^{i\pi/2}\zeta\) \la{re3}\\
\langle   s_{j} , \bar s_{k+1} \rangle (  \zeta) = \langle s_{j},
s_{k}, s_{k+1}, s_{k+2}\rangle(\zeta) &=& \langle s_{j}, s_{j-1},
s_{j-2},s_{k}\rangle (e^{i\pi/2}\zeta) = \langle \bar s_{j-1} , s_k
\rangle ( e^{ i \pi/2} \zeta) \nn \eeqa These two relations are the
important identities which we use in the main text. In deriving this
relation one needs to use that if $U$ is any matrix with unit
determinant, then the product of four solutions $\{ U s_i \} $ is
the same as the product of the four solutions $\{ s_i \}$.

Let us conclude with some remarks about the monodromy when $i \to
i+n$. Notice that if $n \not = 4 k$, then the solutions
\nref{approxbasis} appear to be misaligned after $i \to i+n$. This
is not a problem because they are on different sheets which are
connected by suitable powers of the matrix $C$. When $n = 4 k $ we
can compare $s_{n+1}$ with $s_1$. In general, the $w$ variables are
related by  $w_{n+1} = w_{1} + w_0$. This expresses the fact that
there is a logarithmic branch cut.
 This implies that there is a $\zeta$ dependent
monodromy when we relate $s_{ 4k + a } \sim s_a  e^{ - i^{-a}
w_0/\zeta -i^a  w_0 \zeta}$. If $n \not = 4k$, then we can choose
the origin of the $w$ plane so that there is no $\zeta$ dependent
monodromy. Let us now worry about constant parts. These could arise
as follows. In the $z$ plane we can have a gauge connection whose
integral at large $z$ , $\oint A$, is non-zero (and proportional to
the diagonal matrix ${\rm diag}(1,-1,1,-1)$). In writing
\nref{BCcon} we have made a gauge transformation that sets it to
zero. However, this gauge transformation is not globally well
defined and it appears as an extra gauge transformation that we need
to do when we relate $s_{n+1}$ with $s_1$. When $n$ is odd no such
transformation is possible. In fact, it is easy to check that
  a change in normalization of the solutions of the form $s_i \to \gamma^{(-1)^i} s_i$ and $\bar s_i
\to \gamma^{(-1)^{i+1} } \bar s_i$  can remove any constant from the
relation between $s_{n+1}$ and $s_1$. In the case that $n$ is even,
there can be a constant piece in the formal monodromy.
 We can take it to
 have the form
$s_{n+i} = \mu^{(-1)^i} s_i $. This is present both for $n= 4k $ and
$n= 4k +2$.
In section \ref{monodromies}, these properties were derived directly from the $Y$-system and the analytic properties.

\section{Explicit form of $T$ and $Y$-functions} \la{explicit}

In this appendix we summarize all $T$ and $Y$ functions of the
$AdS_5$ problem. We use the shift identities  (\ref{useful2}) to
bring the various inner products to the expressions below.  
 In this appendix we use
the notation $\< s_a \bar s_b \>=\< s_a  s_{b-1} s_b s_{b+1} \>$ and
$\< \bar s_b s_a\>=\< s_{b-1} s_b s_{b+1} s_a\>$ and drop commas inside angle brackets.
\subsection{$T$-functions} \la{explicitT}
\beqa
\!\!\!\!\!\!\!\!&&T_{2,2k}^-=\<s_{-k-2}s_{-k-1}s_{k}s_{k+1}\>\,\, , \,\, T_{2,2k+1}=\<s_{-k-2}s_{-k-1}s_{k+1}s_{k+2}\>  \nn \\
\!\!\!\!\!\!\!\!&&T_{1,4l}=\<\bar s_{-2l-1}s_{2l+1}\>\,\, , \,\, T_{1,4l+1}^-=\< s_{-2l-2}\bar s_{2l+1}\> \,\, , \,\, T_{1,4l+2}=\<s_{-2l-2}\bar s_{2l+2}\>   \,\, , \,\, T_{1,4l+3}^-=\<\bar s_{-2l-3}s_{2l+2}\>\nn \\
\!\!\!\!\!\!\!\!&&T_{3,4l}=\< s_{-2l-1} \bar s_{2l+1}\>\,\, , \,\,
T_{3,4l+1}^-=\<\bar s_{-2l-2} s_{2l+1}\>\,\, , \,\,T_{3,4l+2}=\<\bar
s_{-2l-2} s_{2l+2}\>   \,\, , \,\,  T_{3,4l+3}^-=\< s_{-2l-3}\bar
s_{2l+2}\> \nn \eeqa
Note the presence of some shifts in the left hand side. Note that $T_{1,s}$ and $T_{3,k}$ only differ
by $s_i \leftrightarrow \bar s_i$.

\subsection{$Y$-functions} \la{explicitY}
\beqa \!\!\!\!\!\!\!\!\!\!&&
Y_{2,2k}= \widehat Y_{2,2k} = \frac{\left\langle s_{-k-2} s_{-k-1} s_{k+1} s_{k+2}\right\rangle  \left\langle s_{-k-1} s_{-k} s_{k} s_{k+1}\right\rangle}{\left\langle \bar s_{-k-1} s_{k+1}\right\rangle  \left\langle s_{-k-1}\bar s_{k+1} \right\rangle} \,\, ,  \la{YAppendix} \\
\!\!\!\!\!\!\!\!\!\!&&
   Y_{2,2k+1}^-= \widehat Y_{2,2k+1} = \frac{\left\langle s_{-k-3} s_{-k-2} s_{k+1} s_{k+2}\right\rangle  \left\langle
   s_{-k-2} s_{-k-1} s_{k} s_{k+1}\right\rangle }{\left\langle \bar s_{-k-2} s_{k+1}\right\rangle
   \left\langle s_{-k-2}\bar s_{k+1}\right\rangle } \nn \,\, ,   \\
\!\!\!\!\!\!\!\!\!\!&& Y_{1,4l+2}^-=\frac{\left\langle \bar s_{-2
l-3} s_{2 l+2}\right\rangle  \left\langle \bar s_{-2 l-2} s_{2
l+1}\right\rangle }{\left\langle s_{-2 l-4} \bar s_{-2 l-2}
\right\rangle  \left\langle   s_{-2 l-3}s_{-2 l-2}s_{2 l+1}s_{2
l+2}\right\rangle }\,\,\, \,\, \,\,\,\,, \,\,
   Y_{1,4l}^-=\frac{\left\langle s_{-2 l-2}\bar s_{2 l+1}\right\rangle  \left\langle s_{-2 l-1} \bar s_{2 l}\right\rangle }{\left\langle s_{-2 l-2} s_{-2 l-1}s_{2 l}s_{2 l+1}\right\rangle  \left\langle
   s_{2 l-1} \bar s_{2 l+1} \right\rangle } \nn \\
\nn  \!\!\!\!\!\!\!\!\!\!&&
 Y_{1,4l+3}=\frac{\left\langle \bar s_{-2 l-3} s_{2 l+3}\right\rangle  \left\langle \bar s_{-2 l-2} s_{2 l+2}\right\rangle }{\left\langle s_{-2 l-4}\bar s_{-2 l-2} \right\rangle  \left\langle
   s_{-2 l-3}s_{-2 l-2}s_{2 l+2}s_{2 l+3}\right\rangle }\,\, \,\, , \,\,
   Y_{1,4l+1}=\frac{\left\langle s_{-2 l-1} \bar s_{2 l+1} \right\rangle  \left\langle s_{-2 l-2}\bar s_{2 l+2} \right\rangle }{\left\langle s_{2 l} \bar s_{2 l+2} \right\rangle  \left\langle
   s_{-2 l-2}s_{-2 l-1}s_{2 l+1}s_{2 l+2}\right\rangle } \\
   \!\!\!\!\!\!\!\!\!\!&& Y_{3,4l+2}^-=\frac{\left\langle s_{-2 l-3}\bar{s}_{2 l+2}\right\rangle  \left\langle s_{-2 l-2}\bar{s}_{2 l+1}\right\rangle
   }{\left\langle s_{-2 l-3}s_{-2 l-2}s_{2 l+1}s_{2 l+2}\right\rangle  \left\langle s_{2 l}\bar{s}_{2
   l+2}\right\rangle }
   \,\,\, \,\, \,\,\,\,\,\,\,\, \,\,\,   , \,\, Y_{3,4l}^-=\frac{\left\langle \bar{s}_{-2 l-2}s_{2 l+1}\right\rangle  \left\langle \bar{s}_{-2 l-1}s_{2 l}\right\rangle
   }{\left\langle s_{-2 l-2}s_{-2 l-1}s_{2 l}s_{2 l+1}\right\rangle  \left\langle s_{-2 l-3}\bar{s}_{-2
   l-1}\right\rangle } \nn  \\
 \nn  \!\!\!\!\!\!\!\!\!\!&& Y_{3,4l+3}=\frac{\left\langle s_{-2 l-3}\bar{s}_{2 l+3}\right\rangle  \left\langle s_{-2 l-2}\bar{s}_{2 l+2}\right\rangle
   }{\left\langle s_{-2 l-3}s_{-2 l-2}s_{2 l+2}s_{2 l+3}\right\rangle  \left\langle s_{2 l+1}\bar{s}_{2
   l+3}\right\rangle } \, , \,\,
    Y_{3,4l+1}=\frac{\left\langle \bar{s}_{-2 l-1}s_{2 l+1}\right\rangle  \left\langle \bar{s}_{-2 l-2}s_{2 l+2}\right\rangle
   }{\left\langle s_{-2 l-2}s_{-2 l-1}s_{2 l+1}s_{2 l+2}\right\rangle  \left\langle s_{-2 l-3}\bar{s}_{-2
   l-1}\right\rangle }
\eeqa We also define the functions $\widehat Y_{a,s}$ to be given by
the right hand side of \nref{explicitY} without any extra shifts. In
other words, $\widehat Y_{a,s}$ are defined as above with all the
products defined at the same value of $\zeta$. In particular, we have included the explicit
definitions of $Y_{2,s}$. These are related to spacetime cross ratios by replacing
 $s_i \to \bar \lambda_i$ and $\bar s_i \to \lambda_i$, where $\lambda_i$ are Hodges' momentum
 twistor variables \cite{Hodges}, see section \ref{TwistorCR}.

\section{Asymptotic form of the $Y$ functions for the $AdS_5$ case }
\label{AdSfiveAsymptotics}

\begin{figure}[htb]
\begin{center}
\includegraphics[angle=0,width=16cm]{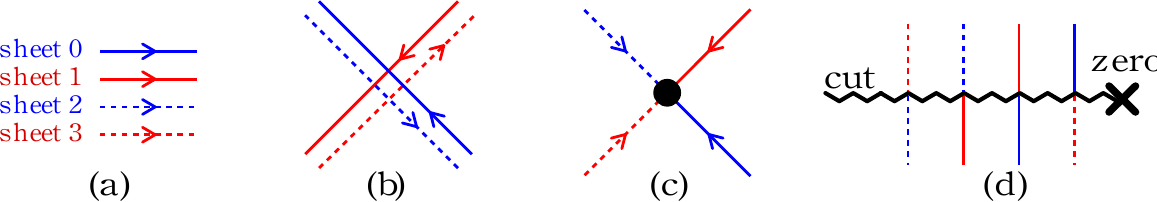}
\end{center}
\caption{ WKB lines for the $AdS_5$ case. We have four sheets. (a)
Shows the conventions for naming the WKB lines for each of the four
sheets. (b) Through a generic point we have four WKB lines passing.
The red lines and the blue lines are orthogonal to each other . The
line for the 1 sheet and the 3 sheet have opposite orientations and
are on top of each other, here we have separated them for ease in
visualization. (c) We evaluate an inner product of four solutions by
bringing together four of the lines. The order of the lines
indicates the sign, but we will not keep track of it. (d) When
various WKB lines pass through a cut they change sheets according to
the pattern here, the orientation of the line is preserved. If we
have several zeros, the effects of this cuts add up. }
\label{WKBfivepreliminary}
\end{figure}

In this appendix, we explain how to obtain the asymptotic form of the
$Y$ functions for small $\zeta$. For small $\zeta$ we again
diagonalize $\Phi_z \sim P^{1/4} \diag{1,-i,-1,i}$. We can then use
a WKB approximation for the study of the solutions. At leading order
the solutions are simply given by $\psi_a = \exp [ - i^{-a} \int
P(z)^{1/4}dz ]$. We will follow WKB lines where the change of phase
is real $ {\rm Im}( i^{-a} P(z)^{1/4} \dot z/\zeta )= 0$. This
defines a family of lines. Here it is useful to think about this
family of lines as lines on the Riemann surface, rather than lines
on the $z$ plane. Of course, the Riemann surface is simply $x^4 =
P(z)$, and we have the differential $x dz = P(z)^{1/4} d z$.
Different sheets of the Riemann surface are associated to different
solutions of the linear problem. We label these sheets as $x_a$, $a=
0,1,2,3$, in a cyclic way. We also have that $x_a = i^{-a} x_0$. The
WKB lines have $Im( x_a \dot z/\zeta ) =0$. They are oriented pointing
towards the direction in which the solution increases. We label the
WKB lines for different sheets by different types of lines in the
$z$ plane, see figure \ref{WKBfivepreliminary}(a). Of course, lines
for the 1st and 3rd sheet coincide  are oriented in the opposite
way. Similarly for lines $0$ and $2$. Note, that we also want to
keep track of the $\zeta$ independent part which comes from the
connection $A$. The $A_{\bar z}$ component of the connection can be
diagonalized and set to zero. Only the diagonal components of
$A_{z}$ are relevant for the small $\zeta$ WKB approximation. These diagonal
components are constrained by the $Z_4$ projection condition to be
of the form \beqa \la{diagonalA} A_{z} = \alpha_z {\rm
diag}(1,-1,1,-1) + ~~{\rm off~diagonal~terms } \eeqa The off
diagonal components are not relevant for us, since we are neglecting
higher order terms in the small $\zeta$ expansion. Again we can
think of $\alpha_z$ as a one form on the Riemann surface. The
Hitchin equations imply that $d \alpha =0$. When we go around a zero
of the Polynomial $P \sim (z - z_0)$ we change sheets $x_{i} \to
x_{i-1}$, see figure \ref{WKBfivepreliminary}(d). Moreover, going
around a zero, also $\alpha_z$ has a $Z_2$ branch cut, so that
$\alpha_z \to -\alpha_z$. In other words, we really should think in
terms of a $U(1)$ gauge field on the Riemann surface. The Riemann
surface has a $Z_4$ symmetry $x \to i x$ and $\alpha$ is constrained
to be odd under this symmetry.

The small solutions $s_i$ correspond to solutions that are
associated to one of the sheets of the Riemann surface. In the large
$z$ region the sheet of the Riemann surface associated to solution
$s_k$ is simply $ k \, {\rm mod} (4)$.

We will draw WKB lines associated to an inner product $\langle s_i,
s_j , s_k ,s_l \rangle $ as four lines that are incoming if the
product is in the numerator of the expression we want to evaluate.
Each of the lines meeting at the inner produce lives on one of the
four sheets, see figure \ref{WKBfivepreliminary}(c). The lines start
from each of the asymptotic regions associated to the corresponding
Stokes sector and they end on the common point where the product is
evaluated. We do this with WKB lines. Once we identify these lines
we can move the point around on the Riemann surface in order to
simplify the final expression. If the inner product is in the
denominator, then we reverse the orientation of all the lines,
without changing the sheet numbers.

We take a real polynomial with all its zeros on the real axis and with
$P(z)>0 $ for sufficiently large real $z$.
We find it convenient to run all the cuts towards the left of the
zero of the Polynomial $P$. The solution $s_k$ lives in sheet $k$
(modulo four). It is represented as a line coming in from infinity in
the corresponding Stokes sector. These lines represent integration contours on the
Riemann surface where we integrate both $P^{1/4} dz = x dz $ as well as $\alpha$.

\begin{figure}[htb]
\begin{center}
\includegraphics[angle=0,width=9cm]{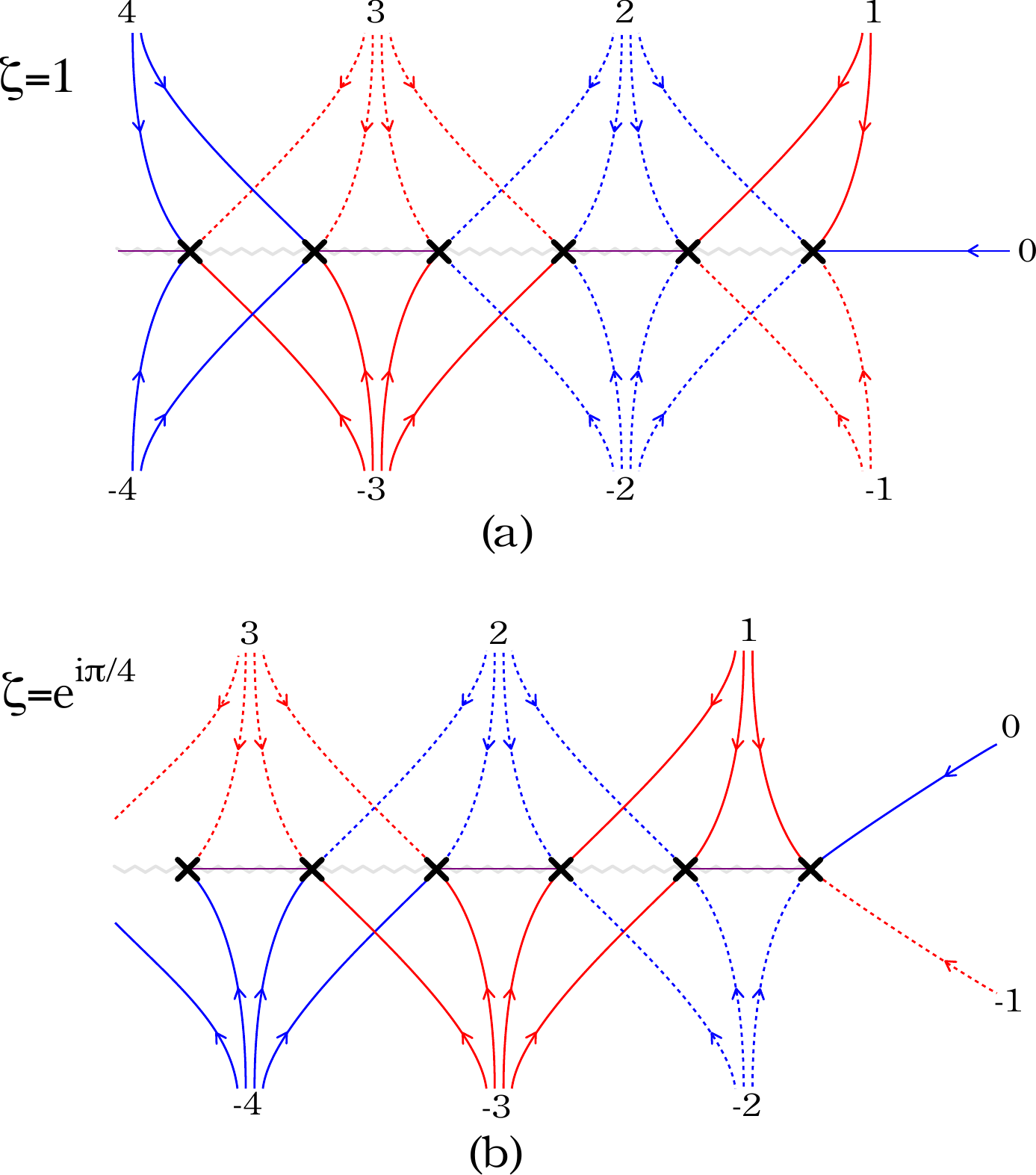}
\end{center}
\caption{ Schematic picture of the patterns of WKB lines for a
Polynomial whose zeros are on the real axis. In (a) we show the
pattern when the phase of $\zeta$ is 0 and in (b) we show it when
the phase of $\zeta$ is $\pi/4$. Lines of different colors are
supposed to be crossing at right angles. We have five lines ending
at each zero. The purple line goes between two different zeros. (Its
precise type depends on whether we put it above or below the cut).
The $k^{\rm th}$ Stokes sector corresponds to a WKB line starting on
sheet $k$. Notice that we have only drawn the lines that end at
zeros. These lines are not useful for evaluating inner products but
they are lines that separate different families of WKB lines. (The
figures in this appendix are best viewed on a color monitor). }
\label{WKBfivezetaone}
\end{figure}

With all these preliminaries we are ready to start evaluating some
cross ratios.
We will evaluate some cross ratios when the phase of $\zeta $ is 1
and some where the phase of $\zeta =e^{i \pi/4}$. The patterns of
WKB lines for our choice of polynomial for these two cases are given
in figure \ref{WKBfivezetaone}. We could also alternatively evaluate
them all at $\zeta = e^{ i \pi/8}$ where the pattern of WKB lines is
something intermediate between the patterns in figure
\ref{WKBfivezetaone}.

It is convenient to redefine the $Y$ functions so that they all
correspond to some cross ratios evaluated at the same value of
$\zeta $. In other words, we define $\widehat Y_{a,s}$ which are
given precisely by the expressions in the right hand side of
formulas (\ref{YAppendix}) , but with no shifts of the $\zeta$
argument in the left hand side. For instance, $\widehat
Y_{2,2k}(\zeta) = Y_{2, 2k}(\zeta)$ and $\widehat Y_{2, 2 k+1}
(\zeta) = Y_{2,2 k+1}^-$.

\begin{figure}[htb]
\begin{center}
\includegraphics[angle=0,width=15cm]{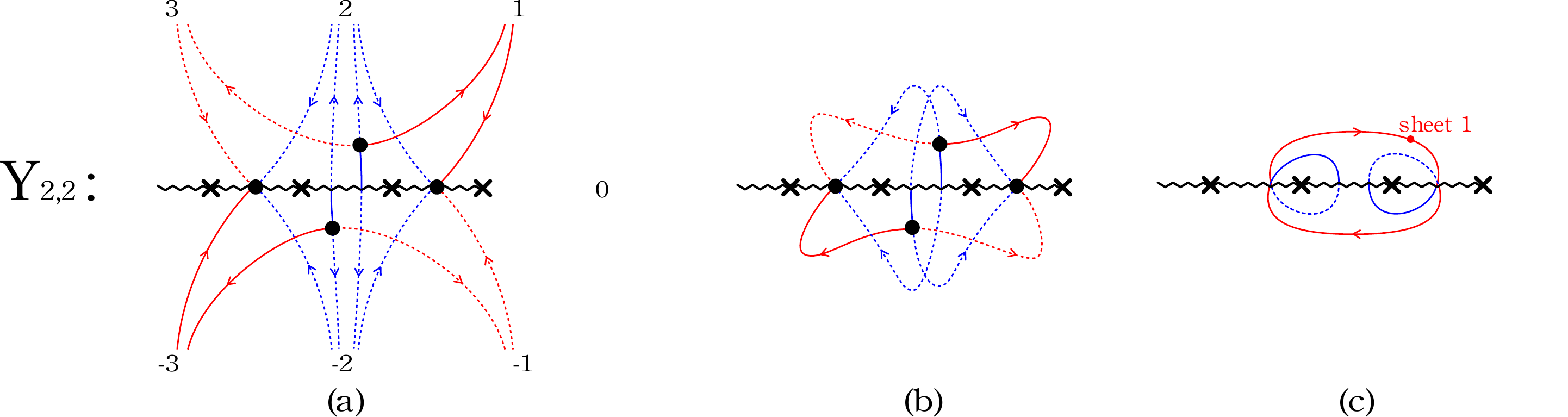}
\end{center}
\caption{ (a) We show the WKB lines that we use to evaluate the
cross ratio $Y_{2,2}$. Each black dot corresponds to each inner
product appearing in the cross ratio $Y_{2,2}$, see \nref{Ytwotwo}.
The inner products that are in the numerator have incoming lines and
the ones in the denominator have outgoing lines. We have set some
dots right on the cuts, we could move them slightly up or down, and
then we would indeed have four different types of lines ending on
them. In (b) we have recombined the lines that we had in (a). In (c)
we have moved the black dots around and we have ``annihilated''
them. We see the final cycle, $\gamma_{2,2}$ corresponding to
$Y_{2,2}$.
Note that the lines going between the two zeros are on the same
sheet and have opposite orientations. Note also that $\alpha$ would
be continuous through the cut going between the 2nd and 3rd zero.
Thus there is no contribution from $\alpha$ to $Y_{2,2}$. For that
reason, there is no constant term in \nref{Ytwoexp}. }
\label{WKBfiveYtwotwo}
\end{figure}
Let us start with
\beqa \la{Ytwotwo} \widehat Y_{2,2} = Y_{2,2} = {
\langle -3,-2,2,3 \rangle \langle -2,-1,1,2 \rangle \over \langle
-3,-2,-1, 2 \rangle \langle -2, 1,2,3 \rangle }
\eeqa Here we are
using a short hand notation $\langle k,l,m,n \rangle \equiv \langle
s_k,s_l,s_m,s_n \rangle$. The WKB lines necessary to evaluate this
quantity are displayed in figure \ref{WKBfiveYtwotwo}(a).  By reconnecting each pair of lines going in and out of an asymptotic region, we arrive at figure \ref{WKBfiveYtwotwo}(b). Then, by bringing the four dots together and reconnecting lines on the same sheet and in the same orientation, we arrive to the closed contour drawn in figure \ref{WKBfiveYtwotwo}(c).
\begin{figure}[htb]
\begin{center}
\includegraphics[angle=0,width=8cm]{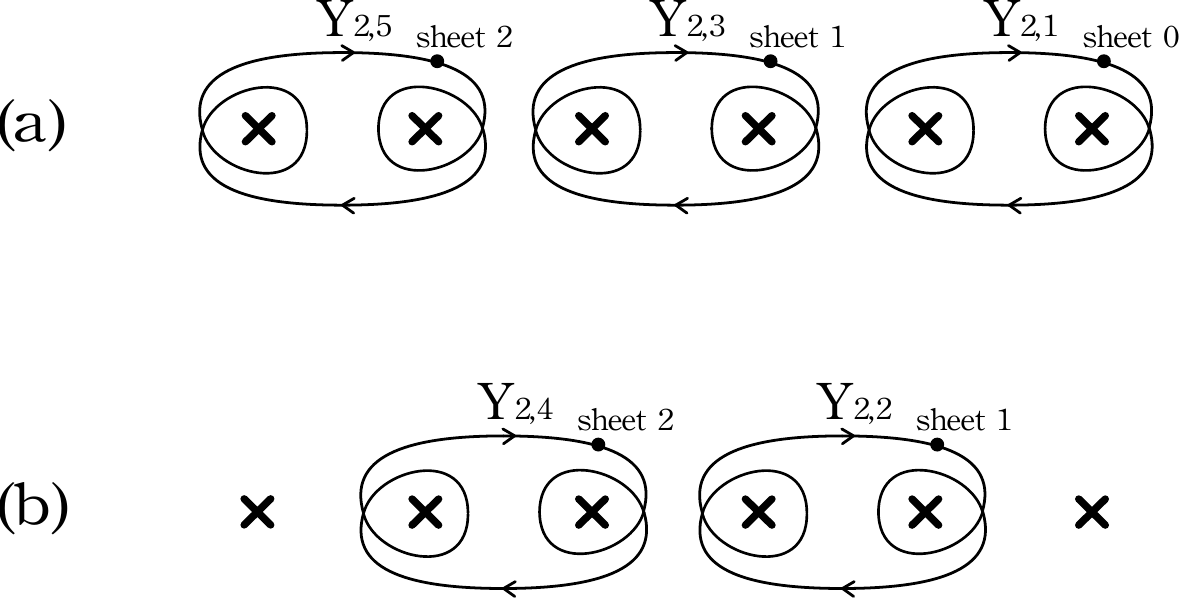}
\end{center}
\caption{ In (a) and (b) we show the cycles $\gamma_{2,s}$ that
correspond to the various $Y_{2,s}$ functions. We have indicated the
sheet where the point with the dot is. Of course there are cuts
emanating from the zeros, etc. } \label{Ytwocycles}
\end{figure}

The procedure necessary to evaluate any of the $Y_{2,s}$ is
basically the same and we get the contours displayed in figure
\ref{Ytwocycles}. To evaluate $Y_{2,2k+1}$ we need to evaluate
$\widehat Y_{2,2k+1}$ at $i\zeta$. The manipulations are identical,
up to a relabeling of the solutions. For $\zeta$ real we can use the
WKB lines that are associated to the pattern in
\ref{WKBfivezetaone}(b). When we draw the contours in figure
\ref{Ytwocycles} we also note on which sheet they start. It is
enough to label the sheet on which any point on the contour is,
which determines the rest. It turns out that there is no
contribution from the gauge connection $\alpha$ for this contour.
We can see this by shrinking the little loops around the two zeros and then
noticing that the two lines that go between the two zeroes in figure \ref{Ytwocycles}(c) are
on the same sheet but have opposite orientation. In addition, we see that $\alpha$ is continuous
across the cut joining the second and third zero in figure \ref{Ytwocycles}(c). This implies that
the $\alpha$ dependent contribution vanishes.

In summary, this means that the $Y_{2,s}$ have the following behavior for small
$s$. \beqa \label{Ytwoexp} \log \widehat Y_{2,s} = { Z_{2 s } \over
\zeta } + o(\zeta ) ~~~~~~~~~~~~~ Z_{2 ,s} = - \oint_{\gamma_{2,s}}
x dz \eeqa where $\gamma_{2,s}$ are the cycles denoted in figure
\ref{Ytwocycles}. Notice that the contours all look the same, except
for the sheet they start on. ( As a first approximation the reader
can ignore the subtlety about the sheet number where the contour
starts.)
The sheets change in such a way that $\log Y_{2,s}(\zeta =
e^{\theta} ) \sim - { m_{2,s} e^{-\theta } \over 2 } $ with
$m_{2,s}$ real for our choice of the polynomial. Here $m_{2,s}$ are
basically the $Z_{2,s}$ up to possible factors of $e^{i \pi/4}$ (for
$s $ odd), see \ref{massZAdS5}. In other words, if we consider one of the contours in
figure \ref{Ytwocycles}, then the four different choices of the
sheet it starts on would give four possible cycles on the Riemann
surface. These would yield $Z_{2s},~i Z_{2,s}, - Z_{2,s}, -i
Z_{2,s}$. The asymptotic behavior of $Y_{2,s}$ for $\zeta$ small and
positive is governed by the cycle that renders $ Y_{2,s}$
exponentially suppressed.

\begin{figure}[htb]
\begin{center}
\includegraphics[angle=0,width=16cm]{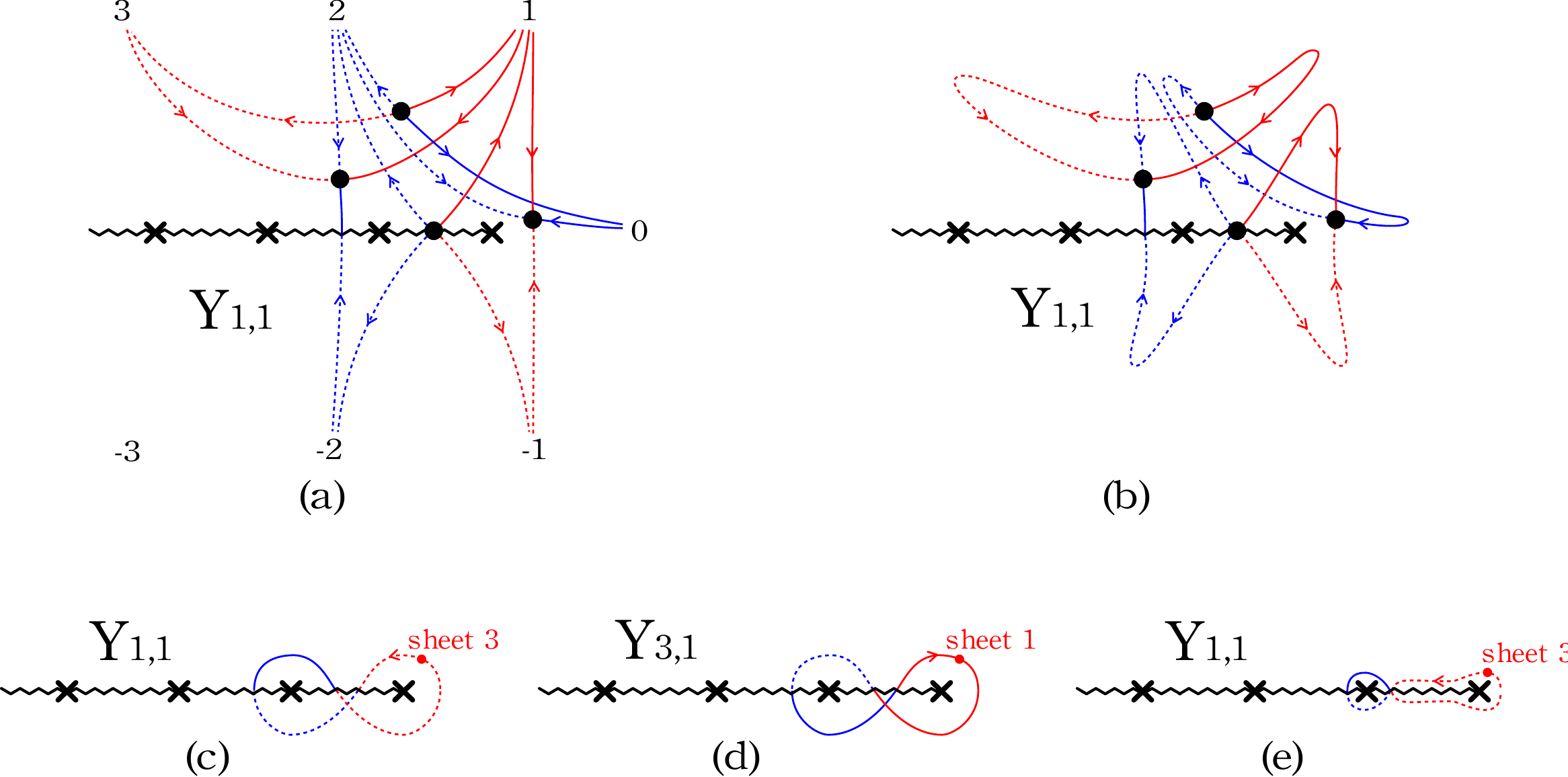}
\end{center}
\caption{ In (a) we show the WKB lines that are used for evaluating
$Y_{1,1}$. In (b) we have reconnected the lines. In (c) we get the
final cycle $\gamma_{1,1}$, which is the figure eight cycle
indicated. In (d) we see the cycle we would have obtained if we had
followed the same procedure for $Y_{3,1}$. We see that the cycles
for $Y_{1,s}$ and $Y_{3,s}$ differ by a change in orientation and by
a shift in sheet by two units. This does not affect the $\zeta$
dependent piece, but it changes the sign of the $\zeta$ independent
piece coming from the gauge connection $\alpha$. }
\label{WKBfiveYoneone}
\end{figure}

\begin{figure}[htb]
\begin{center}
\includegraphics[angle=0,width=8cm]{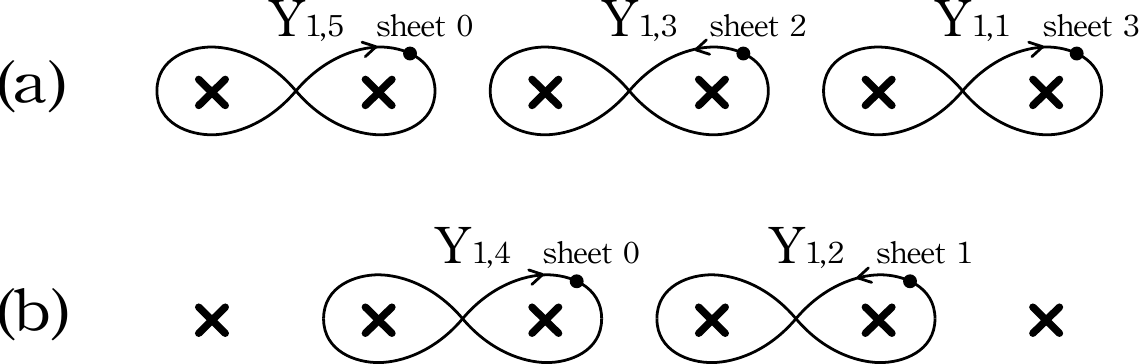}
\end{center}
\caption{ In (a) and (b) we show the cycles $\gamma_{1,s}$ that
correspond to the various $Y_{1,s}$ functions. We have indicated the
sheet where the point with the dot is. Of course, there are cuts
emanating from the zeros, etc. } \label{Yonecycles}
\end{figure}

We can now evaluate one of the $Y_{1,s}$. Let us evaluate, \beqa
Y_{1,1} = { \langle - 2 , { \bar 2 } \rangle \langle -1 , \bar 1
\rangle \over \langle -2,-1,1,2 \rangle \langle 0,1,2,3 \rangle } =
{ \langle -2,1,2,3 \rangle \langle -1,0,1,2 \rangle \over \langle
-2,-1,1,2 \rangle \langle 0,1,2,3 \rangle }\nn \eeqa The corresponding
lines are drawn in figure \ref{WKBfiveYoneone}(a). They can be
reconnected as in figure \ref{WKBfiveYoneone}(b). And finally lead
to the figure eight contour in figure \ref{WKBfiveYoneone}(c). It
can be seen that there is an $\alpha$ dependent contribution to this
contour. We can deform the contour to the one in figure
\ref{WKBfiveYoneone}(e). We see that the upper and lower parts of
the contours really add, due to the properties of $\alpha$ as we
cross a cut. Namely, $\alpha$ changes sign across the cut going
between the first and second zero.
This means that the upper and lower contributions of the contour add
up and give a (generically) non zero answer for $a_{a,s}$. Finally,
in figure \ref{Yonecycles} we draw the cycles $\gamma_{1,s}$ for
$s=1,\dots,5$. They all  have the same eight shape and differ by the
sheet, orientation and zeros they encircle.

We can similarly evaluate \beqa Y_{3,1} = { \langle - \bar 2 , { 2 }
\rangle \langle - \bar 1 , \bar 1 \rangle \over \langle -2,-1,1,2
\rangle \langle 0,1,2,3 \rangle } = { \langle -3,-2,-1, 2 \rangle
\langle -2,-1,0,1 \rangle \over \langle -2,-1,1,2 \rangle \langle
0,1,2,3 \rangle }\nn \eeqa Notice that this differs from $Y_{1,1}$ by
an overall bar. This exchanges the top of the diagram in figure \ref{Yonecycles}(a)
with the
bottom.
The final contour is shown in figure \ref{WKBfiveYoneone}(e). This
differs from the one for $Y_{1,1}$ in figure \ref{WKBfiveYoneone}(c)
by the orientation and a change of sheets by two units. This
combined operation does not change the integral of $ x dz$, which
contributes to the $1/\zeta $ term. However,  it does reverse the
sign of the contribution due to the gauge field $\alpha$.

Let us make a comment on the result for $Y_{a,s}$ as we vary $a$ for
a given $s$. Notice that $Y_{1,s}$ is given by a figure eight
contour, and so is $Y_{3,s}$, but with a different figure eight
contour, e.g. see figure \ref{WKBfiveYoneone}(c) and (d).
In principle, the Riemann surface has four figure eight
contours which start on different sheets. Our course, the sum is
zero. In addition, we can have the sum of two consecutive figure
eight contours. This sum can be deformed into the figure ``double
eight" contour that gives the result for $Y_{2,s}$. This implies the
pattern of masses that we get for $Y_{a,s}$ for a given
$s$.\footnote{ This is a result of the $Z_4$ symmetry of the Riemann
surface. If we were to break that symmetry we would no longer have
this pattern. }

In summary, we get \beqa \la{hYfin} \log \widehat Y_{a,s} = {
Z_{\gamma_{a,s}} \over \zeta} + ( a-2) c_{s} + o(\zeta ) ~;~~~~~~~
c_s = - \oint_{\gamma_{1,s}} \alpha \eeqa

We also have that \beqa \label{zreal} Z_{\gamma_{1,s} } =
Z_{\gamma_{3,s}} ~,~~~~~~~~~~~~ Z_{\gamma_{2,s} } = ( 1 - (-1)^s i )
Z_{\gamma_{1,s} } \eeqa The last equality follows from the fact that
the contours for $\widehat Y_{2,s}$ are a sum of two figure eight
contours on neighboring sheets. Once we take into account the shifts
in $\zeta$ in the definitions of $\widehat Y_{a,s}$ we see that
\nref{zreal} give the relations between the masses that we found in
the $Y$-system.

It is interesting to compute the intersection form of all the cycles
associated to the various Y functions. We get the result in figure
\ref{PB}. This result is used in section \ref{Areatofreeenergy}.
\begin{figure}[t]
\center\includegraphics[width=160mm]{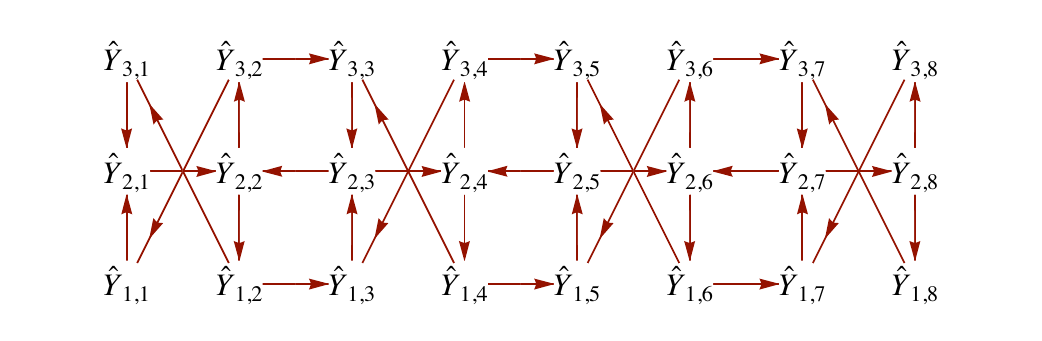} \caption{This figure
shows the intersection form for all the cycles associated to the $Y$
functions. If an arrow points from $Y_A$ to $Y_B$ we have $\langle
\gamma_{A}, \gamma_{B} \rangle =1$ otherwise the intersection
vanishes. } \la{PB}
\end{figure}

The analysis of the large $\zeta$ behavior is very similar to what
we did above. We now diagonalize $\Phi_{\bar z} \to \bar P(\bar
z)^{1/4} ( 1, i,-1,i ) $. We have the same pattern of WKB lines and
the same cycles associated to the $\widehat Y_{a,s}$. But now we
will be integrating $\bar x d \bar z$ to obtain $\bar
Z_{\gamma_{a,s}}$. In addition, there is a contribution from the
gauge connection $A$. In this case we can set $A_{z} =0$ and we
focus on the diagonal part of the $A_{\bar z}$ connection, which is
given in terms of a one form $\bar \alpha_{\bar z }$ on the Remann
surface. We have then an expansion of the form \beqa \log \widehat
Y_{\gamma_{a,s}} = { \bar Z_{\gamma_{a,s} } \zeta } + (a-2) \bar c_s
+ o(\zeta )\nn \eeqa

The constant parts $c_s$ in \nref{hYfin} and $\bar c_s$ are, in
principle, different. They are constrained by the $Y$- system
equation and we expect that, in the end, we have only one constant
per value of $s$ which is actually a free parameter. This is most
clear for low temperatures (or large values of the polynomial $P$).
In this case, we can approximately simultaneously diagonalize
$\Phi_z$ and $\Phi_{\bar z}$. As a result, $\alpha$ and $\bar
\alpha$ coincide and $c_s = \bar c_s$ (see next appendix). As
explained in the main text, in the general case, the average of
these constants is a free parameter while the difference is
determined by the equations.


Let us make a comment on the validity of the derivation of the WKB
analysis. In this case, a given WKB curve gives a good approximation
to the cross ratio in a region of size $\pi/2$ in $\zeta$ around the
value of $\zeta$ for which the WKB curve exists. This is good enough
for us to derive the $Y$-system, since we need displacements only of
$\pi/4$. Note that the WKB curves exist for a range of $\zeta $ of
total angle $\pi/2$ centered around the values where we did the
computation. Thus, the total validity of the WKB analysis is a range
of $\zeta $ of angle $\pi$ centered on real $\zeta$, for which we
did the analysis. These remarks are valid as long as the polynomial
has its zeros along the real axis.

When we move the zeros away from the real axis, the quantities
$Z_{a,s}$ become more general complex numbers, though still
constrained as in \nref{zreal}. Thus we have $2(n-5)$ complex
parameters which correspond to the motion of the zeros of the
polynomial. (There are other $(n-5)$ parameters in the constants
$c_s + \bar c_s$). Now, as we move the zeros of the polynomial we
will find that at some point some of the WKB lines will cease to
exist. However, having derived the integral equation in one region in
parameter space (i.e. for real masses), we can move to other regions by
analytic continuation. In doing so, we have to change the integral equation
according to an easily derived pattern, as explained in appendix \ref{WallCrossing}.

\section{Reality conditions for the $Y$ functions }
\label{realco}

The reality conditions for the $Y$ functions depend on the signature
in which the solutions are embedded. More precisely, for
$SO(p,6-p)$, they depend on the parity of $p$. For $p$ even, which
corresponds to a boundary of $AdS$ with $(3,1)$ or $(1,3)$
signature, the reality conditions in a suitable choice of gauge have the form
\beqa\la{conjugate} &&A_{\bar z}^\dagger =-B A_z B^{-1}
~,\qquad\Phi_{\bar z}^\dagger =B\Phi_z B^{-1} ~,~~~~~~~~~~ ({\cal
A}( \zeta))^{\dagger} = - B {\cal A}(- { 1 \over \zeta^*}) B^{-1}
\eeqa
For a suitable $B$ for each case. For $p$ odd, which for instance
corresponds to the case of $(2,2)$ signature, the reality conditions
have instead the form
\beqa\la{star} &&A_{\bar z}^* =B A_z B^{-1} ~,\qquad\Phi_{\bar z}^*
=B\Phi_z B^{-1} ~,~~~~~~~~~~ ({\cal A}( \zeta))^* = B {\cal A}({ 1
\over \zeta^*}) B^{-1}
\eeqa
For a suitable $B$ for each case. The difference between different
parities of $p$ comes from the fact that conjugation interchanges
the spinorial and anti-spinorial representations for $p$ even, but
it doesn't for $p$ odd. In addition, we also have the $Z_4$
projection condition
 \beqa \la{zfourcon} {\cal A}
(i\zeta) & = & - C {\cal A}^t(\zeta) C^{-1} ~,~~~~~~~~~~~{\cal A}(-
\zeta) = U^{-1} {\cal A}(\zeta) U ~,~~~~~~ U = C^t C^{-1} \eeqa
where the second equality arises by applying the first one twice.

Let us first focus in the case in which $p$ is even. If
$\psi(\zeta)$ is a solution to $(d + {\cal A}(\zeta) ) \psi =0$,
then by complex conjugating the above equation and using
\nref{conjugate} \nref{zfourcon} we get that \beqa ( d - { \cal A
}^t(1/\zeta^*) ) D (\psi(\zeta) )^* =0 ~,~~~~~~~~~~~ D = C^{-1} C^t
B^t\nn \eeqa This implies that $ D (\psi(\zeta) )^* \propto \bar
\psi(1/\zeta^*) $. If we take $\psi$ to be a small solution $s_k$,
then we see that $ (s_k(\zeta))^* \propto D^{-1} \bar
s_k(1/\zeta^*)$. Note that $D$ will drop out when we compute inner
products, and any constant of proportionality drops out when we
compute cross ratios. Thus the effect of performing a conjugation is
to replace $s_k$ by $\bar s_k$ in all formulas. This implies that
\beqa \(Y_{a,s}(\zeta)\)^*=Y_{4-a,s}(1/\zeta^*)\nn \eeqa In particular,
for large $\theta=\log \zeta$ we see that \beq
\[\log\(Y_{1,s}/Y_{3,s}\)(+\infty)\]^*=-\log\(Y_{1,s}/Y_{3,s}\)(-\infty)\nn
\eeq which implies that $c_s = - ( \bar c_s)^*$, or $C_s$ purely
imaginary and $D_s$ purely real.

For $p$ odd the situation is different. By complex conjugating $(d +
{\cal A}(\zeta) ) \psi(\zeta) =0$, and using (\ref{star}), we simply
obtain $(\psi(\zeta))^* \propto \psi(1/\zeta^*)$. This implies $
(s_k(\zeta))^* \propto s_k(1/\zeta^*)$ and hence
\beqa \(Y_{a,s}(\zeta)\)^*=Y_{a,s}(1/\zeta^*)\nn \eeqa
implying the $Y's$ are real when $\zeta$ is a phase. In this case
$c_s^* = \bar c_s$, and $C_s$ is real and $D_s$ is purely imaginary.

\section{Components of the full area} \la{areacomponents}

As seen in the body of the paper, in order to compute scattering
amplitudes at strong couplings, we need to compute the area of
minimal surfaces in $AdS$, given by
\begin{equation}
A=\int d^2z Tr\[\Phi_z \Phi_{\bar z} \] \nn
\end{equation}
Since for solutions relevant to scattering amplitudes, $Tr\[\Phi_z
\Phi_{\bar z} \] \sim (P \bar P)^{1/4}$, this area diverges and
needs to be regularized. This can be conveniently done by dividing
the area in different contributions (we refer the reader to
\cite{Alday:2009yn,Alday:2009dv} for the details). Below we give the
results for the case in which $n \neq 4k$, where the treatment is
simpler \footnote{The case $n=4k$ is subtle, since the $w-$plane
possesses a monodromy at infinity. This was explicitly treated in
\cite{Alday:2009yn} for $AdS_3$ kinematics. We expect similar
results for the $AdS_5$ case, but we have not worked out the
details. It would be interesting to work them out for $n=4 k$ in
$AdS_5$.}
\begin{eqnarray}
A&=&A_{reg}+A_{periods}+A_{cutoff} \nn  \\
A_{reg}&=&\int d^2z \left( Tr\[\Phi_z \Phi_{\bar z} \] -4(P \bar P)^{1/4} \right) \nn \\
A_{periods}&=&4 \int d^2z  (P \bar P)^{1/4} -4 \int_{\Sigma_0} d^2w=4 \int_{\Sigma} d^2w -4 \int_{\Sigma_0} d^2w \nn  \\
A_{cutoff}&=&4 \int_{\sigma_0,z_{AdS}>\epsilon}d^2w \nn
\end{eqnarray}
where we have defined the $w-$plane by $dw=P(z)^{1/4}dz$. $A_{reg}$
is the non-trivial function that is computed by the free energy of
the TBA equations and is the central quantity of this paper.

In $A_{periods}$, $\Sigma$ denotes the surface defined by the
polynomial $P(z)$, while $\Sigma_0$ is a reference surface with a
single branch point at the origin and the same structure at
infinity. When computing it, it is important to use a cut-off in the
$w-$plane, such as $|w| \leq \Lambda $, with $\Lambda \gg 1$. It can
be evaluated in terms of the periods of the Riemann surface
$x^4=P(z)$ and we give its explicit expression below.

When computing $A_{cut-off}$ we impose a physical cut-off, namely
that the radial $AdS$ coordinate should be larger that certain small
$\epsilon$. It can be conveniently written as the sum of two
contributions
\begin{equation}
A_{cut-off}=\frac{1}{8} \sum_i \log^2(\epsilon^2
\bx_{i,i+2}^2)+A_{BDS-like} \nn
\end{equation}
the first term is the standard divergent term, expected for
light-like Wilson loops/amplitudes \cite{Korchemskaya:1992je,BDS}.
The second term is
\begin{eqnarray}
A_{BDS-like}&=&-\frac{1}{8} \sum_{\ell=1}^n \left( \log^2
\bx_{i,i+2}^2
+ \sum_{k=0}^{2K} (-1)^{k+1} \log \bx_{i,i+2}^2 \log \bx_{i+2k+1,i+2k+3}^2 \right),~~~n=4K+2 \nonumber \\
&=& -\frac{1}{4} \sum_{\ell=1}^n \left( \log^2 \bx_{i,i+2}^2 +
\sum_{k=0}^{2K} (-1)^{k+1} \log \bx_{i,i+2}^2 \log
\bx_{i+2k+1,i+2k+3}^2 \right),~~~n=4K+2\pm 1 \nonumber
\end{eqnarray}
$A_{BDS-like}$ is a finite term which obeys the conformal Ward
identities of broken conformal invariance \cite{DrummondAU} .
Actually, it is the unique solution of the anomalous Ward identities
that can be written only in terms of next to nearest distances
$\bx_{i,i+2}^2$.

It is customary to subtract the one loop result $A_{BDS}$ written
down in \cite{BDS}, with the appropriate overall factor. As both,
$A_{BDS-like}$ and $A_{BDS}$ satisfy the Ward identities, their
difference is a function of the cross-ratios. In practise, in order
to express $A_{BDS}- A_{BDS-like}$ in terms of cross-ratios, one
simply starts with any non-next to nearest distance $\bx_{i,j}^2$
appearing in $A_{BDS}$ and write it in terms of the unique
cross-ratio $c_{i,j}$ which involves $\bx_{i,j}^2$ and next to
nearest distances $\bx_{i,i+2}^2$. The strong coupling answer,
however, organizes more naturally as described above.

\subsection{Expression for $A_{periods}$}
\la{AperiodsSec}

As already mentioned $A_{periods}$ can be evaluated in terms of the
periods of the Riemann surface $x^4=P(z)$. More precisely, it is
given by
\begin{equation}
\label{Aper} A_{periods}=-\frac{i}{2} w_{\gamma,\gamma'} Z_\gamma
\bar Z_{\gamma'}
\end{equation}
where $\gamma$ denotes the collective pair $(a,s)$ and
$w_{\gamma,\gamma'}$ is the inverse of the intersection form of the
cycles. Such inverse exists for the case in which $n$ is odd. Using
the relation between the different periods and the relation between
these and the masses
\begin{eqnarray}
Z_{1,s}=Z_{3,s} \equiv Z_s,~~~Z_{2,s}=(1-(-1)^s i)Z_s  \nn \\
\la{massZAdS5} m_{2s+1}=-2 Z_{2s+1},~~~~~m_{2s}=-2
e^{-\frac{i\pi}{4}} Z_{2s}
\end{eqnarray}
we can write $A_{periods}$ in terms of the masses of the TBA
equations \nref{Integraleq}. We obtain
\begin{eqnarray}
A_{periods}=\mathcal{K}_{ij}m_i \bar{m}_j \nn
\end{eqnarray}
The form of $\mathcal{K}$ depends on the parity of $(n-1)/2$. For instance,
for the case $n=4k+5$, $\mathcal{K}$ is a matrix of $4k$ by $4k$ and is equal
to
\begin{eqnarray}
\label{Kmatrix}
\mathcal{K}=1_k \otimes \mathcal{K}_1+ \mathcal{K}_2 \otimes \mathcal{K}_3+\mathcal{K}_2^T \otimes \mathcal{K}_3^T,~~~~~(\mathcal{K}_2)_{i,j}=\theta(j-i)(-1)^{i-j+1},~~~\theta(0)=0\\
\mathcal{K}_1 =-\frac{1}{\sqrt{2}}\left(  \begin{matrix} 0&
\frac{1}{2}&\frac{1}{\sqrt{2}}&\frac{1}{2} \cr
\frac{1}{2}&\frac{1}{\sqrt{2}}&1&\frac{1}{\sqrt{2}} \cr
\frac{1}{\sqrt{2}}&1&\frac{1}{\sqrt{2}}&\frac{1}{2}\cr
\frac{1}{2}&\frac{1}{\sqrt{2}}&\frac{1}{2}&0
\end{matrix}\right),~~~~ \mathcal{K}_3=\frac{1}{2\sqrt{2}} \left(
\begin{matrix} 0&1&\sqrt{2} &0 \\ 0 &\sqrt{2} &2&\sqrt{2} \\ 0&1
&\sqrt{2} &1 \\0&0&0&0 \end{matrix}\right) \nn
\end{eqnarray}
The case $n=4k+3$ has very similar expressions if we write $\mathcal{K}$ in
terms of a $4k$ by $4k$ matrix whose last two rows and columns have
to be chopped off. The expression coincides with that in
(\ref{Kmatrix}), with different matrices $\mathcal{K}_1$ and $\mathcal{K}_3$
\begin{eqnarray}
\mathcal{K}_1= \frac{1}{2\sqrt{2}} \left(\begin{matrix} \sqrt{2}&1&0&-1 \\ 1&\sqrt{2}&0&-\sqrt{2}\\
0&0&0&-1\\-1&-\sqrt{2}&-1&0 \end{matrix}\right),~~~~\mathcal{K}_3 =-\frac{1}{2\sqrt{2}}\left(\begin{matrix} \sqrt{2} &1&0&-1 \\
2 & \sqrt{2} &0&-\sqrt{2} \\ \sqrt{2} &1&0&-1 \\ 0&0&0&0
\end{matrix}\right) \nn
\end{eqnarray}
The case $n=4k+2$ can be regarded as a limit of the case $n=4k+3$,
in which we take one of the zeroes of the polynomial very far away.
The correct prescription is to start with $A_{periods}^{4k+3}$,
subtract $\frac{1}{2}(m_{4k-2} -\kappa)(\bar m_{4k-2}- \bar \kappa)$
and then take $|m_{4k-2}|$ very large. $\kappa$ is then uniquely
fixed by the requirement that linear divergences cancel. Again, the
result can be written in exactly the form (\ref{Kmatrix}), with
different matrices $\mathcal{K}_{1,3}$ and then chopping away the last three
rows and columns
\begin{eqnarray}
\mathcal{K}_1= -\frac{1}{4} \left(\begin{matrix} -1&0&1&\sqrt{2} \\ 0&0&\sqrt{2}&2 \\
1&\sqrt{2}&1&\sqrt{2}\\ \sqrt{2}&2& \sqrt{2}&0 \end{matrix}\right),~~~~\mathcal{K}_3 =-\frac{1}{4}\left(\begin{matrix} 1&0&-1&-\sqrt{2} \\
\sqrt{2} & 0 & -\sqrt{2}&-2 \\ 1 &0&-1&-\sqrt{2} \\ 0&0&0&0
\end{matrix}\right) \nn
\end{eqnarray}
In particular, it can be checked that we reproduce the correct
answer for the hexagon case \cite{Alday:2009dv}.

The result for $A_{periods}$ is particularly simple in the case of
$AdS_3$. It can be computed by using $A_{periods}^{AdS_3}=-i
w_{rs}Z^r \bar Z^s$, but in this case the inverse matrix $w_{r,s}$
is much simpler. For $n=2\hat n$, gluons, with $\hat n$ odd, and
defining $\tilde{m}_{2k}=m_{2k}$, $\tilde{m}_{1}=m_1$ and
$\tilde{m}_{2k+1}=m_{2k-1}+m_{2k+1}$ for $k=1,...$, we obtain
\begin{equation} \label{Aperiods3}
A_{periods}^{AdS_3}=-\frac{1}{4} \sum_{k=1}^{\frac{\hat n-3}{2}}
\left( \tilde{m}_{2k-1} \bar{\tilde{m}}_{2k}+\tilde m_{2k} \bar{
\tilde{m}}_{2k-1} \right),~~~~~
\end{equation}
where we have used the relation between the periods $Z^r$ and the
masses $m_r$ from section 3.

\section{Direct computation of the regularized area} \la{direct}

In this appendix we consider the regularized area $A_{reg}$ for
a particular class of regular polygons which can be embedded
either in $AdS_3$ or $AdS_4$ and correspond to special radially
symmetric solutions (in a sense which will be clear
momentarily). We will then compare such results with the answer
for the free energy of the Y-system in the high temperature
limit.

We will be interested in the case of $(2,2)$ signature, since
in this case we can embed both kind of solutions. Strings on
$AdS_4$ can be described in terms of the usual holomorphic
function $P(z)$ and two fields $\alpha$ and $\beta$. We can
choose a gauge in which the connection becomes
\begin{equation}
A_z=\frac{1}{4}\left(\begin{matrix} -\partial \alpha - \partial
\beta \sigma_3 & 0 \\ 0 & \partial \alpha - \partial \beta \sigma_3
\end{matrix}\right), ~~~~\Phi_z= -\left(\begin{matrix} 0 & e^{-1/2
\alpha} \sqrt{2} P(z)^{1/2} \left( \begin{smallmatrix} 0 & e^{-1/2
\beta}  \cr e^{1/2 \beta}  & 0   \end{smallmatrix} \right) \\
\frac{e^{1/2 \alpha}}{\sqrt{2}}  &0  \end{matrix}\right) \nn
\end{equation}
And $A_{\bar z}=-A_{z}^\dagger$, $\Phi_{\bar
z}=\Phi_z^\dagger$. We have written the connection in terms of
two by two blocks. We will consider a symmetric configuration
in which all the zeroes of $P(z)$ are together, namely the
holomorphic function is a homogeneous polynomial $P(z) \sim
z^{n-4}$. In this limit, all its periods, and hence the masses
entering the TBA equations, vanish. On the other hand, for this
particular case, it is consistent with the equations of motion
and boundary conditions to set $\alpha$ and $\beta$ to be
functions of the radial coordinate only, $|z|$ or $|w| \equiv
\rho$. It is convenient to write the equations they satisfy in
the $w-$plane, defined by $dw=P(z)^{1/4}dz$
\begin{eqnarray*}
\hat \alpha''(\rho)+\frac{\hat \alpha'(\rho)}{\rho}-8e^{\hat \alpha}+8 e^{-\hat \alpha} \cosh \beta=0\\
\beta''(\rho)+\frac{\beta'(\rho)}{\rho}-8 e^{-\hat \alpha} \sinh
\beta=0
\end{eqnarray*}
where we have defined the shifted field $\hat
\alpha=\alpha-\frac{1}{2} \log |P(z)|-\log 2$. For the
scattering of $n$ gluons we will consider two different
configurations. The first configuration, which exists only for
$n$ even, corresponds to regular polygons that can be embedded
into $AdS_3$ and hence $\beta=0$ for these. Geometrically, the
span a regular polygon of $n$ sides in the $(x_1,x_2)$ plane
and a segment in the $(t_1,t_2)$ plane, see figure
\ref{twopolygons} (a). They have been considered
\cite{Alday:2009yn} and their area has already been computed
there.

The second configuration corresponds to regular polygons that
can be embedded into $AdS_4$. Geometrically, they have the
shape of a regular polygon of $n$ sides in the $(x_1,x_2)$
plane and a regular polygon in the $(t_1,t_2)$ plane, of $n/2$
sides if $n$ is even, or $n$ sides if $n$ is odd, see figure
\ref{twopolygons} (b).
 \begin{figure}[t]
\center\includegraphics[width=70mm]{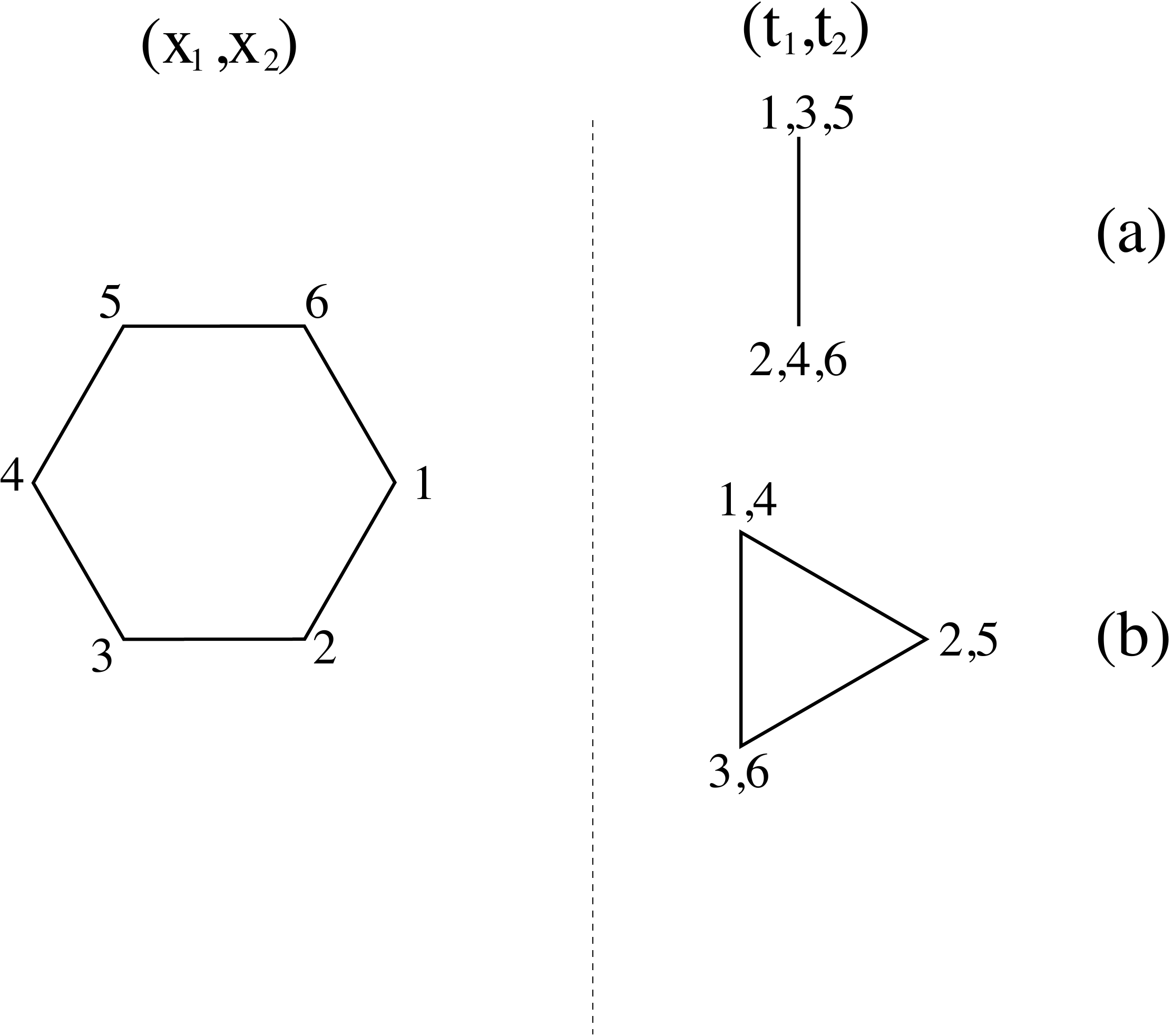} \caption{Two
classes of regular polygons which live in $(2,2)$ signature and can
be embedded in $AdS_3$ (a) and $AdS_4$ (b). Both posses a $Z_n$
symmetry.}
 \la{twopolygons}
\end{figure}
As already mentioned, we expect $\hat \alpha$ and $\beta$ to be
radially symmetric. Furthermore, following the discussion in
appendix B of \cite{Alday:2009dv}, we expect the following
boundary conditions at the origin
\begin{eqnarray*}
\hat \alpha=2\frac{n-4}{n} \log \rho+c_\alpha+... \\
\beta=4\frac{n-4}{n} \log \rho+c_\beta+...
\end{eqnarray*}
and a exponential decay at infinity. Finally, we are interested
in the regularized area:
\begin{eqnarray*}
A_{reg}=2\pi n \int \rho d\rho(e^{\hat \alpha}-1)
\end{eqnarray*}
It is at present unknown how to compute analytically the above
area. Based on our experience with the first class of regular
polygons, we expect a factor of $\pi$, times a simple rational
function of $n$. Furthermore, this expression should vanish for
$n=4$ and, in addition, we also know the correct value for
$n=6$ from the results of \cite{Alday:2009dv}.

We have solved the above equations numerically. The constants
$c_\alpha$ and $c_\beta$ are unknown and are fixed by the
constraint that the solutions decay at infinity. One should
solve a shooting problem, trying several values for these
constants approaching the correct ones by requiring that the
solutions decay exponentially at infinity. We performed such
procedure for $n=7,8$ and $n=10$ and obtained the following
numerical results
\begin{equation}
A_{reg}^{n=7} \approx 4.882,~~~~~A_{reg}^{n=8} \approx
7.067,~~~~~A_{reg}^{n=10} \approx 11.85 \nn
\end{equation}
Note that these expressions are well approximated by
$\frac{87}{56}\pi,\frac{9}{4}\pi$ and $\frac{15}{4}\pi$. Once
the regularized area is computed, in order to compare it to the
free energy, we need to subtract an $n-$dependent constant,
ensuring that the free energy vanishes for the situation in
which all the zeroes are well separated. In this case, the
regularized area approaches the quantity of zeroes, {\it i.e.}
 $n-4$, times the contribution of the regular pentagon, which corresponds to a single zero.

Extracting from \cite{Alday:2009dv} the value of the
regularized area for the regular pentagon,
$A_{penta}=\frac{3}{8}\pi$, we can give the final formula for
the direct computation of the free energy
\begin{eqnarray}
\label{Afree} A_{free}=A_{reg}-(n-4)\frac{3}{8}\pi
\end{eqnarray}
From the numerical computations we can guess a very simple
final formula for the regular polygons of the second class:
\begin{eqnarray}
\label{Afreetwo} A_{free}^{(2)}=\frac{1}{2}\frac{(n-4)(n-5)}{n} \pi
\end{eqnarray}
This vanishes for $n=4$, as expected, gives the correct result
for $n=6$ and agrees very well with the numerics for
$n=7,8,10$.

Let us now plug in (\ref{Afree}) the results for the regular
polygons that can be embedded into $AdS_3$, read off from
\cite{Alday:2009yn}. Substracting the appropriate contribution
we obtain the appropriate quantity to be compared with the free
energy for regular polygons of the first class:
\begin{eqnarray}
A^{(1)}_{free } &=& A_{sinh} - (n-4) A_{pentagon}
 \label{threefromfive}
= - \pi \frac{n-4}{2 n}\
\end{eqnarray}
where $A_{Sinh}$ is the result in \cite{Alday:2009yn}. We see
that \nref{threefromfive}
  vanishes for $n=4$, as
expected, and also agrees with the answer of
\cite{Alday:2009dv} for $n=6$.

\subsection*{$AdS_3$ limit}

In the $AdS_3$ limit the expression for the regularized area of
regular polygons is of course the same as for the polygons we
were discussing above (which could be embedded $AdS_3$).
However, in order to make a direct comparison with the free
energy we need to substract a different contribution.
Considering $ n = 2\hat n$ gluons, the holomorphic polynomial
has $\hat n-2$ zeroes,
  each of which gives a contribution equal to that of the regular hexagon, we obtain:
\begin{eqnarray}
A^{AdS_3}_{free} &=& A_{sinh}-(\hat n-2)\frac{7}{12}\pi= \pi { (n-6)
(n-4) \over 12 n}  \label{threefromthree}
\end{eqnarray}
Note that (\ref{threefromthree}) is different from what we
would obtain from the free energy in the $AdS_5$ case
(\ref{threefromfive}), due to the fact that we subtract
different contributions.

\section{Regular polygons}
\la{RegularPolygons}

In the body of the paper we have solved the TBA equations in
the high-temperature limit, in which constant solutions of the
Y-system are relevant. In this limit, and for an even number of
sides, there is a family of solutions parametrized by a single
parameter $\mu$. The solution interpolates between the two kind
of regular polygons described in appendix \ref{direct}. In the
following we describe the geometrical picture of such polygons.
As we will see, this allows to compute analytic expressions for
the various cross-ratios. These expressions can then be
compared to the results obtained from the Y-system equations
testing both, the geometrical picture and the Y-system
equations.

In addition, when the number of sides is odd, there is a
discrete family of regular polygons that can be embedded into
the boundary of $AdS_4$. We present them in the second
subsection of this appendix, and compare their cross-ratios
with the respective solutions from the Y-system. Finding again
agreement.

\subsection*{Regular polygons with an even number of sides}

In appendix \ref{direct} we have seen that there are two
families of regular polygons with an even number of sides,
which can be embedded into the boundary of $AdS_3$ and the
boundary of
 $AdS_4$ respectively. It is possible to construct a
family of polygons that interpolates between these two in the
boundary of $AdS_5$. For that it is convenient to describe the
boundary of $AdS_5$ in terms of projective coordinates with
$(2,4)$ signature \footnote{These polygons live in a boundary
with $(1,3)$ signature.}.
\begin{equation} \la{zerodistance}
-{\cal Z}_1 \bar{\cal Z}_1-{\cal Z}_2 \bar{\cal Z}_2+{\cal Z}_3
\bar{\cal Z}_3=0
\end{equation}
We propose that the location of the cusps of the polygons under
consideration is
\begin{equation}
\label{cusppos} {\cal Z}_1^p= (-1)^p \ell_1+i~ ,\qquad{\cal
Z}_2^p=\ell_2  e^{\frac{4\pi i p}{n}}~ ,\qquad{\cal Z}_3^p=\ell_3
e^{\frac{2\pi i p}{n}}
\end{equation}
where $\ell_{1,2,3}$ have to be chosen in such a way that
\nref{zerodistance} is obeyed and
  the distance between two
consecutive points is light-like, namely
\begin{equation}
\label{const}
1+\ell_1^2-\ell_3^2+\ell_2^2=0~,\qquad-\ell_1^2+\sin^2\left(\frac{2\pi}{n}
\right) \ell_2^2-\sin^2\left(\frac{\pi}{n} \right) \ell_3^2=0
\end{equation}
This leave us with one free parameter which we can parametrize
in terms of an angle $\phi$ as
\begin{equation}
\ell_1 = \tan \left(\frac{\pi}{n}\right) \tan
\left(\frac{2\pi}{n}\right) \tan{\left(\frac{\phi}{n} \right)} \nn
\end{equation}
where $\phi$ runs from zero to $\frac{n-4}{2}\pi$. This
parameterization is engineered in order to ease the comparison
with the results in the main text. As explained below the
formal monodromy $\mu$ appearing in the main text is given by
$\mu=e^{i \phi}$.

We can interpret the solution as in $(1,3)$ signature, with the
points living on a spatial $S^1$ times a temporal $S^3$. When
$\phi=0$, $\ell_2$ takes its maximal value and $\ell_1=0$. All
the points on the $S^2 \subset S^3$ are located on its equator,
forming a regular polygon of $n/2$ sides, and we recover the
regular polygons embedded into the boundary of  $AdS_4$. When
$\phi=\frac{n-4}{2}\pi$, $\ell_2$ vanishes and $\ell_1$ takes
its maximal value, so the points in the $S^2 \subset S^3$
alternate between the south and north-pole and the solution
reduces to the usual regular polygon that can be embedded into
the boundary of $AdS_3$. For intermediate values of $\phi$, we
interpolate between the two solutions and the corresponding
minimal surface span the full $AdS_5$ \footnote{The full
$AdS_5$ is spanned since the radius of the temporal $S^2$
changes as we go between the two endpoints of the
interpolation.}. Furthermore, note that the solution possesses
a $Z_n$ symmetry. \footnote{More precisely, two consecutive
points are related by an $O(6)$ (but not $SO(6)$) rotation
$X^{i+1}=R X^i$, where $R$ does not depend on $i$.}

In order to compute the cross-ratios corresponding to this
solution, and the relation between $\mu$ used in the body of
the paper and the parametrization used here, it is convenient
to compute the spinors $\lambda_i$ corresponding to this
solution. In other words, we need to give a description of the
solution in terms of twistors. The space-time coordinates can
be parametrised in the spinorial representation as follows
\begin{equation}
X^i_{\alpha,\beta}=\left( \begin{matrix} 0& {\cal Z}_1 & {\cal Z}_2 & {\cal Z}_3 \\
-{\cal Z}_1 & 0 & \bar {\cal Z}_3 & \bar {\cal Z}_2 \\
-{\cal Z}_2 & -\bar {\cal Z}_3 & 0&- \bar{\cal Z}_1 \\
-{\cal Z}_3 & -\bar{\cal Z}_2& \bar{\cal Z}_1 & 0 \end{matrix}
\right) \nn
\end{equation}
we will denote by $\alpha,\beta=1,...,4$ the spinorial indices.
The points $X^i$ satisfy the following conditions
\begin{equation}
\epsilon^{\alpha \beta \gamma \delta} X^i_{\alpha \beta} X^i_{\gamma
\delta}=0~,\qquad\epsilon^{\alpha \beta \gamma \delta} X^i_{\alpha
\beta} X^{i+1}_{\gamma \delta}=0 \nn
\end{equation}
As a consequence, they can be written in terms of spinors
$\lambda^i_\alpha$, such that
\begin{equation}
X^i_{\alpha \beta} =\lambda^{i-1}_\alpha
\lambda^{i}_\beta-\lambda^{i}_\alpha \lambda^{i-1}_\beta \nn
\end{equation}
\begin{figure}[t]
\center\includegraphics[width=70mm]{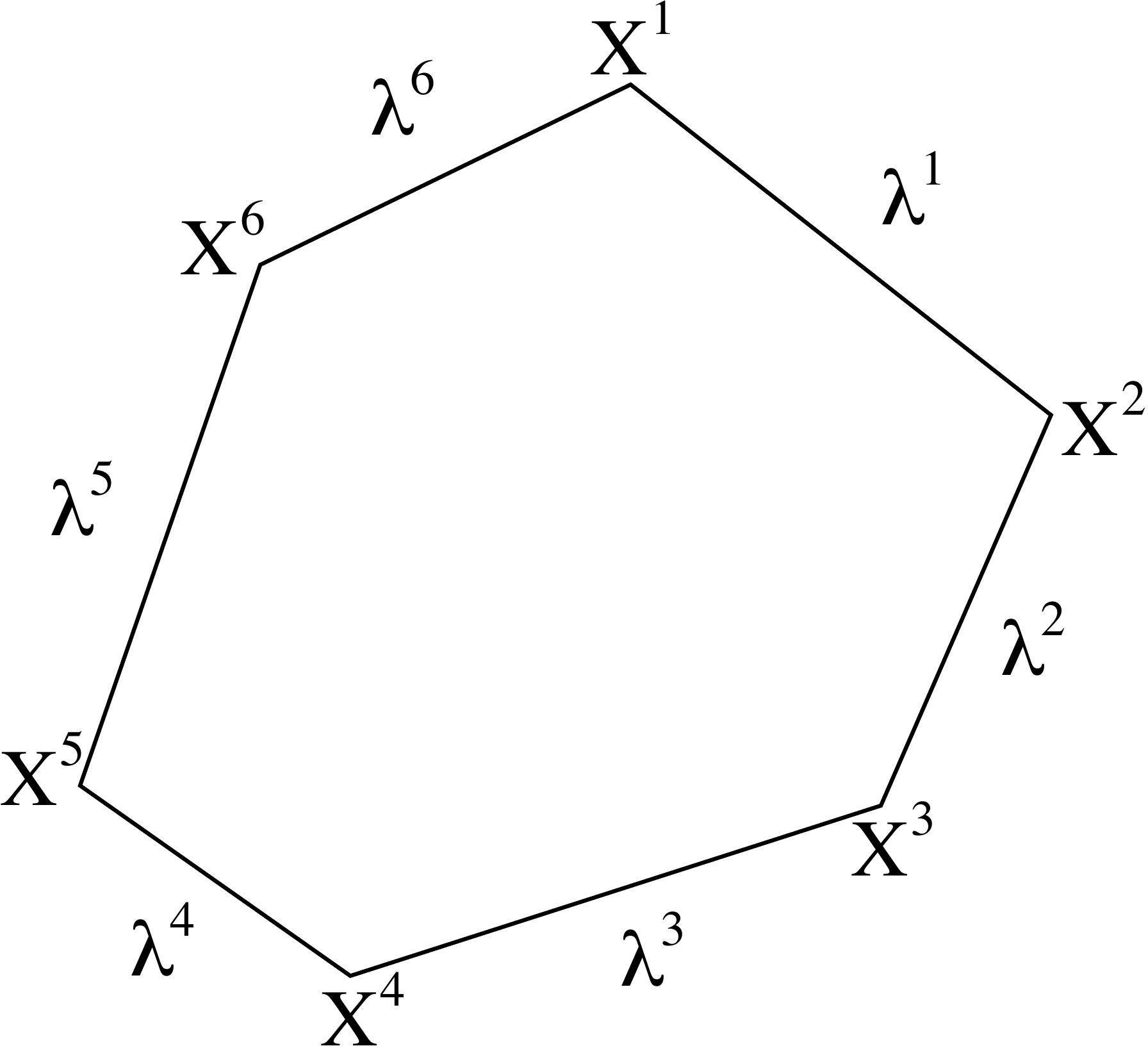} \caption{We can
associate a twistor $\lambda^i$ to each edge of a null polygon, such
that the position of the cusps is given by $\lambda^{i-1} \lambda^i$
} \la{twistors}
\end{figure}
See figure (\ref{twistors}). There is a simple recipe to
determine the spinor $\lambda^i$, since we know it has to
satisfy the following equations
\begin{equation}
\epsilon^{\alpha \beta \gamma \delta} X^i_{\beta \gamma }
\lambda^i_\delta=0~,\qquad\epsilon^{\alpha \beta \gamma \delta}
X^{i+1}_{\beta \gamma } \lambda^i_\delta=0 \nn
\end{equation}
given the points $X^i$ and $X^{i+1}$, each of the equations
above specifies a two dimensional plane and the intersection of
these planes give us $\lambda^i$. Hence $\lambda^i$ will be
completely specified by the above equations up to an overall
normalization factor. Of course, these factors are inessential,
as they will drop out from the computation of any cross-ratio.
We obtain the relatively simple result
\beq 
\lambda^k \propto \{ \kappa_1 \(1- a_k \) (1- b_k)
e^{\frac{3 i \pi k}{n}}, \kappa_2  e^{-\frac{3 i \pi k}{n}},
\kappa_3 (1- a_k) e^{ \frac{i \pi k}{n}}, (1- b_k) e^{- \frac{i
\pi  k}{n}}\} \nn 
\eeq 
where
$$
a_k= (-1)^k \tan  \frac{2\pi}{n}  \tan \frac{\phi}{n}~,\qquad b_k= (-1)^k \tan  \frac{\pi}{n}  \tan \frac{\phi}{n}\,,$$
and $\kappa_i$ are some $k$-independent constants. The spinors
posses a $Z_{n/2}$ symmetry $\lambda^{i+2} \propto U \lambda^i$
with $U=diag(e^{\frac{6 \pi i}{n}},e^{-\frac{6 \pi
i}{n}},e^{\frac{2 \pi i}{n}},e^{-\frac{2 \pi i}{n}})$. Having
the spinors we can compute the invariants
\begin{equation}
\<i,j,k,l\>=\epsilon^{\alpha \beta \gamma \delta} \lambda_\alpha^i
\lambda^j_\beta \lambda^k_\gamma \lambda^l_\delta
\end{equation}
In order to compute the invariants entering in the cross-ratios
$Y_{s,m}$, it is sometimes convenient to have $\bar
\lambda_\alpha^k=\epsilon^{\alpha \beta \gamma \delta}
\lambda_\beta^{k-1} \lambda_\gamma^k \lambda_\delta^{k+1}$. We
obtain
\beq \bar \lambda^k \propto \{\bar \kappa_1  e^{-\frac{3 i \pi
k}{n}}, \bar \kappa_2 \(1+ a_k \) (1+ b_k) e^{\frac{3 i \pi
k}{n}}, \bar \kappa_3(1+ b_k) e^{- \frac{i \pi  k}{n}}, (1+
a_k) e^{ \frac{i \pi k}{n}} \} \nn \eeq
For some constants $\bar \kappa_i$. Finally, using the explicit
expressions for the cross-ratios $Y_{s,m}$ in terms of
invariants, we can obtain analytic expressions for them. These
are ratios of four such terms such that any $\lambda_i$ (or
$\bar \lambda_i$) will appear as many times in the denominator
as in the numerator. In particular, such cross-ratios do not
depend on the $k$ independent constants $\kappa_i,\bar
\kappa_i$. In other words, these constants can be set to one by
a conformal transformation. Re-shuffling the elements of the
resulting spinors $\bar \lambda_i$, we see that they precisely
agree with the ones used to write the solution in the high
temperature limit of the $AdS_5$ Y-system, equation
(\ref{twistor}).

In general the expressions for the $Y$-functions are not
particularly illuminating. In some cases there is significant
simplification. For instance, for the case $n=12$, it greatly
simplifies and we obtain
\begin{equation}
Y_{1,3}^{n=12}=\frac{1}{\sqrt{3}}+\frac{2}{3}(3+\sqrt{3})e^{-i
\phi}\cos\frac{\phi}{3} \nn
\end{equation}
%

Before proceeding, let us note the following   fact. Taking
$\phi \rightarrow -\phi$ effectively interchanges $\lambda
\leftrightarrow \bar \lambda$. From the geometrical point of
view, this simply amounts to take ${\cal Z}_1 \rightarrow -\bar
{\cal Z}_1$ (keeping ${\cal Z}_{2,3}$ fixed ) and is a symmetry
of our configuration. From the point of view of the Y-system,
it interchanges $Y_{1,s} \leftrightarrow Y_{3,s}$ and takes
$\mu \rightarrow \mu^{-1}$, which is also a symmetry of the
equations.

\subsection*{Regular polygons with an odd number of sides}

For an odd number of sides there exist discrete families of
regular polygons that can be embedded into the boundary of
$AdS_4$. These polygons can be conveniently written using
$(2,2)$ signature and are parametrized by the number of sides
$n$ and an extra parameter $r=2,...,(n-1)/2$. The cusps are
located at
\begin{equation}
{\cal Z}_1 = i ~,~~~~~~~~~{\cal Z}_2 = \ell_2 e^{2 \pi i r \over n } ~,~~~~~~~
{\cal Z}_3 = \ell_3 e^{ 2 \pi i \over n } \nn
\end{equation}
$\ell_2$ and $\ell_3$ have to be fixed in such a way that the
distance between consecutive cusps is light-like, and to obey
\nref{zerodistance}.

One can proceed along the lines of the previous subsection,
going to projective coordinates and computing the spinors. The
expression for the spinors are not particularly illuminating,
however, they satisfy the following relation
\begin{equation}
\lambda^{i+1}={\rm diag}(e^{\frac{2\pi i r}{n}},e^{-\frac{2\pi i
r}{n}},e^{\frac{2\pi i (r-1)}{n}},1) \lambda^i \nn
\end{equation}
up to arbitrary constants. This relation, allows  us to write
all the spinors in terms $\lambda^0$, which can be set to
$\lambda^0=(1,1,1,1)$ by conformal transformations (this is
analogous to setting $\kappa_i=1$ in the previous subsection).
In addition, we multiply each spinor by an overall phase $e^{-i
\pi(r-1)\frac{k}{n}}$. We arrive to the very simple result
\begin{equation}
\lambda^{k}=(e^{i \pi(r+1)\frac{k}{n}},e^{-i
\pi(r+1)\frac{k}{n}},e^{i \pi(r-1)\frac{k}{n}},e^{-i
\pi(r-1)\frac{k}{n}}) \nn
\end{equation}

Furthermore, one finds that $\lambda$ and $\bar \lambda$ agree
(up to a multiplication by a constant matrix). The computation
of cross-ratios is now straightforward and one can check that
these are indeed solutions of the Y-system equations.

\end{document}